\documentclass[prd,onecolumn,showpacs,11pt]{revtex4-1}
\usepackage[ngerman]{babel}
\usepackage[utf8]{inputenc}
\usepackage{amsmath, amssymb, amsfonts}
\usepackage{natbib}
\usepackage{hyperref}
\usepackage{color}
\hypersetup{colorlinks=true}

\begin{document}
\title{Holography as a highly efficient RG flow II: An explicit construction}
\author{Nicolas Behr}
\email{nicolas.behr@ed.ac.uk}
\affiliation{LFCS, University of Edinburgh, Informatics Forum, 10 Crichton Street, Edinburgh, EH8 9AB, Scotland, UK}
\affiliation{Maxwell Institute for Mathematical Sciences, Edinburgh, UK}
\author{Ayan Mukhopadhyay}
\email{ayan@hep.itp.tuwien.ac.at}
\affiliation{Institut f\"ur Theoretische Physik, Technische Universit\"at Wien,
        Wiedner Hauptstrasse 8-10, A-1040 Vienna, Austria}

\begin{abstract} 
We complete the reformulation of the holographic correspondence as a \emph{highly efficient RG flow} that can also determine the UV data in the field theory in the strong coupling and large $N$ limit. We introduce a special way to define operators at any given scale in terms of appropriate coarse-grained collective variables, without requiring the use of the elementary fields. The Wilsonian construction is generalised by promoting the cut-off to a functional of these collective variables. We impose three criteria to determine the coarse-graining. The first criterion is that the effective Ward identities for local conservation of energy, momentum, etc. should preserve their standard forms, but in new scale-dependent background metric and sources which are functionals of the effective single trace operators. The second criterion is that the scale-evolution equations of the operators in the actual background metric should be state-independent, implying that the collective variables should not explicitly appear in them. The final required criterion is that the endpoint of the scale-evolution of the RG flow can be transformed to a fixed point corresponding to familiar non-relativistic equations with a finite number of parameters, such as incompressible non-relativistic Navier-Stokes, under a certain universal rescaling of the scale and of the time coordinate. Using previous work, we explicitly show that in the hydrodynamic limit each such highly efficient RG flow reproduces a unique classical gravity theory with precise UV data that satisfy our IR criterion and also lead to regular horizons in the dual geometries. We obtain the explicit coarse-graining which reproduces Einstein's equations. In a simple example, we are also able to construct a low energy effective action and compute the beta function. Finally, we show how our construction can be interpolated with the traditional Wilsonian RG flow at a suitable scale, and can be used to develop new non-perturbative frameworks for QCD-like theories.
\end{abstract}
\maketitle
\tableofcontents
\section{Introduction}
Understanding the holographic duality between field theory and gravity \cite{Aharony:1999ti}, by deciphering precisely how the radial direction of the emergent geometry in one higher dimension captures the effective information about the dual field theory across various energy scales, is certainly crucial for obtaining a concrete formulation of the emergence of spacetime. Through this understanding, for instance, we may be able to decode the quantum dynamics of the horizon from the infrared behaviour of the dual field theory. 

This reconstruction is also important for another reason, namely to learn how to develop new effective frameworks for quantum many body systems, which can generalise effective quantum field theory, and which can describe both perturbative and non-perturbative aspects of the full dynamics. Such frameworks are also necessary to understand effective physics at intermediate energy scales, which are far away from any ultraviolet or infrared fixed point. The challenge is thus to also go beyond well-established examples of the holographic duality, and learn for instance, how to integrate both weak and strong coupling aspects of Quantum Chromodynamics (QCD) at ultraviolet (UV) and infrared (IR) energy scales, respectively. Ultimately one would aim to build a full non-perturbative framework valid at all energy scales, which unlike the lattice \cite{Wilson:1974sk} can allow us to calculate real-time physics directly. Some recent works that have inspired research in this direction include \cite{Heemskerk:2010hk,Balasubramanian:2012hb,Faulkner:2010jy,Lee:2013dln,Lizana:2015hqb}.

In the present work, we complete our reconstruction of the holographic duality as a precise kind of renormalization group (RG) flow  that defines a \textit{constructive field theory} at \textit{strong coupling} and \textit{large $N$}, at least in a special sector of states. In particular, we are able to establish a constructive field theory approach via using a RG flow construction, which defines the field theory at \textit{all} energy scales, and then demonstrate that our construction reproduces the traditional holographic correspondence, so that we can reconstruct the equations of a dual classical gravity theory in certain limits. Our RG flow construction generalises Wilsonian RG, in a manner such that it can have an endpoint at a \textit{finite} scale. By demanding that this endpoint can be transformed to a \textit{simple} non-relativistic endpoint, where the dynamics is given by a \textit{finite} number of parameters (under a universal rescaling of the scale and the time coordinate), we obtain a powerful principle that can determine allowed UV data of the field theory. We explicitly demonstrate here that our constructive field theory approach reproduces the UV data of the holographic duality, which is otherwise determined by the gravity equations via the criterion of regularity of the future horizon. As our RG flow can self-determine the UV data of the field theory, we call it \textit{highly efficient RG flow}.

Our reconstruction of holography as a highly efficient RG flow leads to a natural generalisation as we can build new non-perturbative frameworks for quantum field theories, which could be asymptotically free like QCD, at least in the large $N$ limit. We also outline here how our approach can be generalised to derive such non-perturbative \textit{semi-holographic} frameworks, in which the UV dynamics is described by perturbative methods, the IR dynamics is described by a precise holographic theory, and the interactions between the UV and IR sectors are determined completely by how the Wilsonian RG flow (that captures effective perturbative dynamics for the hard sector) can be interpolated with the highly efficient RG flow (that reconstructs the holographic dynamics for the soft sector) at a suitable intermediate scale.

The \textit{highly efficient RG flow} as distinct from usual Wilsonian RG flow based on integrating out high energy modes can also be motivated as the right type of RG flow to consider for studying time-dependent quantum dynamics. In this case, we would like to average out the features at short time-scales and construct a theory for the coarse-grained time-dependence of observables. What we can directly measure are certain collective variables like the hydrodynamic variables, relaxation modes, etc. and their spectral components, statistical fluctuations and other correlations. In the highly efficient RG flow, we will define quantum operators in terms of these directly measurable collective variables instead of elementary fields -- this is possible for strongly coupled large $N$ theories. The crucial idea behind the construction of the highly efficient RG flow is that the \textit{coarse-grained quantum operators and their local Ward identities assume the same functional form when defined in terms of the coarse-grained collective variables at each scale  in new effective background metrics and sources -- these functional forms are also independent of the specific quantum state being measured}. Furthermore, \textit{the effective background metric and sources are also state-independent functionals of the coarse-grained operators at each scale}. In order to preserve the local Ward identities like energy-momentum conservation, we need to redefine the background -- otherwise it is not possible to capture the effect of high energy modes which should provide driving forces on the soft modes violating the conservation of the energy-momentum tensor projected on the soft sector. The quantum dynamics at each scale gets described by the same set of parameters (like transport coefficients in hydrodynamic limit). It will also be seen that \textit{the number of effective parameters we need to describe low energy physics gets reduced as we run to lower energies} and {at the endpoint of the RG flow we are left with only finite number of parameters}. 

We will focus directly on local operators and their equations of motion, therefore the notion of low energy effective actions and beta functions of couplings should rather be a derived concept. At least for some simple examples, we will be also able to define the analogue of low energy effective action and beta functions of couplings, where the spacetime dependent coupling itself acts as a collective variable parametrising a condensate. The running of the effective parameters as described above however is more crucial to highly efficient RG flow. A more detailed relation with the Wilsonian picture is desirable and will be left to future work.

We summarise our construction briefly below. Before that, we recall that Wilsonian RG flow can be thought of as a method of defining effective composite operators in a \textit{state-independent} manner, allowing us to obtain various observables of the theory at any given scale of observation. In QCD, for instance, the effective energy-momentum tensor operator $t^\mu_{\phantom{\mu}\nu}(\Lambda)$ at the scale of observation $\Lambda$ can be defined via Wilsonian RG flow as a sum of various gauge-invariant composite operators of the asymptotically free theory in the deep UV, with coefficients that are functions of $\Lambda_{\rm QCD}/\Lambda$. This definition is independent of the state and has a systematic expansion in $\Lambda_{\rm QCD}/\Lambda$ for $\Lambda \gg \Lambda_{\rm QCD}$. Indeed, this is how the Wilsonian RG flow allows us to calculate the operator algebra, i.e. the operator product expansion of the effective operators at scale $\Lambda$, in terms of the algebra of the operators in the asymptotically free limit in the deep UV, which thus define an effective theory at scale $\Lambda$. It is therefore expected that any construction that generalises the Wilsonian RG flow should also be able to define the operators that determine the observables in the theory, in a \textit{state-independent manner} at \textit{any} scale of observation.

It is indeed a tremendous challenge to establish such a constructive approach at strong coupling, because the usual description of the dynamics in terms of elementary fields is not possible using currently available methods. The first idea, which we pursue here, is to use the collective variables, which parametrise the expectation values of gauge-invariant operators in all  states in the Hilbert space, to define the coarse-grained effective dynamics across various energy scales. 

In particular, we assume that there exists a subspace of the Hilbert space in the strong coupling and large $N$ limit where the only single-trace operator which takes an \textit{non-trivial} expectation value is the energy momentum tensor $t^\mu_{\phantom{\mu}\nu}$ (we can justify this assumption a posteriori, and we also show how we can go beyond this sector of states). Denoting the microscopic operator (to be coarse-grained) by ${t^\mu_{\phantom{\mu}\nu}}^\infty$, we introduce collective variables ${X^A}^\infty$ and parameters ${g^B}^\infty$ which can parametrise the expectation values of ${t^\mu_{\phantom{\mu}\nu}}^\infty$, so that the expectation value of ${t^\mu_{\phantom{\mu}\nu}}^\infty$ is captured by the function $\langle{t^\mu_{\phantom{\mu}\nu}}^\infty\rangle [{X^A}^\infty,{g^B}^\infty]$. We then define coarse-grained collective variables:
\begin{equation}
X^A(\Lambda, x) = \int {\rm d^d}k \, e^{ik\cdot x} \, \Theta\left(1 - \frac{k^2}{\Lambda^2}\right) F^{AB}\left(X^C(\Lambda, x), k\right)\tilde{X}^{B\infty} (k),
\end{equation}
where the cut-off in momentum space has been promoted to a function of the coarse-grained variables via the functionals $F^{AB}\left(X^C(\Lambda, x), k\right)$, so that the above equations determine the coarse-grained collective variables $X^A(\Lambda)$ recursively. The purpose of the coarse-graining is to define the coarse-grained operator $t^\mu_{\phantom{\mu}\nu}(\Lambda)$ via its expectation value, which can be declared to be the function $\langle t^\mu_{\phantom{\mu}\nu}\rangle [X^A(\Lambda,x),g^B(\Lambda)]$. This is possible if we can determine the coarse-graining functional $F^{AB}\left(X^C(\Lambda, x), k\right)$ and the parameters $g^B(\Lambda)$ appropriately. We use three criteria to determine these:
\begin{enumerate}
\item \textbf{High efficiency:} There must exist a $g_{\mu\nu}(\Lambda)[X^A(\Lambda,x),g^B(\Lambda)]$ at each $\Lambda$ such that \begin{displaymath}\nabla_{(\Lambda)\mu}t^\mu_{\phantom{\mu}\nu}(\Lambda) = 0\end{displaymath} is satisfied for all $\Lambda$ with $\nabla_{(\Lambda)}$ being the covariant derivative built out of it. Indeed as we are projecting out hard degrees of freedom, the effective energy-momentum tensor of the soft part will not be conserved because of the driving force arising from the interaction with the hard sector. The high efficiency principle assures that all these driving forces, arising from the harder degrees of freedom projected out by the coarse-graining, can be absorbed into a redefinition of the background metric.
\item \textbf{Upliftability to operator dynamics:} Although  $\langle t^\mu_{\phantom{\mu}\nu}(\Lambda)\rangle$ is a functional of $X^A(\Lambda)$ and $g^B (\Lambda)$, its first-order scale evolution equation (in the \textit{fixed} flat Minkowski space background) should be independent of the state, meaning $\partial \langle t^\mu_{\phantom{\mu}\nu}(\Lambda)\rangle/\partial \Lambda$ should be a (non-linear) functional of $\langle t^\mu_{\phantom{\mu}\nu}(\Lambda)\rangle$ and $\Lambda$ only, explicitly determining how it \textit{mixes} with the effective multi-trace operators built from its products. It implies that the coarse-grained collective variables should not explicitly appear in the scale-evolution equations of the effective operators , although they have been used to define them. This principle mimics the known procedure of defining effective composite operators at any given scale in a state-independent way via Wilsonian RG flow as discussed above.

Furthermore, we also impose the criterion that the effective Ward identity 
\[\nabla_{(\Lambda)\mu}t^\mu_{\phantom{\mu}\nu}(\Lambda) = 0\]
can be rewritten as an operator equation of the form 
\[\partial_{\mu}t^\mu_{\phantom{\mu}\nu}(\Lambda) = \text{sum of multi-trace operators built from products of $t^\mu_{\phantom{\mu}\nu}(\Lambda)$}.\] This implies that the effective metric $g_{\mu\nu}(\Lambda)$ can also be rewritten as a functional of $\langle t^\mu_{\phantom{\mu}\nu}(\Lambda)\rangle$ and $\Lambda$ only (in the fixed flat Minkowski space background).

\item \textbf{Good endpoint behaviour:} The endpoint of the RG flow, where the expectation values of the effective operator typically blow up and the effective background metric becomes non-invertible, should be transformable to a \textit{non-relativistic} fixed point with a \textit{finite} number of parameters, under the universal rescaling of the scale and of the time coordinate: 
\begin{equation}\label{rescale-intro}
\frac{1}{\Lambda_{\rm IR}} - \frac{1}{\Lambda} = \frac{\xi}{\tilde{\Lambda}}, \quad t = \frac{\tau}{\xi}, \quad \text{and $\xi \rightarrow 0$ with $\tilde{\Lambda}, \tau$ held fixed.}
\end{equation}
For instance, this fixed point takes the form of incompressible non-relativistic Navier-Stokes equations in the hydrodynamic limit. The rescaling to a fixed point can be realised only if we impose bounds on how the scale-dependent effective operators behave near the endpoint, thus determining all the integration constants of the first order RG flow and also the UV data. 
\end{enumerate}
Using these criteria, we determine the \textit{unique} coarse graining functions $F^{AB}\left(X^C(\Lambda, x), k\right)$ and the UV data $g^{B\infty}$ such that we can reproduce the dual geometries, which are described by Einstein's equations and have regular future horizons. This reproduction is possible because the effective background metric $g_{\mu\nu}(\Lambda)$ leads to a bulk $(d+1)-$metric, which is a solution of Einstein's equations without naked singularities in a specific gauge. We achieve this explicitly in a special dynamical limit corresponding to hydrodynamic evolution. Furthermore, our method can be readily applied to show that for each pure classical gravity theory, there exist only unique choices of $F^{AB}\left(X^C(\Lambda, x), k\right)$ and $g^{B\infty}$ such that all the appropriately regular dual solutions are reproduced.

The above criteria will be defined more precisely in Section \ref{derivation}. In short, the first two criteria ensure that the RG flow can be recast in the form of $(d+1)-$classical gravity equations with full $(d+1)-$diffeomorphism invariance. In particular, it follows that $g_{\mu\nu}(\Lambda)$ is the induced metric on hypersurfaces $r = \Lambda^{-1}$ on precise kinds of radial foliations in the dual geometry where the bulk metric satisfies precise $(d+1)-$classical gravity equations.

The third criterion establishes a unique correspondence between a highly efficient RG flow and a dual classical gravity theory. In other words, for every choice of $F^{AB}\left(X^C(\Lambda, x), k\right)$ and $g^B(\Lambda)$ which satisfy all our three criteria and define the effective operator $t^\mu_{\phantom{\mu}\nu}(\Lambda)$, there will be a unique dual classical gravity theory. The UV parameters ${g^B}^\infty$ are also determined by the third criterion. Furthermore, the third criterion also reproduces the same values of these UV parameters, which are otherwise determined from the regularity of the horizon on the gravity side in the traditional holographic correspondence. The special example to be worked out in full details will be the hydrodynamic limit, where the collective variables $X^A$ will be the hydrodynamic variables $u^\mu$ (local velocity) and $T$ (local temperature), and the parameters $g^B$ will be the transport coefficients (which are infinitely many and capture the asymptotic dynamics of thermalisation in the derivative expansion). In particular, we will explicitly see how the RG flow construction determines the first and second order transport coefficients, whose values are usually determined  \cite{Rangamani:2009xk} via solving the long wavelength limit of classical gravity. 

At first sight, it may look somewhat surprising that by imposing conditions on the IR end point we can determine the UV data. Nevertheless, this is possible because the IR end point is not a fixed point, although it transforms into one under the rescaling (\ref{rescale-intro}) mentioned. Throughout this paper, we will benefit a lot from the first part of this work \cite{Behr:2015yna} and also \cite{Kuperstein:2011fn,Kuperstein:2013hqa} by one of the authors, where it has been demonstrated how gravity equations themselves can be rephrased as \textit{first} order RG flow, without explicitly solving for the bulk metric, and how this RG flow encodes regularity properties of the geometry. Particularly, it has been explicitly demonstrated in \cite{Kuperstein:2013hqa} that when Einstein's equations are rephrased as  \textit{first} order RG flow equations of transport coefficients, the third criterion mentioned above determines the values of first and second order transport coefficients to be exactly those as in \cite{Rangamani:2009xk}, which are required for the regularity of the horizon. 

It is nevertheless not clear if the third criterion will be valid in an arbitrary theory of gravity to ensure regularity of the emergent horizon when it is reconstructed as a RG flow. Perhaps, there are theories of gravity where the regularity of the horizon cannot be rephrased as an appropriate criterion for the endpoint behaviour of a RG flow, which reconstructs the bulk spacetime in terms of effective operator algebra at various energy scales. We conjecture that such theories of gravity cannot make sense as consistent dual field theories. We plan to investigate this issue further in the future.

Although the first criterion stated above is intended to recover dual classical gravity equations with full $(d+1)-$diffeomorphism invariance and is very physical and elegant from the field-theoretic viewpoint, admittedly it lacks an independent motivation. Intuitively, however, as discussed in the first part of this work \cite{Behr:2015yna}, this principle leads to emergence of $(d+1)-$diffeomorphism symmetry, resulting in the feature that the scale evolution of the RG flow is generated by a Hamiltonian, which is essentially the $(d+1)-$Hamiltonian constraint of the emergent gravity theory. In fact this observation has been the key to the Hamiltonian formulation of holographic renormalisation \cite{deBoer:1999xf}, which is essential to regularise the gravitational action \cite{Akhmedov:1998vf,Henningson:1998ey,Balasubramanian:1999re,deHaro:2000xn,Bianchi:2001kw} and define the holographic correspondence precisely. Our first criterion is thus an elegant and consistent way of generalising Wilsonian RG flow so that the RG flow is still generated by a Hamiltonian (for related ideas see \cite{Dolan:1993cf,Erdmenger:2001ja,Nakayama:2013wda,Nakayama:2015ita}) -- we plan to investigate this matter further in the future. We also believe that this criterion is fundamentally connected with efficient averaging of quantum dynamics, such that it retains as much information as possible regarding long range entanglement, as in \textit{entanglement renormalisation} (see \cite{PhysRevLett.99.220405,0912.1651}).  We will  have more to say about this in Section \ref{final}.

Our construction of the highly efficient RG flow can be performed, not only in the fixed background Minkowski space, but also in an arbitrary fixed background metric. This automatically leads to determination of the correlation functions of the theory, hence indeed our RG flow construction is demonstrably a constructive field theory approach at strong coupling and large $N$.

Although for most of this paper, we have focused on the pure gravity sector and its dual, in Section \ref{beta} we also consider the vacuum state of the CFT deformed by a relevant coupling in the UV. We find that the highly efficient RG flow construction can be extended to give a definition of the beta function which satisfies familiar identities of the Wilsonian RG flow. Nevertheless, there are subtle differences which require further exploration.

In Section \ref{semihol}, we will outline how this RG flow construction can be generalised by interpolating it with a weakly coupled description in the UV. This will lead to us a concrete proposal for a novel non-perturbative semiholographic framework for quantum field theories such as QCD.

The organisation of the paper is as follows. In Section \ref{Exact}, we outline the details of the coarse-graining in the hydrodynamic limit. In Section \ref{Flow}, we review briefly how gravity can be rephrased in the form of first order RG flow equations. In Section \ref{Derive}, we show explicitly how our criteria for constructing the RG flow recover the flow equations that reconstruct gravity, while self-determining the UV data of the theory. In Section \ref{beta}, we discuss the beta function in the highly efficient RG flow. In Section \ref{Outlook}, we discuss further extensions of our approach. We also comment towards the end regarding extensions of our approach to finite $N$, and moreover on the relation of this work with some other approaches in the literature.

\section{The coarse-graining procedure for generalising Wilsonian RG flow}\label{Exact}

\subsection{Preliminary considerations and notations}

Our first assumption is that we can consider a sector of states at strong coupling and large $N$ in which $t^\mu_{\phantom{\mu}\nu}$ is the \textit{only} single-trace operator which takes an \textit{independent} expectation value. This means that the expectation values of other operators are either vanishing, or are algebraic functionals of the expectation value of $t^\mu_{\phantom{\mu}\nu}$, such that they can be obtained directly from the latter without requiring us to solve any additional dynamical equation. This sector of states form a subspace of the full Hilbert space of the theory at strong coupling and large $N$. In this sector,  $t^\mu_{\phantom{\mu}\nu}$ will mix under the RG flow only with multi-trace operators built out of it. Furthermore, due to large $N$ factorisation, the RG flow equation for $t^\mu_{\phantom{\mu}\nu}$ can be thought of as a \textit{classical} equation, meaning that the expectation value $\langle t^\mu_{\phantom{\mu}\nu}\rangle$ and the operator $t^\mu_{\phantom{\mu}\nu}$ both follow the same equation.

Our assumption, which is stated above, can be justified a posteriori via our construction -- we need to show that we can indeed obtain a consistent operator algebra at all energy scales and determine all necessary UV data to define this sector of states. The motivation for making this assumption is of course to reconstruct the \textit{pure gravity} sector of the holographic correspondence, where the bulk dynamics involves the graviton only -- such a consistent truncation is always expected to be possible in a classical gravity theory which maps to a dual field theory holographically in the strong coupling and large $N$ limit.

We will aim to define the RG flow via use of collective variables, as mentioned in the Introduction. The easiest limit to consider is the hydrodynamic limit, where $\langle t^\mu_{\phantom{\mu}\nu}\rangle$ can be expressed in terms of hydrodynamic variables. Furthermore, the operator algebra of $t^\mu_{\phantom{\mu}\nu}(\Lambda)$ constructed by this method will be state-independent, and will thus be valid far away from the hydrodynamic limit. Therefore, even an extension of our RG flow construction away from the hydrodynamic limit should also reproduce the same RG flow equation and operator algebra for $t^\mu_{\phantom{\mu}\nu}(\Lambda)$. In Section \ref{Outlook1}, we will discuss how a complete reconstruction of the \textit{pure gravity sector} by incorporating non-hydrodynamic variables can be achieved.

As a many-body system undergoes thermalisation, the time-dependent behaviour of the microscopic data can be captured by an \textit{exact asymptotic hydrodynamic expansion} at late time. This expansion for the energy-momentum tensor involves an infinite number of transport coefficients, each multiplying tensors built out the hydrodynamic variables and with a fixed number of derivatives.  The expansion parameter is the ratio of the length scale of variation of these variables to the thermal wavelength. It is to be noted that this gives an exact asymptotic series for time-dependent dynamics, and by itself does not involve any coarse-graining.

At weak coupling and low density, this expansion can be obtained from the \textit{normal solutions} of Boltzmann equation, which can be constructed using the Chapman-Enskog method \cite{opac-b1097179}. These special solutions of the Boltzmann equation can be mapped exactly to a hydrodynamic expansion involving an arbitrarily large number of transport coefficients. The values of the latter can thus be obtained from microscopic data up to a certain degree of approximation. It can also be shown that the \textit{normal solutions} of the Boltzmann equation, and hence hydrodynamics, provide an asymptotic expansion for an \textit{arbitrary} solution of the Boltzmann equation. It is to be noted that no coarse-graining of the Boltzmann equation is involved in the construction of normal solutions -- the hydrodynamic variables actually capture all details of the microscopic quasiparticle distribution function, for which it gives a generic asymptotic time series describing how the latter approaches the thermal Boltzmann distribution.

At strong coupling and large $N$, the exact asymptotic hydrodynamic expansion can be obtained from the fluid/gravity correspondence for holographic quantum many-body systems \cite{Rangamani:2009xk}. This correspondence amounts to a one-to-one correspondence between hydrodynamics in the dual field theory, and long wavelength solutions of classical gravity in which a black brane is formed. Once again, the derivative expansion for the energy-momentum tensor involves infinitely many transport coefficients, each of which is determined by imposing the regularity of the future horizon in the dual geometries perturbatively, order by order in this expansion. It also gives the generic asymptotic time-series series for various observables as the system undergoes thermalisation, since the entire bulk geometry, and thus the dual state, is described by the hydrodynamic variables.

The hydrodynamic variables can be defined following Landau and Lifshitz. The $d-$velocity $u^\mu$ is defined as the time-like eigenvector of $\langle t^\mu_{\phantom{\mu}\nu}\rangle$, meaning that it is the local velocity of energy transport. Thus
\begin{equation}
\langle t^\mu_{\phantom{\mu}\nu} \rangle u^\nu = -\epsilon u^\mu, \quad \text{with} \quad u^\mu g_{\mu\nu}u^\nu = -1.
\end{equation}
Above $g_{\mu\nu}$ is the non-dynamical background metric on which the field theory lives (and should not be confused with the effective metric $g_{\mu\nu}(\Lambda)$ which has been mentioned in the Introduction and to be discussed extensively later). In order for the hydrodynamic limit to exist, $g_{\mu\nu}$ has to be weakly curved -- the average radius of curvature of the background metric should be larger than the thermal wavelength. The local temperature $T$ is defined by the assuming that the equation of state is valid locally, whence assuming that the energy density $\epsilon(T)$ as a function of the local temperature is the same as that in thermal equilibrium.  

The exact asymptotic hydrodynamic expansion of $\langle t^\mu_{\phantom{\mu}\nu} \rangle$ involves an expansion in covariant derivatives of $u^\mu$ and $T$, and the Riemann curvature of $g_{\mu\nu}$ and its covariant derivatives. The scale of variation of all these quantities are assumed to be larger than the thermal wavelength. 

Let us denote hydrodynamic scalars constructed out of derivatives of the above kind as $\mathcal{S}^{(n,m)}$, where $n$ denotes the number of derivatives, and $m$ is a counting index. Similarly, let us denote transverse hydrodynamic vectors constructed out of derivatives of the above kind  as ${\mathcal{V}^{\mu}}^{(n, m)}$ -- being transverse means that they are orthogonal to $u^\mu$ so that ${\mathcal{V}^{\mu}}^{(n, m)}u_\mu = 0$. Also we denote transverse symmetric traceless hydrodynamic tensors as ${\mathcal{T}^{\mu\nu}}^{(n, m)}$ -- these satisfy ${\mathcal{T}^{\mu\nu}}^{(n, m)}u_\nu = 0 $ and ${\mathcal{T}^{\mu\nu}}^{(n, m)}g_{\mu\nu} = 0$. 

At first order in derivatives, these are
\begin{equation}
\mathcal{S}^{(1)} = \nabla\cdot u, \quad {\mathcal{V}^{\mu}}^{(1)} = (u\cdot \nabla) u^\mu, \quad {\mathcal{T}^{\mu\nu}}^{(1)} = \sigma^{\mu\nu},
\end{equation}
with $\sigma^{\mu\nu}$ being the shear of $u^\mu$ and is defined as below
\begin{equation}
\sigma^{\mu\nu}= \frac{1}{2}\Delta^{\mu\alpha}\Delta^{\nu\beta}\left(\nabla_\alpha u_\beta + \nabla_\beta u_\alpha\right) - \frac{1}{(d-1)}\Delta^{\mu\nu}\Delta^{\alpha\beta}\nabla_\alpha u_\beta,
\end{equation}
with
\begin{equation}
\Delta^{\mu\nu} = u^\mu u^\nu + g^{\mu\nu},
\end{equation}
denoting the projector in the direction orthogonal to $u^\mu$. One may consider another scalar $(u\cdot\nabla)\ln T$, however it is related to $\nabla\cdot u$ by the hydrodynamic equation of motion at the leading order in derivatives. As it makes sense to consider the operator $t^\mu_{\phantom{\mu}\nu}$ only on-shell, meaning when it satisfies its equation of motion, without loss of generality we need to consider only those independent scalars which are not related equations of motion. This is also why we need not consider the transverse scalar $\Delta^{\mu\nu}\nabla_\nu \ln T$ which can be related to the acceleration $(u\cdot \nabla)u^\mu$ by hydrodynamic equation of motion at leading order in derivatives.

At second order in derivatives, some of the independent scalars, vectors and tensors are
\begin{eqnarray}
\mathcal{S}^{(2, i)} &:=& \{ (\nabla\cdot u)^2, \quad \sigma_{\mu\nu}\sigma^{\mu\nu}, \quad R,  \quad\text{etc.} \},\nonumber\\
{\mathcal{V}^{\mu}}^{(2,i)} &:=& \{\sigma^{\mu\nu}\nabla_{\nu}\ln T,  \quad (\nabla\cdot u) (u\cdot\nabla)u^\mu,  \quad\Delta^{\mu\nu}u^\rho R_{\nu\rho},  \quad \text{etc.}\},  \nonumber\\
{\mathcal{T}^{\mu\nu}}^{(2,i)} &:=& \{(\nabla\cdot u) \sigma^{\mu\nu},  \quad \sigma^{\mu}_{\phantom{\mu}\rho}\sigma^{\rho\nu} - \frac{1}{(d-1)} \Delta^{\mu\nu}\sigma^{\alpha\beta}\sigma_{\alpha\beta}, \nonumber\\&& \Delta^{\mu\alpha}\Delta^{\nu\beta}R_{\alpha\beta} - \frac{1}{(d-1)}\Delta^{\mu\nu}\Delta^{\alpha\beta}R_{\alpha\beta}, \quad \text{etc.}\}.
\end{eqnarray}
For a complete list, we refer the reader to \cite{Rangamani:2009xk}. 

Let us denote the number of independent hydrodynamic scalars, transverse vectors and transverse traceless symmetric tensors with $n$ derivatives as $n_{\rm s}$, $n_{\rm v}$ and $n_{\rm t}$ respectively.

With the above notations, we can write $\langle t^{\mu}_{\phantom{\mu}\nu}\rangle$ in the exact asymptotic hydrodynamic expansion as follows:
\begin{eqnarray}\label{emhydro}
\langle t^{\mu}_{\phantom{\mu}\nu}\rangle &=& \epsilon (T) u^\mu u_\nu + P(T) \Delta^\mu_{\phantom{\mu}\nu} - \sum_{n_1}^\infty \sum_{m = 1}^{n_{\rm s}}\gamma_{\rm s}^{(n,m)}(T) \, \mathcal{S}^{(n,m)}\Delta^{\mu}_{\phantom{\mu}\nu}- \nonumber\\&&- \sum_{n =1}^\infty \sum_{m = 1}^{n_{\rm t}}\gamma_{\rm t}^{(n,m)}(T) \, {\mathcal{T}^{\mu}_{\phantom{\mu}\nu}}^{(n,m)}.
\end{eqnarray}
The parameters $\gamma_{\rm s}^{(n,m)}$ will be called the scalar transport coefficients and $\gamma_{\rm t}^{(n,m)}$ will be called the tensor transport coefficients. At first order in derivatives, the scalar transport coefficient associated with $(\nabla\cdot u)$ is the bulk viscosity $\zeta(T)$ and the tensor transport coefficient associated with $\sigma^{\mu\nu}$ is the shear viscosity $\eta(T)$. As $u^\mu$ is an eigenvector of $\langle t^{\mu}_{\phantom{\mu}\nu}\rangle$, terms of the form $u^\mu {\mathcal{V}_\nu}^{(n,m)} + u_\nu {\mathcal{V}^\mu}^{(n,m)}$ cannot appear in (\ref{emhydro}). Similarly terms of the form $u^\mu u_\nu \mathcal{S}^{\rm (n.m)}$ cannot appear as they will modify the eigenvalue.

The hydrodynamic equations for evolution of $u^\mu$ and $T$ simply follow from the Ward identity for energy-momentum conservation, namely
\begin{equation}
\nabla_\mu \langle t^{\mu}_{\phantom{\mu}\nu}\rangle = 0.
\end{equation}

Let us briefly consider the case when expectation values of other single-trace operators are also non-trivial in the hydrodynamic limit. For instance, a scalar operator $O$ and a vector operator $V^\mu$ will have the exact asymptotic hydrodynamic expansion of the forms:
\begin{eqnarray}
\langle O\rangle &=& O^{(0)}(T) + \sum_{n=1}^\infty \sum_{m=1}^{n_{\rm s}} O^{(n,m)}(T) \, \mathcal{S}^{(n,m)}, \nonumber\\
\langle V^\mu \rangle &=& V^{(0)}(T) \, u^\mu + \sum_{n=1}^\infty \sum_{m=1}^{n_{\rm s}} V_{\rm s}^{(n,m)}(T) \, \mathcal{S}^{(n,m)} u^\mu + \sum_{n=1}^\infty \sum_{m=1}^{n_{\rm v}} V_{\rm v}^{(n,m)}(T) \, {\mathcal{V}^\mu}^{(n,m)}.
\end{eqnarray}
Thus expectation values of all operators have an exact asymptotic hydrodynamic expansion which can be expressed in terms of the hydrodynamic scalars, vectors and tensors, with coefficients (like the transport coefficients) being functions of local temperature $T$ which are determined by the microscopic theory. In presence of other non-vanishing global charges, we have to add the local  charge densities or the local chemical potentials as \textit{additional} hydrodynamic variables. In this case, all coefficients of the asymptotic hydrodynamic expansion are functions of the temperature and local charge densities/local chemical potentials also. 

The Ward identity for conservation of energy and momentum is also modified in presence of external sources that couple to the other operators. For instance, we get 
\begin{equation}\label{WIscalar}
\nabla_\mu \langle t^{\mu}_{\phantom{\mu}\nu}\rangle = \langle O\rangle \,\nabla_\nu J
\end{equation}
in presence of the external source $J$ that couples to the scalar operator $O$.

From the above discussion, it is clear that under the \textit{highly efficient RG flow} which we are going to construct the expectation value of the coarse-grained operator $t^{\mu}_{\phantom{\mu}\nu}(\Lambda)$ also takes the same hydrodynamic form (\ref{emhydro}) at each scale $\Lambda$, but with scale-dependent $u^\mu(\Lambda)$ and $T(\Lambda)$ in a scale-dependent effective metric $g_{\mu\nu}(\Lambda)$. Also the energy density, pressure and transport coefficients become scale-dependent. Therefore,
\begin{eqnarray}\label{emhydroscale}
\langle t^{\mu}_{\phantom{\mu}\nu}(\Lambda)\rangle &=& \epsilon \left(T(\Lambda), \Lambda\right) u^\mu(\Lambda) u_\nu(\Lambda) + P \left(T(\Lambda), \Lambda\right) \Delta^\mu_{\phantom{\mu}\nu}(\Lambda) -\nonumber\\&&
- \sum_{n_1}^\infty \sum_{m = 1}^{n_{\rm s}}\gamma_{\rm s}^{(n,m)} \left(T(\Lambda), \Lambda\right) \, \mathcal{S}^{(n,m)}(\Lambda)\Delta^{\mu}_{\phantom{\mu}\nu}(\Lambda)- \nonumber\\&&- \sum_{n =1}^\infty \sum_{m = 1}^{n_{\rm t}}\gamma_{\rm t}^{(n,m)} \left(T(\Lambda), \Lambda\right) \, {\mathcal{T}^{\mu}_{\phantom{\mu}\nu}}^{(n,m)}(\Lambda).
\end{eqnarray}
By $\Delta^\mu_{\phantom{\mu}\nu}(\Lambda)$, $\mathcal{S}^{(n,m)}(\Lambda)$, etc. we mean these are constructed out of $u^\mu(\Lambda)$, $T(\Lambda)$ and $g_{\mu\nu}(\Lambda)$.

On the other hand, if we simply know the coarse-grained $u^\mu(\Lambda)$ and $T(\Lambda)$, implying that we also know their equations of motion, we may be able to also construct both the coarse-grained operator $t^{\mu}_{\phantom{\mu}\nu}(\Lambda)$ and the effective metric $g_{\mu\nu}(\Lambda)$. This is possible only if the equations of motion for $u^\mu(\Lambda)$ and $T(\Lambda)$ can be written in the form $\nabla_{(\Lambda)\mu}\langle t^{\mu}_{\phantom{\mu}\nu}(\Lambda)\rangle = 0$ with $\nabla_{(\Lambda)}$ being the covariant derivative constructed from an appropriate $g_{\mu\nu}(\Lambda)$. We will see that this is possible if the scale evolutions of $u^\mu(\Lambda)$ and $T(\Lambda)$ are highly constrained.

\subsection{General procedure for coarse-graining}\label{coarse-grain}

By coarse-graining a field variable, we mean replacing its exact microscopic value by its average over a neighbourhood in spacetime. The size of the neighbourhood sets the scale of the coarse-graining. This procedure can be generalised easily and for very good reasons. The size of the neighbourhood over which the averaging is done can be kept fixed on average, but it could be modulated over space and time depending on the interesting features of the behaviour of the microscopic field in the neighbourhood of the point. Thus the scale of coarse-graining can be promoted to a functional of the microscopic field itself in order to retain information of microscopic dynamics in a more efficient manner, while still averaging out information which is unnecessary for studying the dynamics at the given overall scale of resolution.

Let us first propose the following general coarse-graining of the hydrodynamic variables $u^\mu(\Lambda)$ and $T(\Lambda)$. The simplest way of constructing a coarse-graining is to work in Fourier space and cut out modes which vary faster than $\Lambda^{-1}$, and then modulate this cut-off recursively as functions of  $u^\mu(\Lambda)$ and $T(\Lambda)$ in the derivative expansion. Thus we arrive at the most general Ansatz to define $u^\mu (\Lambda)$ and $T(\Lambda)$ such that it is consistent with the derivative expansion. In the flat Minkowski background these take the form:
\begin{eqnarray}\label{cguT}
u^\mu (\Lambda, x) &=& \int {\rm d}^d k\,\, e^{ik\cdot x}\,\,\Theta\left(1-\frac{k^2}{\Lambda^2}\right) \Bigg[v^{(0)}\left(T^\infty(x)/\Lambda\right)\delta^\mu_{\,\,\nu} + i v^{(1)}_{{\rm v}}\left(T^\infty(x)/\Lambda\right)u^\alpha(\Lambda, x)\frac{k_\alpha}{T^\infty(x)}\delta^\mu_{\,\, \nu}+ \nonumber\\ &&\qquad\qquad\qquad \qquad\qquad\quad+  i v^{(1)}_{{\rm s}}\left(T^\infty(x)/\Lambda\right)u^\mu(\Lambda, x)\frac{k_\nu}{T^\infty(x)} +\nonumber\\ &&\qquad\qquad\qquad \qquad\qquad\quad +\mathcal{O}\left(\frac{k^2}{{T^\infty}^2(x)}, \partial u(\Lambda) \frac{k}{T^\infty(x)}, \partial T(\Lambda) \frac{k}{T^\infty(x)}\right)\Bigg]\tilde{u}^{\nu\,\infty}(k),\nonumber\\
T(\Lambda, x) &=&\int {\rm d}^d k\,\, e^{ik\cdot x}\,\,\Theta\left(1-\frac{k^2}{\Lambda^2}\right) \Bigg[w^{(0)}\left(T^\infty(x)/\Lambda\right) + i w^{(1)}_{{\rm s}}\left(T^\infty(x)/\Lambda\right)u^\alpha(\Lambda,x)\frac{k_\alpha}{T^\infty(x)}+\nonumber\\&&\qquad\qquad\qquad \qquad\qquad+ \mathcal{O}\left(\frac{k^2}{{T^\infty}^2(x)}, \partial u(\Lambda) \frac{k}{T^\infty(x)}, \partial T(\Lambda) \frac{k}{T^\infty(x)}\right)\Bigg]{\tilde{T}}^\infty(k)\,.
\end{eqnarray}
up to first order in the derivative expansion. Above $\tilde{u}^\infty$ and $\tilde{T}^\infty$ denote the Fourier transforms of the exact microscopic UV variables. Clearly the above equations have a recursive structure, as the right hand sides depend on $u^\mu(\Lambda)$ and $T(\Lambda)$ also. Up to this order, we obtain five coarse-graining functions, namely $v^{(0)}, v_{\rm v}^{(1)}, v_{\rm s}^{(1)}, w^{(0)}$ and $w_{\rm s}^{(1)}$, which are to be determined here using our general criteria.

There is also an overall cut-off function in momentum space in (\ref{cguT}), which cuts out momentum modes which are larger than $\Lambda^2$. Instead of being a theta function, this can be another function such as $\tanh (\Lambda^2/k^2)$ which cuts down modes $k^2 \gg \Lambda^2$, while flattens out at $k^2 \ll \Lambda^2$, so that it necessarily changes rapidly when $k^2\approx \Lambda^2$. We will show that any choice of the overall cut-off function will lead to the same evolution in the hydrodynamic sector, as this overall function will affect only how hydrodynamic modes mix with non-hydrodynamic modes along the RG flow. As non-hydrodynamic modes do not mix with hydrodynamic modes under RG flow \footnote{The RG flow evolution is expected to be upper triangular -- the light modes can mix with heavy modes along the flow but the reverse should not happen.}, we can indeed project the evolution of hydrodynamic modes to the hydrodynamic sector. Nevertheless, if we go away from the hydrodynamic limit, the choice of the overall cut-off function should be fixed by our general criteria -- we will leave this investigation to the future.

It can be easily noted that the function $v_{\rm v}^{(1)}$ in (\ref{cguT}) is related to the acceleration vector $(u\cdot \partial) u^\mu$, where the derivative is now promoted to a spatially and temporally modulated operation in Fourier space. Similarly, the functions $v_{\rm s}^{(1)}$ and $w_{\rm s}^{(1)}$ are related to the scalar $(\partial \cdot u)$. Since these are the only independent vectors and scalars on-shell at first order, this is the most general form of coarse-graining where the cut-off has been modulated as a functional of $u^\mu(\Lambda)$ and $T(\Lambda)$ up to first order in derivatives. This will become more explicit when we write the flow equations in differential form. It is to be noted that the derivative expansion at higher orders will count both derivatives of $u^\mu(\Lambda)$ (and $T(\Lambda)$) and also the powers of $k/T^\infty(x)$.

The coarse-graining procedure can be readily generalised to higher order in derivatives. For each hydrodynamic vector ${\mathcal{V}^\mu}^{\rm (n,m)}$, we can construct a correction to the overall cut-off in (\ref{cguT}), such that it is linear in $\tilde{u}^{\mu\,\infty}(k)$ but not necessarily linear in $u^\mu(\Lambda, x)$ and $T(\Lambda, x)$. There is also an associated coefficient $v_{\rm v}^{(n,m)}(T^\infty(x)/\Lambda)$. Similarly for each scalar $\mathcal{S}^{(n,m)}$, we will obtain a correction to the modulation of the cut-off for $u^\mu(\Lambda)$ which is associated with a function $v_{\rm s}^{(n,m)}(T^\infty(x)/\Lambda)$, and a correction that modulates the cut-off for $T(\Lambda)$ associated with $w_{\rm s}^{(n,m)}(T^\infty(x)/\Lambda)$. 

Some of these hydrodynamic vectors and scalars are associated with the curvature of the background metric. In order to see these explicitly, we need to put the conformal field theory (CFT) on a conformally flat metric $\eta_{\mu\nu}e^{2\sigma(x)}$ instead of the flat Minkowski background. As for example, the coarse-graining function that modulates the cut-off for $u^\mu(\Lambda)$, and which is associated with the scalar $R$ is:
\begin{equation}
\Big[ ... v_{\rm s}^{(2,i)}\left(\frac{T^\infty(x)}{\Lambda}\right)e^{-2\sigma}\left(2(d-1) \Box\sigma -(d-2)(d-1)\partial_\alpha\sigma\partial^\alpha\sigma\right)\delta^\mu_{\phantom{\mu}\nu} + ...\Big] \tilde{u}^{\nu\, \infty}(k),
\end{equation}
with $i$ being the order in which $R$ appears in the list of independent second-order hydrodynamic scalars. The multiplying $\sigma-$dependent function is nothing but the Ricci scalar $R$ of the fixed background metric $\eta_{\mu\nu}e^{2\sigma}$. 

\paragraph{Flow equations in the differential form:} So far we have defined the coarse-graining that leads to the construction of $u^\mu(\Lambda)$ and $T(\Lambda)$ in the integral form. We will show that the coarse-graining construction that defines $u^\mu(\Lambda)$ and $T(\Lambda)$  can also be cast to the form of first order differential equations for these variables yielding their scale evolution. The latter formulation will be more suitable for two purposes, namely for a better intuitive understanding of the construction and also for deriving the emergence of gravity.

Let us define the following immediately useful variables $u^{\mu(0)}(\Lambda)$ and $T^{(0)}(\Lambda)$ as follows:
\begin{eqnarray}\label{u0T0}
u^{\mu(0)} (\Lambda, x) &=& \int {\rm d}^d k\,\, e^{ik\cdot x}\,\,\Theta\left(1-\frac{k^2}{\Lambda^2}\right) v^{(0)}\left(T^\infty(x)/\Lambda\right) \tilde{u}^{\mu\,\infty}(k),\nonumber\\
T^{(0)}(\Lambda, x) &=&\int {\rm d}^d k\,\, e^{ik\cdot x}\,\,\Theta\left(1-\frac{k^2}{\Lambda^2}\right) w^{(0)}\left(T^\infty(x)/\Lambda\right)\tilde{T}^{\infty}(k)\,,
\end{eqnarray} 
which fulfill the following useful identities \footnote{Since $v^{(0)}$ and $w^{(0)}$ are functions of $T^\infty/\Lambda$, it follows that $\partial v^{(0)}/\partial \ln \Lambda = -\partial v^{(0)}/\partial \ln T^\infty$ and similarly for $w^{(0)}$. These relations have been used below.}:
\begin{eqnarray}
\int {\rm d}^d k\,\, e^{ik\cdot x}\,\,\Theta\left(1-\frac{k^2}{\Lambda^2}\right)(ik_{\nu}) \tilde{u}^\mu(k)
&=&
\frac{1}{v^{(0)}}
\partial_{\nu}u^{\mu(0)} +u^{\mu(0)} \frac{\frac{\partial \ln v^{(0)}}{\partial \ln \Lambda}}{\frac{\partial \ln T^{(0)}(\Lambda)}{\partial \ln T^\infty}}\partial_\nu \ln T^{(0)}(\Lambda) \, ,
\nonumber\\
\int {\rm d}^d k\,\, e^{ik\cdot x}\,\,\Theta\left(1-\frac{k^2}{\Lambda^2}\right) (ik_{\nu})\tilde{T}(k) &=& \frac{1}{w^{(0)}}
\partial_{\nu}T^{(0)} + T^{(0)}\frac{\frac{\partial \ln w^{(0)}}{\partial \ln \Lambda}}{\frac{\partial \ln T^{(0)}(\Lambda)}{\partial \ln T^\infty}}\partial_\nu \ln T^{(0)}(\Lambda).\nonumber\\&&
\end{eqnarray}
Using the above identities, it is straightforward to see from (\ref{cguT}) that in real space \footnote{Here, we have used recursively that $u^\mu = u^{\mu(0)} + \mathcal{O}(\partial)$ at leading order in derivatives.}
\begin{eqnarray}\label{useful1}
u^\mu (\Lambda) &=& u^{\mu(0)} (\Lambda) +\frac{1}{T^\infty} \left(\frac{v^{(1)}_{{\rm v}}}{v^{(0)}}\right)\left(u^{(0)}(\Lambda)\cdot \partial\right)u^{\mu(0)} (\Lambda) +\frac{1}{T^\infty} \left(\frac{v^{(1)}_{{\rm s}}}{v^{(0)}}\right)\left(\partial\cdot u^{(0)}(\Lambda)\right)u^{\mu(0)} (\Lambda)+ \nonumber\\ && +\frac{1}{T^\infty}\left(v_{\rm v}^{(1)} + v_{\rm s}^{(1)}\right)\frac{\frac{\partial \ln v^{(0)}}{\partial \ln \Lambda}}{\frac{\partial \ln T^{(0)}(\Lambda)}{\partial \ln T^\infty}}u^{\mu(0)}(\Lambda)\left(u^{(0)}(\Lambda)\cdot\partial\right)\ln T^{(0)} (\Lambda)+
 \mathcal{O}(\partial^2), \nonumber\\
T^{(0)} (\Lambda) &=&T (\Lambda) -\frac{T(\Lambda)}{T^\infty}w^{(1)}_{{\rm s}} \left(\frac{1}{w^{(0)}}+\frac{\frac{\partial \ln w^{(0)}}{\partial \ln \Lambda}}{\frac{\partial \ln T(\Lambda)}{\partial \ln T^\infty}}\right)\left(u(\Lambda)\cdot\partial\right)\ln T (\Lambda)  + \mathcal{O}(\partial^2)\,.
\end{eqnarray}
We can also invert the above relations as follows:
\begin{eqnarray}\label{useful2}
u^{\mu(0)} (\Lambda) &=& u^{\mu} (\Lambda) -\frac{1}{T^\infty} \left(\frac{v^{(1)}_{{\rm v}}}{v^{(0)}}\right)\left(u(\Lambda)\cdot \partial\right)u^{\mu} (\Lambda) -\frac{1}{T^\infty} \left(\frac{v^{(1)}_{{\rm s}}}{v^{(0)}}\right)\left(\partial\cdot u(\Lambda)\right)u^{\mu} (\Lambda)- \nonumber\\ && -\frac{1}{T^\infty}\left(v_{\rm v}^{(1)} + v_{\rm s}^{(1)}\right)\frac{\frac{\partial \ln v^{(0)}}{\partial \ln \Lambda}}{\frac{\partial \ln T(\Lambda)}{\partial \ln T^\infty}}u^{\mu}(\Lambda)\left(u(\Lambda)\cdot\partial\right)\ln T (\Lambda)+
 \mathcal{O}(\partial^2), \nonumber\\
T^{(0)} (\Lambda) &=& T (\Lambda) -\frac{T^{(0)}(\Lambda)}{T^\infty}w^{(1)}_{{\rm s}} \left(\frac{1}{w^{(0)}}+\frac{\frac{\partial \ln w^{(0)}}{\partial \ln \Lambda}}{\frac{\partial \ln T(\Lambda)}{\partial \ln T^\infty}}\right)\left(u(\Lambda)\cdot\partial\right)\ln T (\Lambda)  + \mathcal{O}(\partial^2)\,.
\end{eqnarray}
Clearly, by differentiating (\ref{u0T0}) with respect to $\Lambda$, we obtain
\begin{eqnarray}\label{diff0}
\frac{\partial u^{\mu(0)}(\Lambda)}{\partial \Lambda} &=& \frac{\partial(\ln v^{(0)})}{\partial \Lambda} u^{\mu(0)}(\Lambda) \nonumber\\ && + \int {\rm d}^d k\,\, e^{ik\cdot x}\,\,\delta\left(1-\frac{k^2}{\Lambda^2}\right)\,\frac{k^2}{\Lambda^3}\,\, {v^{(0)}\left(T^\infty(x)/\Lambda\right)} u^\mu(k),\nonumber\\
\frac{\partial T^{(0)}(\Lambda)}{\partial \Lambda} &=& \frac{\partial(\ln w^{(0)})}{\partial \Lambda} T^{(0)}(\Lambda) \nonumber\\ && + \int {\rm d}^d k\,\, e^{ik\cdot x}\,\,\delta\left(1-\frac{k^2}{\Lambda^2}\right)\,\frac{k^2}{\Lambda^3}\,\, {w^{(0)}\left(T^\infty(x)/\Lambda\right)} T(k)\,.
\end{eqnarray}
We observe that although $u^{\mu(0)}(\Lambda)$ and $T^{(0)}(\Lambda)$ are hydrodynamic variables in the sense of containing only slowly varying modes, their derivatives with respect to $\Lambda$ contain fast varying pieces, particularly the terms involving delta functions which have support at $k^2 = \Lambda^2$. By definition slow variables are those which have maximum support at $k^2\ll \Lambda^2$ and the fast variables are those which have minimum support for $k^2 \gg  \Lambda^2$. Therefore we should invent a regularisation procedure by which we can remove the non-hydrodynamic contributions which are non-perturbative in the derivative expansion. Let us write $:\cdots:$ to denote a term where such non-hydrodynamic terms are removed. Then it follows that
\begin{eqnarray}\label{useful3}
:\frac{\partial u^{\mu(0)}(\Lambda)}{\partial \Lambda}: &=& \frac{\partial\ln v^{(0)}}{\partial \Lambda} u^{\mu(0)}(\Lambda),\nonumber\\
:\frac{\partial T^{(0)}(\Lambda)}{\partial \Lambda}: &=& \frac{\partial\ln w^{(0)}}{\partial \Lambda} T^{(0)}(\Lambda).
\end{eqnarray}
It is to be noted that on the gravity side too, such non-hydrodynamic terms are implicitly present as they give non-perturbative completion to the asymptotic hydrodynamic series. In particular, these involve non-hydrodynamic quasi-normal modes. Therefore, non-hydrodynamic terms on the field-theory side should be removed, and then they should be compared with expressions obtained from gravity. So, we adopt the procedure for removal of non-hydrodynamic terms as denoted by $:\cdots:$ which amounts to simply removing terms where the overall cut-off function is differentiated. As the overall cut-off function is not differentiated, we will get the same flow equations for any choice of the overall cut-off function (whose $\Lambda-$derivative will be sharply localised at $k^2 = \Lambda^2$), as we have mentioned before.

We make another important observation that when we differentiate the right hand side of (\ref{useful1}), then we may get terms which involve product of a non-hydrodynamic piece and a hydrodynamic piece, but never terms which involve the product of two non-hydrodynamic pieces. This is because the non-hydrodynamic terms always originate form $\partial u^{\mu(0)}/\partial\Lambda$, and the latter can appear only once in each term when the  right hand side of (\ref{useful1}) is differentiated with respect to $\Lambda$. Note when a fast piece like $\sin(k_1t)$ with $k_1  \gg  \Lambda$ is multiplied with a slow piece like $\sin(k_2 t)$ with $k_2 \ll \Lambda$, the result is proportional to $\cos\left((k_1 - k_2)t\right) -\cos\left((k_1 + k_2)t\right)$. We note that both terms are fast. Thus fast times slow terms are always fast terms. However a product of two fast terms may involve slow terms. Since we never get a product of two non-hydrodynamic pieces, we can simply remove the delta function piece in $\partial u^{\mu(0)}/\partial\Lambda$ and then evaluate the differentiation of right hand side of (\ref{useful1}) in order to obtain all the slow pieces of the latter. 

A straightforward integration of (\ref{diff0}) gives us
\begin{equation}\label{useful4}
T^{(0)}(\Lambda) = w^{(0)}\left(T^\infty(x)/\Lambda\right)\,\, T^\infty, \quad u^{\mu(0)}(\Lambda) = v^{(0)}\left(T^\infty(x)/\Lambda\right)\,\, u^{\mu\infty}\,,
\end{equation}
where we have used $w^{(0)} = v^{(0)}=1$, $u^\mu(\Lambda) = u^{\mu\infty}$, and $T(\Lambda)= T^\infty$ at $\Lambda = \infty$, and we have removed the non-hydrodynamic contributions. From here it follows that
\begin{equation}\label{useful5}
\frac{\partial \ln T^{(0)}(\Lambda)}{\partial \ln T^\infty} = 1 - \frac{\partial \ln w^{(0)}}{\partial \ln \Lambda}.
\end{equation}
Similar manipulations give us
\begin{gather}\label{useful6}
\begin{aligned}
:\partial_\mu\frac{\partial \ln T^{(0)}(\Lambda)}{\partial \Lambda}: &=
-\left(\frac{\frac{\partial}{\partial \Lambda} \omega}{1-\omega}\right)\partial_\mu \ln T^{(0)}(\Lambda)\,,\quad \omega:=\frac{\partial \ln w^{(0)}}{\partial \ln \Lambda}\\
	\\
:\partial_\nu \frac{\partial u^{\mu(0)}(\Lambda)}{\partial\Lambda}:&=
	\frac{\partial \ln v^{(0)} }{\partial \Lambda}\partial_\nu u^{\mu(0)}
	-\left(\frac{\frac{\partial}{\partial \Lambda}\nu }{1-\omega}\right)u^{\mu(0)}\partial_\nu
\ln T^{(0)}\,,\quad \nu:=\frac{\partial \ln v^{(0)}}{\partial \ln \Lambda}\,.
\end{aligned}
\end{gather}
Since $u^{\mu\infty}$ and $T^{\infty}$ follow conformal hydrodynamic equations,
\begin{equation}
\left(u^{\infty}\cdot \partial\right)\ln T^{\infty} = - \frac{1}{3}\left(\partial\cdot u^{\infty}\right) + \mathcal{O}(\partial^2).
\end{equation}
It follows from (\ref{useful4}) that
\begin{equation}\label{useful7}
(u^{\mu(0)}\cdot\partial) \ln T^{(0)}(\Lambda) = 
- \frac{1}{3}
\frac{(1- \omega)}{1+\frac{1}{3} \nu}(\partial\cdot u^{\mu(0)})+\mathcal{O}(\partial^2).
\end{equation}
Using (\ref{useful1}), (\ref{useful2}), (\ref{useful3}), (\ref{useful4}), (\ref{useful5}), (\ref{useful6}) and (\ref{useful7}) we finally obtain
\begin{gather}\label{uTevol1}
\begin{aligned}
\partial_{\Lambda}u^{\mu}&=
\left(\partial_{\Lambda}\ln v^{(0)}\right)u^{\mu}+\frac{1}{T^{\infty}}\left(\frac{\partial_{\Lambda}v^{(1)}_{{\rm v}}}{v^{(0)}}\right)(u\cdot\partial)u^{\mu}\\
&+\frac{1}{T^{\infty}}\bigg(\frac{\partial_{\Lambda}v^{(1)}_{{\rm s}}}{v^{(0)}}\\
&-\frac{1}{3}\left(\frac{\nu}{1+\frac{1}{3}\nu}\right)\left[
\frac{\partial}{\partial {\Lambda}}
+\left(
(\partial_{\Lambda}\ln v^{(0)})+\left(1-\frac{1}{v^{(0)}}\right)(\partial_{\Lambda}\ln\nu)
\right)\right](v^{(1)}_{{\rm v}}+v^{(1)}_{{\rm s}})\bigg)
u^{\mu}(\partial\cdot u)\\
\partial_{\Lambda}T(\Lambda)&=
(\partial_{\Lambda}\ln w^{(0)})T(\Lambda)\\
&+T(\Lambda)(\partial\cdot u)\tau^{(1)}\tau^{(2)}\bigg[
(\partial \ln (v^{(0)}w^{(1)}_{{\rm s}}))(1+\omega(w^{(0)}-1))\\
&-(1+\omega)w^{(0)}(\partial_{\Lambda}w^{(0)})+(w^{(0)}-1)\Lambda\frac{\partial^2_{\Lambda}w^{(0)}}{w^{(0)}}
\bigg]+\mathcal{O}(\partial^2)\\
\tau^{(1)}&:=\frac{1}{T^{\infty}}\left(\frac{w^{(1)}_{{\rm s}}}{w^{(0)}}\right)\,,\;
\tau^{(2)}=-\frac{1}{3}\frac{1}{1+\frac{1}{3}\nu}\,,\;
\nu:=\frac{\partial \ln v^{(0)}}{\partial \ln \Lambda}\,,\;
\omega:=\frac{\partial \ln w^{(0)}}{\partial \ln \Lambda}\,.
\end{aligned}
\end{gather}
We will call the above equations differential RG flow equations. Clearly, we can construct this differential form for the general coarse-graining up to any arbitrary order in the derivative expansion, using the manipulations described above. The structure will be as follows:
\begin{eqnarray}\label{uTevol}
:\frac{\partial u^{\mu} (\Lambda)}{\partial \Lambda}: &=& a^{(0)}(\Lambda) u^\mu(\Lambda) + \sum_{n=1}^\infty \sum_{m=1}^{n_{\rm s}}a^{(n,m)}_{\rm s}(\Lambda)\, \mathcal{S}^{(n,m)}_{\rm cf}(\Lambda)\, u^\mu(\Lambda) +\nonumber\\&&+\sum_{n=1}^\infty \sum_{m=1}^{n_{\rm v}}a^{(n,m)}_{\rm v}(\Lambda) \,{\mathcal{V}^\mu_{\rm cf}}^{(n,m)}(\Lambda)\, , \nonumber\\
:\frac{\partial T (\Lambda)}{\partial\Lambda}: &=& b^{(0)}(\Lambda)  + \sum_{n=1}^\infty \sum_{m=1}^{n_{\rm s}}b^{(n,m)}_{\rm s}(\Lambda) \mathcal{S}^{(n,m)}_{\rm cf}(\Lambda).
\end{eqnarray}
Above $\mathcal{S}^{(n,m)}_{\rm cf}(\Lambda)$ and ${\mathcal{V}^\mu_{\rm cf}}^{(n,m)}(\Lambda)$ are the hydrodynamic scalars and vectors constructed from $u^\mu(\Lambda)$ and $T(\Lambda)$ in a fixed conformally flat background metric $\eta_{\mu\nu}e^{2\sigma}$. At each order in derivative expansion, we get $2n_{\rm s}+ n_{\rm v}$ coefficients, namely $a^{(n,m)}_{\rm s}$, $a^{(n,m)}_{\rm v}$ and $b^{(n,m)}_{\rm s}$, which are as many as the coarse-graining functions which appear in (\ref{cguT}). It is to be noted that, since $u^\mu(\Lambda)\eta_{\mu\nu}e^{2\sigma} u^\nu(\Lambda) \neq 1$ for finite values of $\Lambda$, the acceleration $(u(\Lambda)\cdot \partial) u^\mu(\Lambda)$ is not orthogonal to $u^\mu(\Lambda)$. This is similarly the case for each member in ${\mathcal{V}^\mu_{\rm cf}}^{(n,m)}(\Lambda)$. We should keep this caveat in mind. Nevertheless it is still true that these give us a basis of independent vectors which are not proportional to $u^\mu(\Lambda)$ at each $\Lambda$.

The leading order coefficients in (\ref{uTevol}) are as in (\ref{uTevol1}), and are computed explicitly as below:
\begin{eqnarray}\label{uTevol-coeffs}
a^{(0)} &=&\frac{\partial\ln v^{(0)}}{\partial \Lambda} \,,\nonumber\\
a^{(1)}_{\rm s}&=& \frac{1}{T^{\infty}}\bigg(\frac{\partial_{\Lambda}v^{(1)}_{{\rm s}}}{v^{(0)}}\nonumber\\
&&-\frac{1}{3}\left(\frac{\nu}{1+\frac{1}{3}\nu}\right)\left[
\frac{\partial}{\partial {\Lambda}}
+\left(
(\partial_{\Lambda}\ln v^{(0)})+\left(1-\frac{1}{v^{(0)}}\right)(\partial_{\Lambda}\ln\nu)
\right)\right](v^{(1)}_{{\rm v}}+v^{(1)}_{{\rm s}})\bigg)\,,\nonumber\\
a^{(1)}_{\rm v}&=& \frac{1}{T^{\infty}}\left(\frac{\partial_{\Lambda}v^{(1)}_{{\rm v}}}{v^{(0)}}\right)\,,\nonumber\\
b^{(0)} &=&(\partial_{\Lambda}\ln w^{(0)})T(\Lambda) \,, \quad {\rm and}\nonumber\\
b^{(1)}_{\rm s}&=& T(\Lambda)\tau^{(1)}\tau^{(2)}\bigg[
(\partial_\Lambda \ln (v^{(0)}w^{(1)}_{{\rm s}}))(1+\omega(w^{(0)}-1))\nonumber\\
&&-(1+\omega)w^{(0)}(\partial_{\Lambda}w^{(0)})+(w^{(0)}-1)\Lambda\frac{\partial^2_{\Lambda}w^{(0)}}{w^{(0)}}
\bigg]+\mathcal{O}(\partial^2), \quad{\text{with}}\nonumber\\
\tau^{(1)}&:=&\frac{1}{T^{\infty}}\left(\frac{w^{(1)}_{{\rm s}}}{w^{(0)}}\right)\,,\;
\tau^{(2)}=-\frac{1}{3}\frac{1}{1+\frac{1}{3}\nu}\,,\;
\nu:=\frac{\partial \ln v^{(0)}}{\partial \ln \Lambda}\,,\;
\omega:=\frac{\partial \ln w^{(0)}}{\partial \ln \Lambda}\,.
\end{eqnarray}

We will now see how we can directly derive similar flow equations from the $(d+1)-$dimensional gravity equations. This will be useful to show how we can reconstruct the coarse-graining described here as $(d+1)-$dimensional gravity equations.

\section{Flow equations from gravity}\label{Flow}

\subsection{The flow equations in UV expansion}\label{gravity-flow-UV-expansion}

In the previous Section, we have taken the first steps of constructing the RG flow in the field theory, in the limit when the exact asymptotic hydrodynamic expansion is valid. Our goal is to see how gravity equations emerge from the coarse-graining (\ref{cguT}) of ${u^\mu}^\infty$ and $T^\infty$, which in turn defines the coarse-grained operator $t^\mu_{\phantom{\mu}\nu}(\Lambda)$ constructively. 

In order to achieve this, we need to independently work on the gravity side also. This is because the traditional AdS/CFT correspondence merely relates asymptotic charges and quantities in gravity to observables in field theory. It is necessary to understand how classical gravity equations can be naturally recast as flow equations of the type (\ref{uTevol}) involving field-theoretic observables only, which will contain complete information of the \textit{emergent} spacetime. This has been the focus of the first part \cite{Behr:2015yna} of this work. We summarize the necessary key results of  \cite{Behr:2015yna} very briefly below. 

The first important proposition of \cite{Behr:2015yna} is as follows: let us consider the $d-$dimensional scale evolution of $t^\mu_{\phantom{\mu}\nu}(\Lambda)$ of the form:
\begin{equation}\label{schematic1}
\frac{\partial}{\partial \Lambda}t^\mu_{\phantom{\mu}\nu}(\Lambda) = F^\mu_{\phantom{\mu}\nu}[t^\mu_{\phantom{\mu}\nu}(\Lambda),\Lambda ],
\end{equation}
in the \textit{fixed} background metric $g^{\rm (b)}_{\mu\nu}$, such that there exists a background metric $g_{\mu\nu}(\Lambda)$ which is a functional of $t^\mu_{\phantom{\mu}\nu}(\Lambda)$ and $\Lambda$ in the same \textit{fixed} background metric $g^{\rm (b)}_{\mu\nu}$ taking the form
\begin{equation}\label{schematic2}
g_{\mu\nu}(\Lambda) = G_{\mu\nu}[t^\mu_{\phantom{\mu}\nu}(\Lambda),\Lambda ],
\end{equation}
at each $\Lambda$, and in which $t^\mu_{\phantom{\mu}\nu}(\Lambda)$ satisfies the local conservation equation
\begin{equation}\label{WILambda}
\nabla_{(\Lambda)\mu} t^\mu_{\phantom{\mu}\nu}(\Lambda) = 0
\end{equation}
with $\nabla_{(\Lambda)}$ being the covariant derivative constructed from $g_{\mu\nu}(\Lambda)$. Note that $g_{\mu\nu}(\Lambda)$ has to coincide with the fixed background metric $g^{\rm (b)}_{\mu\nu}$ at $\Lambda = \infty$ in which the functionals $F$ and $G$ are constructed, because ${t^\mu_{\phantom{\mu}\nu}}^\infty$ should satisfy $\nabla_{\rm(b)\mu}{t^\mu_{\phantom{\mu}\nu}}^\infty = 0$, with $\nabla_{\rm (b)}$ being the covariant derivative constructed from $g^{\rm (b)}_{\mu\nu}$. We claim that it follows that $g_{\mu\nu}(\Lambda)$ then gives a bulk metric in the Fefferman-Graham gauge:
\begin{equation}\label{FGmetric}
{\rm d}s^2 = \frac{l^2}{r^2}\left({\rm d}r^2 + g_{\mu\nu}(r,x) {\rm d}x^2\right),
\end{equation}
which solves the equations of a \textit{pure} $(d+1)-$classical gravity theory with \textit{full} $(d+1)-$diffeomorphism invariance and a negative cosmological constant determined by the asymptotic curvature radius $l$, and with $r$ identified with $\Lambda^{-1}$ (i.e. $r = \Lambda^{-1}$). In this paper, for the sake of convenience, we will choose the fixed background metric $g^{\rm (b)}_{\mu\nu}$ to be flat $\eta_{\mu\nu}$ or conformally flat $e^{2\sigma(x)}\eta_{\mu\nu}$.

At this point, it may be useful to provide an example. Let us consider the RG flow equation in the flat Minkowski space $\eta_{\mu\nu}$ in $4$ spacetime dimensions:
\begin{eqnarray}\label{t-rg-example}
\frac{\partial t^\mu_{\phantom{\mu}\nu}(\Lambda)}{\partial \Lambda} &=& \frac{1}{\Lambda^3}\cdot\frac{1}{2} \Box t^{\mu}_{\phantom{\mu}\nu}(\Lambda)- \frac{1}{\Lambda^5}\cdot\left(\frac{1}{4}\, \delta^\mu_{\phantom{\mu}\nu}
{t^\alpha_{\phantom{\alpha}\beta}}(\Lambda)
{t^\beta_{\phantom{\beta}\alpha}}(\Lambda) - \frac{7}{128}\,\Box^2 {t^\mu_{\phantom{\mu}\nu}}(\Lambda)\right)-
\nonumber\\&&+\frac{1}{\Lambda^5}\, \log\, \Lambda \cdot \frac{1}{32}\cdot \Box^2 {t^\mu_{\phantom{\mu}\nu}}(\Lambda)
+\mathcal{O}\left(\frac{1}{\Lambda^7}\, \log\, \Lambda\right).
\end{eqnarray}
For the above RG flow, we can indeed construct the following $g_{\mu\nu}(\Lambda)$:
\begin{eqnarray}\label{g-example}
g_{\mu\nu}(\Lambda) &=& \eta_{\mu\nu} +\, \frac{1}{\Lambda^4}\cdot\frac{1}{4} \eta_{\mu\alpha}{t^\alpha_{\phantom{\alpha}\nu}}(\Lambda)
+\,\frac{1}{\Lambda^6}\cdot \frac{1}{24}\eta_{\mu\alpha}\Box {t^\alpha_{\phantom{\alpha}\nu}}(\Lambda)+
\nonumber\\&&
+ \frac{1}{\Lambda^8} \cdot \left(\frac{1}{32}\,\eta_{\mu\alpha} {t^\alpha_{\phantom{\alpha}\rho}}(\Lambda)
{t^\rho_{\phantom{\rho}\nu}}(\Lambda) -\frac{7}{384}\, \eta_{\mu\nu}{t^\alpha_{\phantom{\alpha}\beta}}(\Lambda)
{t^\beta_{\phantom{\beta}\alpha}}(\Lambda) +\frac{11}{1536}\,\eta_{\mu\alpha}\Box^2 {t^\alpha_{\phantom{\alpha}\nu}}(\Lambda)\right)-
\nonumber\\&& + \frac{1}{\Lambda^8}\,\log \, \Lambda\cdot \frac{1}{516} \cdot \eta_{\mu\alpha}\Box^2 {t^\alpha_{\phantom{\alpha}\nu}}(\Lambda)+ \mathcal{O}\left(\frac{1}{\Lambda^{10}}\, \log\, \Lambda\right), 
\end{eqnarray}
as a functional of $t^\mu_{\phantom{\mu}\nu}(\Lambda)$ and $\Lambda$ in the flat Minkowski space background at each $\Lambda$ such that, when it is considered as an effective background metric, the scale-dependent Ward identity (\ref{WILambda}) is satisfied ( given that in the UV $\partial_\mu {t^\mu_{\phantom{\mu}\nu}}^\infty = 0)$. Furthermore, the $5-$dimensional bulk metric (\ref{FGmetric}) then satisfies Einstein's equations with the cosmological constant set to $-6/l^2$, and with $r = \Lambda^{-1}$. The derivation of the above equations is given in \cite{Behr:2015yna}. The \textit{log} term in (\ref{t-rg-example}) is related to the conformal anomaly.

It is to be noted that the Ward identity (\ref{WILambda}) can be recast as an effective operator equation. For example, in the above case (\ref{WILambda}) reduces to
\begin{eqnarray}
\partial_\mu t^\mu_{\phantom{\mu}\nu}(\Lambda) &=& \frac{1}{\Lambda^4}\cdot\left(\frac{1}{16}\partial_\nu \left(t^\alpha_{\phantom{\alpha}\beta}(\Lambda)t^\beta_{\phantom{\beta}\alpha}(\Lambda)\right)-\frac{1}{8}t^\mu_{\phantom{\mu}\nu}(\Lambda)\partial_\mu\, {\rm Tr}\,t(\Lambda) \right)+\nonumber\\&&
+\frac{1}{\Lambda^6}\cdot\left(\frac{1}{48}t^\alpha_{\phantom{\alpha}\beta}(\Lambda)\partial_\nu\Box t^\beta_{\phantom{\beta}\alpha}(\Lambda)-\frac{1}{48}t^\mu_{\phantom{\mu}\nu}(\Lambda)\partial_\mu\Box\, {\rm Tr}\,t(\Lambda) \right)
+\mathcal{O}\left(\frac{1}{\Lambda^8}\right).
\end{eqnarray}
We can now see that the usual Ward identity is broken at a finite scale by contributions due to multi-trace operators built from $t^\mu_{\phantom{\mu}\nu}(\Lambda)$. Therefore, the scale-dependent effective background $g_{\mu\nu}(\Lambda)$ as given by (\ref{g-example}) serves to absorb these multi-trace contributions in a manner such that the effective Ward identity preserves its form (\ref{WILambda}) at each scale.

The second important proposition of \cite{Behr:2015yna} is that although for a given classical gravity theory, there exists different choices of the functional $F^\mu_{\phantom{\mu}\nu}$ in (\ref{schematic1}) (and therefore the associated functional $G_{\mu\nu}$ in (\ref{schematic2})), such that we reproduce the same $g_{\mu\nu}(\Lambda)$ and hence the same bulk metric, \textit{only} unique choices of $F^\mu_{\phantom{\mu}\nu}$ and therefore the associated functional $G_{\mu\nu}$ (up to an overall numerical constant for normalisation of $t^\mu_{\phantom{\mu}\nu}$) leads to the good endpoint behaviour, where the endpoint can be transformed to a non-relativistic fixed point with finite number of parameters under the rescaling (\ref{rescale-intro}), for appropriate values of UV data. In the context of Einstein's gravity, these unique choices are given by (\ref{t-rg-example}) and (\ref{g-example}) for $F^\mu_{\phantom{\mu}\nu}$ and $G_{\mu\nu}$ respectively. This claim has not been completely proven in \cite{Behr:2015yna} -- we will establish this completely in the following subsection.

The above propositions for rewriting classical gravity theory as a highly efficient RG flow works also when we include higher derivative corrections to Einstein's gravity, provided these corrections are treated perturbatively. In what follows in the remainder of this subsection, we identify how to relate the scale $\Lambda$, the effective operator $t^\mu_{\phantom{\mu}\nu}(\Lambda)$ and the effective metric $g_{\mu\nu}(\Lambda)$ to quantities on the gravity side in a gauge-independent manner. We will repeatedly use the requirements that these identifications should be state-independent and that $l$, the asymptotic curvature radius of the dual spacetime which has no direct interpretation in the field theory should not appear explicitly either in (\ref{schematic1}) or in (\ref{schematic2}). Since diffeomorphism invariance of the dual classical gravity equations, of which the bulk metric (\ref{FGmetric}) is a solution, is a necessary and sufficient criterion for the high efficiency of the RG flow, as demonstrated in \cite{Behr:2015yna}, we will utilise the unique map between gravity and highly efficient RG flow established in \cite{Behr:2015yna} to explicitly construct the latter.

Let us first identify the radial coordinate $r$ in the dual geometry with the scale $\Lambda$ of the RG flow. Requiring that the identification is independent of, (i) the field theory state (therefore also of the gravity solution and the gauge fixing of the diffeomorphism symmetry on the gravity side), (ii) $l$, the asymptotic curvature radius of the geometry and also (iii) the field-theory coordinates, we obtain the general rule $r=\Lambda^{-1}$ from simple dimensional analysis. 

The immediate question is whether a \textit{unique} object exists on the gravity side, such that it can be defined on the hypersurfaces $r = {\rm constant}$ for any choice of hypersurface foliation of the $(d+1)-$dimensional geometry, and which can be naturally identified with $t^\mu_{\phantom{\mu}\nu}(\Lambda)$ of the field theory RG flow at $r = \Lambda^{-1}$. Most importantly, this $t^\mu_{\phantom{\mu}\nu}(\Lambda)$ should satisfy (\ref{WILambda}) for all $\Lambda$, with $\nabla_{(\Lambda)}$ being the covariant derivative constructed from an appropriate metric $g_{\mu\nu}(\Lambda)$. Obviously, this $t^\mu_{\phantom{\mu}\nu}(\Lambda)$ has to agree with the holographic $\langle {t^\mu_{\phantom{\mu}\nu}}^\infty\rangle$, the expectation value of the microscopic operator in the field-theory state dual to the geometry at $r =  0$ (i.e. $\Lambda = \infty$), which is the asymptotic boundary. In \cite{Behr:2015yna}, we have shown that indeed such a  $t^\mu_{\phantom{\mu}\nu}(\Lambda)$ exists uniquely for any $(d+1)-$dimensional classical theory of gravity, up to an overall multiplication constant, provided
\begin{enumerate}
\item the scale evolution equations for $t^\mu_{\phantom{\mu}\nu}(\Lambda)$ are independent of $l$, the asymptotic curvature radius of the dual geometry,
\item assumes a form independent of the choice of gauge-fixing of the diffeomorphism symmetry and is also independent of the gravity solution, and
\item it has an appropriate infrared end point, such that it becomes an appropriate fixed point, like incompressible non-relativistic Navier-Stokes equations, under an appropriate rescaling of the scale evolution equations to be explicitly stated later.
\end{enumerate}
The first requirement implies that $l$ appears in $t^\mu_{\phantom{\mu}\nu}(\Lambda)$ only in the combination $l^{d-1}/ (16\pi G_N)$ (which is identified as the $N^2$ of the large $N$ limit) with $G_N$ the $(d+1)-$dimensional Newton's constant, or via $\alpha^\prime_{(i)}/l^2$, where $\alpha^\prime_{(i)}$ with dimensions of length squared are the coupling constants associated with higher derivative corrections to Einstein's equations (and are identified with the couplings of the dual conformal field theory). All the three requirements together imply that $t^\mu_{\phantom{\mu}\nu}(\Lambda)$ in any theory of classical gravity can be written as:
\begin{eqnarray}\label{tgravity1}
t^\mu_{\phantom{\mu}\nu}(r=\Lambda^{-1}) = \left(\frac{l}{r}\right)^d \left( {T^\mu_{\phantom{\mu}\nu}}^{\rm ql}+ {T^\mu_{\phantom{\mu}\nu}}^{\rm ct}\right).
\end{eqnarray}
Above, ${T^\mu_{\phantom{\mu}\nu}}^{\rm ql}$ is the unique quasi-local stress tensor of the classical gravity theory. In the case of Einstein's gravity, it is given by
\begin{equation}\label{BY}
{T^\mu_{\phantom{\mu}\nu}}^{\rm ql} = - \frac{1}{8\pi G_N}\left(K^\mu_{\phantom{\mu}\nu} - K\,
 \delta^\mu_{\phantom{\mu}\nu}\right)\, ,
\end{equation} 
and is also known as the Brown-York stress tensor. The quasi-local stress tensor has one less radial derivative of the metric than the classical gravity equations. It can be shown that by virtue of the $(d+1)-$Bianchi identities of the gravity equations, which follow from $(d+1)-$diffeomorphism invariance, such a ${T^\mu_{\phantom{\mu}\nu}}^{\rm ql}$ should exist in any classical gravity theory such that it satisfies $\nabla_{(\gamma)\mu}{T^\mu_{\phantom{\mu}\nu}}^{\rm ql} = 0$ along all hypersurfaces $r = \text{constant}$, and with $\nabla_{(\gamma)}$ being the covariant derivative constructed from $\gamma_{\mu\nu}$, the induced metric on this hypersurface \cite{Deruelle:2007pt} (see also \cite{Balcerzak:2007da}). $ {T^\mu_{\phantom{\mu}\nu}}^{\rm ct}$ is the sum of possibly infinite number of counter-terms ${T^\mu_{\phantom{\mu}\nu}}^{\rm ct}_{(n)}$ which \textit{identically} satisfy $\nabla_{(\gamma)\mu}{T^\mu_{\phantom{\mu}\nu}}^{\rm ct}_{(n)} = 0$. Clearly ${T^\mu_{\phantom{\mu}\nu}}^{\rm ql}$ and each counterterm ${T^\mu_{\phantom{\mu}\nu}}^{\rm ct}_{(n)}$ can be multiplied by a function of $r$ only, so that $\nabla_{(\gamma)\mu}{T^\mu_{\phantom{\mu}\nu}}^{\rm ql} = 0$ and $\nabla_{(\gamma)\mu}{T^\mu_{\phantom{\mu}\nu}}_{(n)}^{\rm ct} = 0$ directly imply $\nabla_{(\gamma)\mu}{t^\mu_{\phantom{\mu}\nu}} = 0$ in an arbitrary solution at a fixed $r$ -- these functions of $r$ can be determined simply by requiring that $l$ does not appear explicitly in the scale-evolution equations as mentioned above \cite{Behr:2015yna}. This results in an overall multiplication constant by $(l/r)^d$ as in (\ref{tgravity1}), and we can also similarly conclude that the coefficients of each counter-term should be a numerical constant, which depends on the gravitational coupling constants $\alpha^\prime_{(i)}/l^2$ only. Therefore,
\begin{eqnarray}\label{tgravity2}
{T^\mu_{\phantom{\mu}\nu}}^{\rm ct} &=& - \frac{1}{8\pi G_N}\left(C_{(0)}\Bigg(\alpha^\prime_{(i)}/l^2\right)\cdot \frac{1}{l}\cdot \delta^\mu_{\phantom{\mu}\nu}+C_{(2)}\left(\alpha^\prime_{(i)}/l^2\right)\cdot l\cdot \left(R^\mu_{\phantom{\mu}\nu}[\gamma] -\frac{1}{2}R[\gamma]\delta^\mu_{\phantom{\mu}\nu} \right)+\cdots \Bigg).
\end{eqnarray}
The requirement of an appropriate infrared end point uniquely determines the coefficients $C_{(n)}$s \cite{Kuperstein:2013hqa} (more on this in the next subsection). A finite number of counter-terms are usually fixed by UV finiteness of the gravitational action in the AdS/CFT correspondence  \cite{Henningson:1998ey,Balasubramanian:1999re,deHaro:2000xn}. Remarkably, even these counterterms are also determined by our requirement regarding the infrared end point \cite{Kuperstein:2013hqa} as we will elaborate later. This feature is crucial for interpreting gravity as a RG flow -- since the RG flow is first order, therefore  it should be determined completely by the constraints imposed at a specific scale, which in this case is a finite scale where the RG flow naturally ends. Thus we obtain 
\begin{equation}\label{coeffs}
C_{(0)} = d-1, \quad C_{(2)} = -\frac{1}{d-2}, \, \cdots \, ,
\end{equation}
in the case of Einstein's gravity. It is to be noted that the identification (\ref{tgravity1}) makes a choice of overall normalisation of $t^\mu_{\phantom{\mu}\nu}(\Lambda)$ -- this can also be fixed independently from the field theory side by obtaining the two point vacuum correlation function at $\Lambda = \infty$, which can also be obtained from our approach. For the sake of convenience, we use a normalisation from now on which involves a rescaling of (\ref{tgravity1}) by a factor of $(16\pi G_N)/l^{d-1}$.

Finally the unique $g_{\mu\nu}(\Lambda)$ which is a functional of $t^\mu_{\phantom{\mu}\nu}(\Lambda)$ can be obtained by from the following requirements:
\begin{enumerate}
\item $\nabla_{(\Lambda)\mu} t^\mu_{\phantom{\mu}\nu}(\Lambda) = 0$ at any arbitrary $\Lambda$, with $\nabla_{(\Lambda)}$ being the covariant derivative constructed from $g_{\mu\nu}(\Lambda)$,
\item it is independent of the gravity solution, and
\item its scale evolution also  does not depend on $l$, the asymptotic curvature radius explicitly.
\end{enumerate}
It turns out that $g_{\mu\nu}(\Lambda)$ is then related to the induced metric $\gamma$ at $r = \Lambda^{-1}$ via:
\begin{equation}\label{gphys}
g_{\mu\nu}(\Lambda = r^{-1}) = \frac{r^2}{l^2}\gamma_{\mu\nu},
\end{equation}
when the bulk metric satisfies the dual classical gravity equations. This effective physical metric $g_{\mu\nu} (\Lambda)$ also coincides with the fixed background metric (i.e. the boundary metric) at $\Lambda =\infty$. It can be easily obtained by inverting (\ref{tgravity1}) in the case of Einstein's gravity in Fefferman-Graham gauge, which can be systematically done in the UV expansion, i.e. in a power series in $\Lambda^{-1}$ to reproduce (\ref{g-example}). For more details on the derivation of (\ref{t-rg-example}) and (\ref{g-example}) and further clarifications on the various assertions made in this subsection, the reader may wish to consult \cite{Behr:2015yna}. 

Although the identifications (\ref{gphys}) and (\ref{tgravity1}) are independent of the choice of gauge on the gravity side, the scale evolution equations (\ref{schematic1}) is strictly state-independent only when the corresponding gauge choice on the gravity side is Feffermam-Graham gauge. In other gauges, auxiliary non-dynamical variables corresponding to the lapse function and shift vector on the gravity side, appear in (\ref{schematic1}). We will return to this issue in Section \ref{diffeo}.

\subsection{Flow equations for reconstructing fluid/gravity correspondence}\label{Kuperstein-Mukhopadhyay}
The flow equations of the form (\ref{t-rg-example}) give us a reconstruction of the classical gravity (Einstein's) equations in a power series in $\Lambda^{-1}$ -- this power series is typically valid for $\Lambda \gg \Lambda_{\rm IR}$, where $\Lambda_{\rm IR}$ is a solution (i.e. state) dependent scale related to the inverse of the value of the radial coordinate where the corresponding Fefferman-Graham coordinates break down. Typically, this corresponds to the location of the event horizon at very late time. Although this UV expansion is generally valid, it is not useful in the context of hydrodynamics. In order to make sure that the flow reaches the right infrared end point which satisfies our requirement, we need to sum over all orders in $\Lambda^{-1}$ at each order in the derivative expansion.

The \textit{fluid/gravity correspondence} is a map between a class of solutions of the classical gravity equations, that represent the long-wavelength perturbations of the black brane, to non-linear fluid mechanics living in the asymptotic boundary of the spacetime \cite{Rangamani:2009xk}. As mentioned before, this allows us to obtain the exact asymptotic hydrodynamic expansion of the dual strongly coupled QFT in the AdS/CFT correspondence. In order to achieve our goal stated above, we need to follow the method of \cite{Kuperstein:2013hqa}, co-developed by one of the authors, to reconstruct the fluid/gravity correspondence as first order flow equations of the hydrodynamic variables and transport coefficients. The method of \cite{Kuperstein:2013hqa} is based on \cite{Kuperstein:2011fn}, where a precise strategy for obtaining effective hydrodynamic variables at a given scale from the dual classical gravity theory was first developed. The latter work has been inspired by \cite{Bredberg:2010ky}, where the important role of near-horizon dynamics for the leading order hydrodynamic behaviour in the fluid/gravity correspondence has been first emphasised. Here we are going to take an additional step to rewrite classical gravity equations in the stated limit as first order flow equations of hydrodynamic variables, which take the form (\ref{uTevol}) in the fixed background metric. Along the way, we will clarify why the infrared criterion determines the gravitational counterterms (\ref{coeffs}) and thus fixes the scale evolution equation for $t^\mu_{\phantom{\mu}\nu}(\Lambda)$, namely (\ref{t-rg-example}), uniquely.

Our first step is to make the assumption that $t^\mu_{\phantom{\mu}\nu}(\Lambda)$, the scale-dependent stress-tensor (\ref{tgravity1}) on the gravity side at $r = \Lambda^{-1}$, is a relativistic hydrodynamic energy-momentum tensor living in the physical metric $g_{\mu\nu}(\Lambda)$ that is related to the induced metric via (\ref{gphys}) on the \textit{same} hypersurface. This automatically implies that it is also assumed that $g_{\mu\nu}(\Lambda)$ is weakly curved for all values of $\Lambda > \Lambda_{\rm IR}$, where $\Lambda_{\rm IR}$ is related to the final equilibrium temperature, and the effective mean-free path that controls the derivative expansion is also $\Lambda-$dependent. Thus $t^\mu_{\phantom{\mu}\nu}(\Lambda)$ is assumed to take the standard hydrodynamic form (\ref{emhydroscale}) at each scale with scale-dependent transport coefficients and equation of state. This assumption also implies that the effective $u^\mu(\Lambda)$ and $T(\Lambda)$ should satisfy the standard Landau-Lifshitz definitions at each scale.

The second step is to make the assumption that the first order evolution of the effective hydrodynamic variables $u^\mu(\Lambda)$ and $T(\Lambda)$ takes the form: 
\begin{eqnarray}\label{uTevolgrav}
\frac{\partial u^{\mu} (\Lambda)}{\partial \Lambda} &=& \alpha^{(0)}(\Lambda) u^\mu(\Lambda) + \sum_{n=1}^\infty \sum_{m=1}^{n_{\rm s}}\alpha^{(n,m)}_{\rm s}(\Lambda)\, \mathcal{S}^{(n,m)}(\Lambda)\, u^\mu(\Lambda) +\nonumber\\&&+\sum_{n=1}^\infty \sum_{m=1}^{n_{\rm v}}\alpha^{(n,m)}_{\rm v}(\Lambda) \,{\mathcal{V}^\mu}^{(n,m)}(\Lambda), \nonumber\\
\frac{\partial T (\Lambda)}{\partial\Lambda} &=& \beta^{(0)}(\Lambda)  + \sum_{n=1}^\infty \sum_{m=1}^{n_{\rm s}}\beta^{(n,m)}_{\rm s}(\Lambda) \mathcal{S}^{(n,m)}(\Lambda).
\end{eqnarray}
Although, the above seems very similar to the flow equations (\ref{uTevol}), there is a crucial difference. The hydrodynamic scalars and vectors which appear in the latter, namely $\mathcal{S}^{(n,m)}_{\rm cf}$ and ${\mathcal{V}^\mu_{\rm cf}}^{(n,m)}$ are constructed in the \textit{fixed} background conformally flat metric, while those appearing above are constructed in the scale-dependent effective physical metric $g_{\mu\nu}(\Lambda)$, where $u^\mu(\Lambda)$ satisfies the normalisation condition $u^\mu(\Lambda)g_{\mu\nu}(\Lambda) u^\nu(\Lambda) = -1$. This assumption amounts to the statement that the flow equations of the hydrodynamic variables themselves have a systematic derivative expansion at any arbitrary scale in the scale-dependent background metric $g_{\mu\nu}(\Lambda)$. This is also to be justified a posteriori by showing that we recover the classical bulk metric of the fluid/gravity correspondence, and the derivative expansion actually improves in the sense of converging faster, as we evolve with the scale $\Lambda$.

The most remarkable point is that in order to rewrite the classical gravity equations as the first order flow equations (\ref{uTevolgrav}), we do not need to make any assumption about the effective metric $g_{\mu\nu}(\Lambda)$ itself other than it is weakly curved, nor do we need to know it explicitly. For the purpose of concrete demonstration, we will take the case of Einstein's gravity for the rest of this subsection, although our method can be carried over to any classical theory of gravity admitting a black brane solution, which is crucial for the hydrodynamic expansion to make sense. Furthermore, we are going to work in the Fefferman-Graham gauge. We can readily construct the flow equations in any other gauge using the method of our previous work \cite{Behr:2015yna} via appropriate transformations (cf. Section \ref{diffeo}).

With $g_{\mu\nu}(\Lambda)$ in the Fefferman-Graham gauge (\ref{FGmetric}), it is useful to define
\begin{equation}
z^\mu_{\phantom{\mu}\nu} = g^{\mu\rho}\frac{\partial g_{\rho\nu}}{\partial r} = -\Lambda^2 g^{\mu\rho}\frac{\partial g_{\rho\nu}}{\partial \Lambda}.
\end{equation}
The expression (\ref{tgravity1}) for the gravitational $t^\mu_{\phantom{\mu}\nu}(\Lambda)$ with the appropriate counterterm coefficients (\ref{coeffs}) in the Fefferman-Graham gauge can be written as:
\begin{eqnarray}\label{t-z4d}
t^\mu_{\phantom{\mu}\nu} &=& \frac{1}{r^3}\cdot \left(z^\mu_{\phantom{\mu}\nu} - ({\rm Tr}\, z) \,\delta^\mu_{\phantom{\mu}\nu}\right) + \frac{1}{r^2} \left(R^\mu_{\phantom{\mu}\nu}[g] - \frac{1}{2}R[g]\delta^\mu_{\phantom{\mu}\nu}\right)
+ \nonumber\\&& 
+\cdots \, .
\end{eqnarray}
We switch back to using the radial coordinate $r$ instead of $\Lambda$.

The third step is to use simple linear algebra to invert this relation to obtain:
\begin{equation}
\label{t-z}
z^{\mu}_{\phantom{\mu}\nu} = r^{d-1}\Bigg(t^\mu_{\phantom{\mu}\nu} - \frac{\text{Tr} \, t}{d-1}  \delta^\mu_{\phantom{\mu}\nu}\Bigg) - \frac{2r}{d-2} \Bigg(R^{\mu}_{\phantom{\mu}\nu} - \frac{R}{2(d-1)} \delta^\mu_{\phantom{\mu}\nu}\Bigg)+ ... \, .
\end{equation}
and substitute the above in Einstein's equation for the radial evolution of $g_{\mu\nu}$, namely \footnote{The $r\mu$ and $rr$ components of Einstein's equations are constraints. Once these are imposed at one hypersurface, i.e. at a given value of $r$, they are satisfied everywhere. The first (i.e. vector) constraint amounts to energy-momentum conservation, giving the fluid equations at the hypersurface. The second (i.e. scalar) constraint can be ensured by requiring that the boundary stress tensor is Weyl covariant up to the conformal anomaly, which implies that the scalar transport coefficients and some combinations of the tensor transport coefficients should vanish at the boundary. As our criterion for the infrared end point will be remarkably compatible with the scalar constraint, this will be good enough also to ensure that it is satisfied as well.}:
\begin{equation}\label{tensor-equation}
\frac{\partial}{\partial r}{z^\mu_{\phantom{\mu}\nu}} - \frac{d-1}{r} z^\mu_{\phantom{\mu}\nu} + {\rm Tr}\, z\left(\frac{1}{2}z^\mu_{\phantom{\mu}\nu}-\frac{1}{r}\delta^\mu_{\phantom{\mu}\nu}\right)
= 2 \, R^\mu_{\phantom{\mu}\nu}.
\end{equation}
It is understood in the above and from now on all the curvature tensors and other tensors are constructed out of the effective metric $g$. Using the identities
\begin{eqnarray}\label{ids}
\frac{\partial\Gamma^{\mu}_{\nu\rho} }{\partial r}&=& \frac{1}{2}\left(\nabla_\nu z^\mu_{\phantom{\mu}\rho} + \nabla_\rho z^\mu_{\phantom{\mu}\nu}- \nabla^\mu z_{\nu\rho}\right), \\
\frac{\partial R^{\mu}_{\phantom{\mu}\nu\rho\sigma}}{\partial r}
&=& \frac{1}{2} \left(  \nabla_\rho \nabla_\nu z^\mu_{\phantom{\mu}\sigma}  - \nabla_\sigma \nabla_\nu z^\mu_{\phantom{\mu}\rho}- \nabla_\rho \nabla^\mu z_{\nu\sigma} + \nabla_\sigma \nabla^\mu z_{\nu\rho} \right)
\nonumber\\&&  + \frac{1}{2}\left(  R^{\mu}_{\phantom{\mu}\kappa\rho\sigma}
z^\kappa_{\phantom{\mu}\nu} -
R^{\kappa}_{\phantom{\mu}\nu\rho\sigma}
z^\mu_{\phantom{\mu}\kappa} \right) \, , \quad \text{etc.,}
\end{eqnarray}
and using the inversion (\ref{t-z}) recursively we obtain,
\begin{eqnarray}
\label{EMTensorEoM-Full}
&&
\frac{\partial t^\mu_{\phantom{\mu} \nu}}{\partial r} - \frac{2 r^{2-d}}{d-2} \frac{\partial R^\mu_{\phantom{\mu} \nu}}{\partial r} - 
 \dfrac{r^{d-1}}{2(d-1)}\left( \textrm{Tr}\, t + r^{2-d} R \right) \left( t^\mu_{\phantom{\mu} \nu} - \frac{2 r^{2-d}}{d-2} R^\mu_{\phantom{\mu} \nu} \right) +
\nonumber\\
&& 
\,
 + \frac{1}{d-1}\Bigg( - \frac{\partial \textrm{Tr}\, t}{\partial r} + \dfrac{r^{2-d}}{d-2} \frac{\partial R}{\partial r}+ \frac{ \textrm{Tr}\, t}{r} + \nonumber\\ &&
 \qquad\qquad
 +\frac{r^{d-1}}{2(d-1)} 
 \left( \textrm{Tr} \,t + r^{2-d} R \right) \left( \textrm{Tr}\, t - \dfrac{r^{2-d}}{d-2} R \right) \Bigg) \delta^\mu_{\phantom{\mu} \nu} +\nonumber\\ &&
 \qquad\qquad+ \text{terms relevant at third order in derivatives}
 = 0 \, .
\end{eqnarray}
Note, although the above equation is of the form of first order evolution of $t^\mu_{\phantom{\mu} \nu}$, it is not yet useful because it requires the explicit knowledge of the explicit metric $g$ out of which the curvature and other tensors are built.

The fourth step is to do substitute (\ref{uTevolgrav}) and the hydrodynamic form (\ref{emhydroscale}) of $t^\mu_{\phantom{\mu} \nu}(r =\Lambda^{-1})$ in (\ref{EMTensorEoM-Full}) and:
\begin{itemize}
\item obtain the \textit{algebraic equations} for the parameters $\alpha^{(0)}$, $\alpha^{(n,m)}_{\rm s}$, $\alpha^{(n,m)}_{\rm v}$, $\beta^{(0)}$ and $\beta^{(n,m)}_{\rm s}$ appearing in (\ref{uTevolgrav}) as functions of $\epsilon$, $P$ and the transport coefficients $\gamma^{(n,m)}_{\rm s}$ and $\gamma^{(n,m)}_{\rm t}$, solve for them explicitly, and then 
\item obtain the \textit{first order ordinary differential equations} for the radial (scale) evolution of $\epsilon$, $P$ and the transport coefficients $\gamma^{(n,m)}_{\rm s}$ and $\gamma^{(n,m)}_{\rm t}$.
\end{itemize}
Thus the equations of classical gravity can be reduced to just first order flow of equation of state and transport coefficients. While substituting (\ref{uTevolgrav}) and the hydrodynamic form (\ref{emhydroscale}) of $t^\mu_{\phantom{\mu} \nu}(r =\Lambda^{-1})$ in (\ref{EMTensorEoM-Full}), we need to use the identities (\ref{ids}) and the inversion (\ref{t-z}) recursively, to achieve the above goals. However, these identities are not enough as we can readily see that (\ref{EMTensorEoM-Full}) gives us in total $2n_{\rm s} + n_{\rm v} +n_{\rm t}$ equations at $n-$th order in derivative expansion, whereas the unknowns, namely $\alpha^{(n,m)}_{\rm s}$, $\alpha^{(n,m)}_{\rm v}$, $\beta^{(n,m)}_{\rm s}$, $\gamma^{(n,m)}_{\rm s}$ and $\gamma^{(n,m)}_{\rm t}$ are in total $3n_{\rm s} + n_{\rm v} +n_{\rm t}$ in number. The remaining $n_{\rm s}$ equations are obtained from imposing the norm of $u^\mu(\Lambda)$, i.e. by demanding that $u^\mu(\Lambda)g_{\mu\nu}(\Lambda)u^\nu(\Lambda) = -1$. The norm condition cannot be used directly as we do not know $g_{\mu\nu}(\Lambda)$ explicitly yet, however we can impose the norm condition via its radial derivative which implies:
\begin{equation}\label{dnorm}
2\frac{\partial u^\mu}{\partial r}u_\mu + u^\mu z_{\mu\nu}u^\nu = 0, \quad \text{with $z_{\mu\nu}= g_{\mu\rho}z^\rho_{\phantom{\rho}\nu}$.} 
\end{equation}
Substituting $t$ in place of $z$ above, using its explicit hydrodynamic form (\ref{emhydroscale}) and also the flow equation for $u^\mu$ as given in (\ref{uTevolgrav}), we obtain the missing $n_s$ equations at the $n-$th order in the derivative expansion. 

It is not hard to see that since only first order $r-$derivative of $t$ appears in (\ref{EMTensorEoM-Full}) and no $r-$derivative of $t$ appears in (\ref{dnorm}), once we substitute the hydrodynamic form (\ref{emhydroscale}) of $t$ and the flow equations (\ref{uTevol}) in these equations, we should only get algebraic equations for $\alpha^{(n,m)}_{\rm s}$, $\alpha^{(n,m)}_{\rm v}$ and $\beta^{(n,m)}_{\rm s}$, and first order ODEs for $\gamma^{(n,m)}_{\rm s}$ and $\gamma^{(n,m)}_{\rm t}$, as we have claimed. A more detailed algorithm can be obtained in \cite{Kuperstein:2013hqa}.

The fifth step of the algorithm is to solve for the first order non-linear ODEs of $\epsilon$, $P$, $\gamma^{(n,m)}_{\rm s}$ and $\gamma^{(n,m)}_{\rm t}$ by ensuring that the flow equations reach a fixed point corresponding to incompressible non-relativistic Navier-Stokes fluid under the rescalings 
\begin{equation}\label{rescale}
r_{\rm H} - r = \xi \cdot \tilde{r}, \quad t = \frac{\tau}{\xi},
\end{equation}
with $\xi \rightarrow 0$ keeping $\tilde{r}$ and $\tau$ fixed \cite{Kuperstein:2013hqa}. This requirement imposes bounds on the near-endpoint behaviour of of $\epsilon$, $P$ and the transport coefficients and therefore fixes all the integration constants in their first order scale evolution equations, and thus also determine their UV values \cite{Kuperstein:2013hqa}, which as claimed before are exactly those which are necessary for the regularity of the horizon of the bulk geometry. 

We are going to elaborate more about how the infrared criterion determines the UV data in the next section. However, at this point, we can state with an example why it is important to choose the gravitational counterterms as in (\ref{coeffs}), and hence the correct form of the evolution equations (\ref{schematic1}) or equivalently (\ref{EMTensorEoM-Full}) in order to satisfy the infrared criterion. The relevant example can be found by studying the scale-evolution of the tensor transport coefficient proportional to $(\nabla\cdot u)\sigma^\mu_{\phantom{\mu}\nu}$ following \cite{Kuperstein:2013hqa}. In order that the infrared criterion is realised, it is required that this transport coefficient should behave \textit{weaker} than $(r_{\rm H} - r)^{-1}$ near $r = r_{\rm H}$. Nevertheless, it has four bad source terms in the corresponding ODE giving its scale evolution equation which violate this bound, but only two available integration constants to fix, namely one corresponding to the near horizon behaviour of the shear viscosity and another the same for a combination of two other tensor transport coefficients. It is possible to adjust the two integration constants (and thus determine the UV value of shear viscosity and also that of another tensor transport coefficient) to cancel the four dangerous source terms, provided the two gravitational counterterms which determine the ODE for the scale evolution are chosen exactly as in (\ref{coeffs}). Therefore, appropriate UV values for satisfying the infrared criterion can be chosen only if the gravitational counterterms that give the scale evolution are chosen correctly. The same gravitational counterterms thus obviously also determine (\ref{t-rg-example}) or equivalently (\ref{EMTensorEoM-Full}) which lead to the ODEs for the scale evolutions of the transport coefficients.

We thus complete the task of solving the first order flow equations of the equation of state and transport coefficients, and demonstrating that they have complete information of the emergent spacetime, including its regularity. Nevertheless, these flow equations, particularly (\ref{uTevolgrav}) are not yet taking place in the fixed background metric, as noted earlier.

Here we need to add the sixth and final step. We need to make the Anstaz that $g_{\mu\nu}(\Lambda)$ can be reduced to a functional of $u^\mu (\Lambda)$ and $T(\Lambda)$ in the \textit{fixed} background metric $\eta_{\mu\nu} e^{2\sigma}$ as below:
\begin{eqnarray}\label{ghydroscale}
g_{\mu\nu}(\Lambda) &=& -A^{(0)} \left(T(\Lambda), \Lambda\right) u_\mu(\Lambda) u_\nu(\Lambda) + B^{(0)} \left(T(\Lambda), \Lambda\right) \left(u_\mu(\Lambda) u_\nu(\Lambda)+ \eta_{\mu\nu} e^{2\sigma}\right) -\nonumber\\&&
- \sum_{n=1}^\infty \sum_{m = 1}^{n_{\rm s}}A^{(n,m)} \left(T(\Lambda), \Lambda\right) \, \mathcal{S}^{(n,m)}_{\rm cf}(\Lambda)\,  u_\mu(\Lambda)u_\nu(\Lambda) + \nonumber\\&&+\sum_{n=1}^\infty \sum_{m = 1}^{n_{\rm s}}B^{(n,m)} \left(T(\Lambda), \Lambda\right) \, \mathcal{S}^{(n,m)}_{\rm cf}(\Lambda)\,  \left(u_\mu(\Lambda)u_\nu(\Lambda)+ \eta_{\mu\nu} e^{2\sigma}\right)+ \nonumber\\&&+\nonumber\\&&+\sum_{n=1}^\infty \sum_{m = 1}^{n_{\rm v}}C^{(n,m)} \left(T(\Lambda), \Lambda\right) \, \left(u_\mu(\Lambda)\mathcal{V}^{(n,m)}_{\rm cf\, \nu}(\Lambda)+u_\nu(\Lambda)\mathcal{V}^{(n,m)}_{\rm cf\, \mu}(\Lambda)\right)+\nonumber\\&&+ \sum_{n =1}^\infty \sum_{m = 1}^{n_{\rm t}}D^{(n,m)} \left(T(\Lambda), \Lambda\right) \, \mathcal{T}_{\rm cf \,\mu\nu}^{(n,m)}(\Lambda).
\end{eqnarray}
All these unknown functions can be readily solved order by order in derivatives by substituting the above in (\ref{t-z}) and using the already known solutions for $\epsilon$, $P$ and the transport coefficients. This readily gives first order ODEs for the above functions whose boundary conditions are determined by requiring that $g_{\mu\nu}(\Lambda = \infty) = \eta_{\mu\nu}e^{2\sigma}$, the fixed background metric -- or equivalently by the requirement that at $\Lambda=\infty$,  $A^{(0)} =B^{(0)} =1$ and all other functions vanish. The required manipulations are very similar to those described above. After solving all the above functions, we can bring the evolution equations (\ref{uTevolgrav}) into the form (\ref{uTevol}) as we naturally obtained from field theory. We will explicitly do manipulations of this type in the next section, but our starting premises will be entirely different.

We also mention that similar manipulations also help us to verify that the scale evolution equations (\ref{uTevolgrav}) along with the scale evolution equations for $\epsilon$, $P$ and the transport coefficients reproduces the general scale evolution equations (\ref{t-rg-example}) for $t^\mu_{\phantom{\mu}\nu}(\Lambda)$ in the $\Lambda^{-1}$ expansion. In the latter form, the scale-evolution is state-independent as here only $t^\mu_{\phantom{\mu}\nu}(\Lambda)$ and $\Lambda$ appear; while the scale-dependent hydrodynamic variables and transport coefficients do not appear explicitly. Similarly, (\ref{ghydroscale}) can be seen to be a special case of the general UV expansion (\ref{g-example}) of $g_{\mu\nu}(\Lambda)$ as a state-independent functional of $t^\mu_{\phantom{\mu}\nu}(\Lambda)$ and $\Lambda$. 

At this point, the reader may suspect that we are assuming gravity equations in the first place to derive the RG flow, although our goal is to do the reverse. Actually, this is not the case, because the central proposition used by us is that the state-independent RG flow equation (\ref{schematic1}) with the property that a $g_{\mu\nu}(\Lambda)$ for satisfying the effective Ward identity (\ref{WILambda}) exists at each scale, leads to a diffeomorphism invariant pure classical gravity theory in one higher dimension and vice versa. \textit{Nevertheless, in order to obtain such a RG flow practically and also sum it over all orders in $\Lambda^{-1}$ at each order in derivatives in the hydrodynamic expansion, it is useful to know exactly how (\ref{schematic1}) maps to a given diffeomorphism invariant pure classical gravity theory and utilise this map explicitly to construct it.} This is exactly what we have done here. Furthermore, we actually do not need to study the solution of the classical gravity theory as in the traditional holographic correspondence in order to determine the correct UV data -- we actually determine the latter by applying the infrared criterion on the RG flow.

\section{Gravity equations from the coarse-graining}\label{Derive}
\subsection{Derivation of classical gravity equations in Fefferman-Graham gauge}\label{derivation}
Our aim is now to derive the classical equations of $(d+1)-$gravity from the coarse-graining (\ref{cguT}) of the $d-$dimensional field theory, by imposing three simple constraints on the latter. We are going to restrict ourselves to the long wavelength limit (close to thermal equilibrium) in this subsection on both sides of the duality (which is to be established). Our procedure should work not only as a method for obtaining classical gravity in one higher dimension from the coarse-graining of a quantum field theory, but also for formulating an approach to \textit{constructive quantum field theory} at \textit{strong coupling} in the first place.

Let us recall that in the stated limit, the expectation value of the microscopic operator ${t^\mu_{\phantom{\mu}\nu}}^\infty$ is parametrised by $d-$independent variables, namely ${u^\mu}^\infty$ and $T^\infty$. The dynamics of the operator is exactly captured by the asymptotic derivative expansion of the hydrodynamic equations which ${u^\mu}^\infty$ and $T^\infty$ follow -- the field-theoretic inputs are via the transport coefficients. We assume large $N$ factorisation of expectation values of products of ${t^\mu_{\phantom{\mu}\nu}}^\infty$ here. As we are aiming to formulate a constructive field theory approach, our construction should also determine the UV data (in this context the transport coefficients in ${t^\mu_{\phantom{\mu}\nu}}^\infty$ which give the evolution of ${u^\mu}^\infty$ and $T^\infty$, and hence also that of ${t^\mu_{\phantom{\mu}\nu}}^\infty$) and derive the dual $(d+1)-$classical gravity equations. Our construction proceeds by imposing the following restrictions on the coarse-graining functions in (\ref{cguT}) that define the coarse-grained variables $u^\mu(\Lambda)$ and $T(\Lambda)$:
\begin{enumerate}
\item \textbf{High efficiency:} We impose the requirement that a scale-dependent metric $g_{\mu\nu}(\Lambda,x)$  should exist at each scale $\Lambda$ such that, the coarse-grained variables $u^\mu(\Lambda)$ and $T(\Lambda)$ follow standard hydrodynamic equations in this background metric with appropriate $\Lambda$-dependent equation-of-state and transport coefficients. This also implies that we can define the coarse-grained operator $t^\mu_{\phantom{\mu}\nu}(\Lambda)$ in this limit via its expectation value that takes the standard hydrodynamic form (\ref{emhydroscale}) as a functional of $u^\mu(\Lambda)$, $T(\Lambda)$ and $g_{\mu\nu}(\Lambda)$ -- the hydrodynamic equations   simply follows from $\nabla_{(\Lambda)\mu}t^\mu_{\phantom{\mu}\nu}(\Lambda) =0$, with $\nabla_{(\Lambda)}$ being the covariant derivative constructed out of $g_{\mu\nu}(\Lambda)$, and which $t^\mu_{\phantom{\mu}\nu}(\Lambda)$ should satisfy.
\item \textbf{Upliftability to operator dynamics:} Our second requirement is composed of two parts. 
\begin{itemize}
\item Firstly, the coarse-grained operator $t^\mu_{\phantom{\mu}\nu}(\Lambda)$ should satisfy a \textit{first-order} differential-flow-equation that describes its evolution in scale $\Lambda$ in the \textit{fixed} background metric, such that this equation depends only on $t^\mu_{\phantom{\mu}\nu}(\Lambda)$, its spacetime derivatives and $\Lambda$ explicitly. In particular, in the present context, this equation should not depend on $u^\mu(\Lambda)$, $T(\Lambda)$ or $g_{\mu\nu}(\Lambda)$ explicitly, and take the form (\ref{schematic1}) schematically. This equation can be non-linear in $t^\mu_{\phantom{\mu}\nu}(\Lambda)$ -- large $N$ factorisation of expectation values implies $F$ is a classical functional of $t^\mu_{\phantom{\mu}\nu}(\Lambda)$, meaning that $t^\mu_{\phantom{\mu}\nu}(\Lambda)$ can be safely replaced by its expectation value.

\item The effective background metric $g_{\mu\nu}(\Lambda)$ in which $\nabla_{(\Lambda)\mu}t^\mu_{\phantom{\mu}\nu}(\Lambda) =0$ is satisfied is a (non-linear) functional of $t^\mu_{\phantom{\mu}\nu}(\Lambda)$ and its spacetime derivatives only -- in particular it does not depend on the variables $u^\mu(\Lambda)$ and $T(\Lambda)$ explicitly, and should take the form (\ref{schematic2}) schematically. These together imply that the scale-evolution of the operator $t^\mu_{\phantom{\mu}\nu}(\Lambda)$ is \textit{state-independent} as in Wilsonian RG flow. Furthermore, the effective $g_{\mu\nu}(\Lambda)$ serves the purpose of absorbing the multi-trace contributions in the energy-momentum conservation equation. The latter implies that the effective Ward identity $\nabla_{(\Lambda)\mu}t^\mu_{\phantom{\mu}\nu}(\Lambda) =0$ can be written as a state-independent operator equation of the type $\partial_\mu t^\mu_{\phantom{\mu}\nu}(\Lambda) = K[t^\mu_{\phantom{\mu}\nu}(\Lambda), \Lambda]$ at each $\Lambda$.
\end{itemize}

\item \textbf{Good  \textit{endpoint} behavior:} The final requirement to be imposed is that the endpoint of this RG flow at $\Lambda = \Lambda_{\rm IR}$ of the scale-dependent relativistic fluid that describes the effective dynamics of $t^\mu_{\phantom{\mu}\nu}(\Lambda)$, can be rescaled to a fixed point, which is \textit{exactly} described by \textit{non-relativistic incompressible Navier-Stokes equations}. These re-scalings are:
\begin{equation}\label{rescaling}
\frac{1}{\Lambda_{\rm IR}} - \frac{1}{\Lambda} = \frac{\xi}{\tilde{\Lambda}}, \quad t = \frac{\tau}{\xi}, \quad \text{and $\xi \rightarrow 0$ with $\tilde{\Lambda}, \tau$ held fixed.}
\end{equation}
It has been analyzed on very general grounds \cite{Kuperstein:2013hqa} that the rescaling can turn the end-point into a fixed point described by \textit{non-relativistic incompressible Navier-Stokes equations} provided \textit{near $\Lambda = \Lambda_{\rm IR}$}, $\epsilon(\Lambda)$, $P(\Lambda)$ and the scale-dependent transport coefficients behave as follows:
\begin{eqnarray}
\epsilon (\Lambda) \approx \text{const.}, \quad P (\Lambda) \approx (\Lambda - \Lambda_{\rm IR})^{-1}, \quad\text{$\eta(\Lambda)$ and $\zeta(\Lambda)$ are finite}\nonumber\\\ \gamma_{\rm t}^{(n,m)}(\Lambda) < (\Lambda - \Lambda_{\rm IR})^{-k^{(n,m)}_{\rm t}}, \gamma_{\rm s}^{(n,m)}(\Lambda) \leq (\Lambda - \Lambda_{\rm IR})^{-k^{(n,m)}_{\rm s}} \quad \text{for $n \geq 2$}\,,
\end{eqnarray}
 where $k^{(n,m)}_{\rm s}$ and $k^{(n,m)}_{\rm t}$ are appropriate integers. These integers can be obtained following \cite{Kuperstein:2013hqa}, where the explicit values have also been listed for all second-order transport coefficients.
 
 This criterion determines all integration constants in the first order RG flow, and predicts the UV transport coefficients which appear in the parametrisation of expectation value of ${t^\mu_{\phantom{\mu}\nu}}^\infty$. Each set of values will be related to a specific class of quantum field theories obtained via this constructive procedure. Furthermore, based on explicit calculations in \cite{Kuperstein:2013hqa}, we will see that these UV values will be exactly those in \cite{Rangamani:2009xk} which will give regular future horizons in the dual gravity theories. This principle thus completes the reconstruction of the traditional AdS/CFT correspondence in this special limit.
\end{enumerate}

We also note that the scaling (\ref{rescaling}) discussed here, which is necessary to obtain the fixed point, is essentially the same as (\ref{rescale}) on the gravity side in Section \ref{Kuperstein-Mukhopadhyay}, with $r$
replaced by $\Lambda^{-1}$. We will also see below that $\Lambda_{\rm IR}$ corresponds to $r_{\rm H}^{-1}$, the inverse of the horizon radius.

The first and second requirements are central, as we have discussed in Section \ref{gravity-flow-UV-expansion} and also more elaborately in the first part of our work \cite{Behr:2015yna}, to ensure that the RG flow equation for $t^\mu_{\phantom{\mu}\nu}(\Lambda)$ maps to a $(d+1)$ pure classical gravity theory that has full diffeomorphism invariance.  The third principle is necessary to determine the specific dual theory of gravity for a specific set of compatible UV data (of course some sets of UV data need not satisfy this criterion, thus ruling out existence of any gravity dual, and similarly vice versa). Furthermore, the third principle also removes the ambiguities in the scale evolution equation (\ref{schematic1}) corresponding to a specific dual classical gravity theory as discussed before in detail. 

Therefore, we can assume for the moment that the gravity theory to which the RG flow maps to, is Einstein's gravity with a negative cosmological constant, for the purpose of illustration. Although the classical gravity has $(d+1)-$diffeomorphism invariance, the RG flow equation for $t^\mu_{\phantom{\mu}\nu}(\Lambda)$ will still map to this gravity theory in a specific gauge where the $(d+1)-$diffeomorphism symmetry has been fixed by a specific choice of foliations of $d-$dimensional time-like hypersurfaces. The requirements that $u^\mu(\Lambda)$ and $T(\Lambda)$ do not appear explicitly in the flow equations, imply as discussed before, that the gauge fixing is done by imposing Fefferman-Graham gauge, where the lapse function is a function of the radial coordinate only and the shift vector is zero. The case of other gauges will be dealt with in the following subsection. 

Essentially, for the purpose of illustration, our task then boils down to showing that for unique choices of the functions defining $u^\mu(\Lambda)$ and $T(\Lambda)$ in (\ref{cguT}), we can map this coarse-graining procedure to the first order RG flow equation (\ref{t-rg-example}) of $t^\mu_{\phantom{\mu}\nu}(\Lambda)$. The latter is exactly equivalent to Einstein's equations in Fefferman-Graham gauge, and also corresponds to appropriate choices of gravitational counterterms, as shown in the first part of this work \cite{Behr:2015yna} and reviewed in Section \ref{gravity-flow-UV-expansion}. Our task has already been made easy by the following observations:
\begin{enumerate}
\item In Section \ref{coarse-grain}, it has already been shown that the coarse-graining (\ref{cguT}) can be converted into the first order differential equations for the flow of $u^\mu(\Lambda)$ and $T(\Lambda)$, namely (\ref{uTevol}). The coefficients in this flow equation are determined by the coarse-graining functions in (\ref{cguT}), as explicitly shown in (\ref{uTevol-coeffs}). These differential equations live in the \textit{fixed} background metric, which for the sake of convenience has been chosen to be conformally flat.
\item In Section \ref{Kuperstein-Mukhopadhyay}, it has been shown that the gravity equations in this limit can be rewritten as the first order differential equations for the flow of $u^\mu(\Lambda)$ and $T(\Lambda)$, namely (\ref{uTevolgrav}), where the coefficients are functions of the scale dependent equation-of-state and the transport coefficients. Unlike (\ref{uTevol}), these equations are covariant with respect to the scale-dependent effective metric $g_{\mu\nu}(\Lambda)$, which in turn is determined as functionals of $u^\mu(\Lambda)$, $T(\Lambda)$ and the scale-dependent equation-of-state and the transport coefficients in the fixed background metric, as in (\ref{ghydroscale}).
\end{enumerate}
Since the dual gravity equations and the corresponding scale-evolution equations are encoded in the gravitational flow equations (\ref{uTevolgrav}) as shown in Section \ref{Kuperstein-Mukhopadhyay}, we need to show that they imply the flow equations (\ref{uTevol}), and that the relation between these two determine the coarse-graining functions in (\ref{cguT}) uniquely.  Then, it follows that for unique choice of the coarse-graining functions in (\ref{cguT}), we can map this RG flow to the evolution of the operator $t^\mu_{\phantom{\mu}\nu}(\Lambda)$ given by (\ref{t-rg-example}), such that it satisfies $\nabla_{(\Lambda)\mu}t^\mu_{\phantom{\mu}\nu}(\Lambda)= 0$ at each scale in the unique background metric $g_{\mu\nu}(\Lambda)$. The latter is a non-linear functional of $t^\mu_{\phantom{\mu}\nu}(\Lambda)$ as given by (\ref{g-example}), and satisfies Einstein's equations (\ref{tensor-equation}) in the Fefferman-Graham gauge. Thus we are going to complete the derivation of classical gravity equations from the highly efficient RG flow.

In order to relate (\ref{uTevolgrav}) and (\ref{uTevol}), we begin by relating hydrodynamic scalars, vectors and tensors in fixed background metric $\eta_{\mu\nu}$ with those in the effective background metric $g_{\mu\nu}(\Lambda)$. It is convenient to parametrise $g_{\mu\nu}(\Lambda)$ in terms of the \textit{microscopic} hydrodynamic variables as below:
\begin{eqnarray}\label{ghydroscaleInf}
g_{\mu\nu}(\Lambda) &=& - f\Big(\frac{T^\infty}{\Lambda}\Big) {u_\mu}^\infty {u_\nu}^\infty + g\Big(\frac{T^\infty}{\Lambda}\Big){\Delta_{\mu\nu}}^\infty + \nonumber\\&&
+\frac{1}{T^\infty}\Bigg[ {h_{\rm s(1)}\Big(\frac{T^\infty}{\Lambda}\Big)} (\partial\cdot u^\infty){u_\mu}^\infty {u_\nu}^\infty +{h_{\rm s(2)}\Big(\frac{T^\infty}{\Lambda}\Big)} (\partial\cdot u^\infty){\Delta_{\mu\nu}}^\infty+ \nonumber\\&&
\qquad\quad+ {h_{\rm v}\Big(\frac{T^\infty}{\Lambda}\Big)} \left({u_\mu}^\infty (u^\infty\cdot \partial){u_\nu}^\infty +{u_\nu}^\infty (u^\infty\cdot \partial){u_\mu}^\infty\right)+\nonumber\\&&\qquad\quad
+ {h_{\rm t}\Big(\frac{T^\infty}{\Lambda}\Big)}  {\sigma_{\mu\nu}}^\infty\Bigg] +\mathcal{O}(\epsilon^2).
\end{eqnarray}
Although the above parametrisation is different from (\ref{ghydroscale}), it is an exactly equivalent form, and more convenient for manipulations. Firstly, we readily see that if the norm condition $u^\mu(\Lambda)\, g_{\mu\nu}(\Lambda)\, u^\nu(\Lambda) = -1$ is to be satisfied, then
\begin{eqnarray}\label{ULambdaUInf}
u^\mu(\Lambda) &=& \frac{1}{\sqrt{f\Big(\frac{T^\infty}{\Lambda}\Big)}} {u^\mu}^\infty +\frac{1}{2}\frac{h_{\rm s(1)}\Big(\frac{T^\infty}{\Lambda}\Big)}{T^\infty \, f^{\frac{3}{2}}\Big(\frac{T^\infty}{\Lambda}\Big)} (\partial \cdot u^\infty){u^\mu}^\infty + \nonumber\\&& + \frac{1}{T^\infty}k\Big(\frac{T^\infty}{\Lambda}\Big) (u^\infty\cdot \partial){u^\mu}^\infty +
\mathcal{O}(\epsilon^2).
\end{eqnarray}
Note the $k_{\rm v}$ in the vector piece above in the second line cannot be determined by  functions appearing in $g_{\mu\nu}(\Lambda)$ using the norm condition. We can also write similarly,
\begin{eqnarray}\label{TLambdaTinf}
T(\Lambda) = T^\infty\, {l\left(\frac{T^\infty}{\Lambda}\right)} + {m\left(\frac{T^\infty}{\Lambda}\right)}\,(\partial\cdot u^\infty) + O(\epsilon^2).
\end{eqnarray}
Although the functions $k$, $l$, and $m$ can be determined from what follows, we will not need to use them explicitly to solve coarse-graining functions at zeroth and first orders in derivatives. Nevertheless the explicit forms are required at higher orders, therefore we will mention below how these can be determined also.
 
 We need to implement the \textit{high efficiency} criterion that $u^\mu(\Lambda)$ and $T(\Lambda)$ satisfies the Euler equations at the leading order, therefore
\begin{eqnarray}\label{Euler-cov}
\left(u^\alpha(\Lambda) \nabla_\alpha\right) u^\mu(\Lambda) &=& - \Delta^{\mu\nu}(\Lambda)\nabla_\nu \, \ln \, T(\Lambda) + \mathcal{O}(\epsilon), \nonumber\\
 \left(u^\alpha(\Lambda) \nabla_\alpha\right) \ln \, T(\Lambda) &=& - c_s^2(\Lambda) \left(\nabla_\alpha u^\alpha(\Lambda)\right)+ \mathcal{O}(\epsilon). 
\end{eqnarray}
Above and from now on, by $\nabla$ we will actually mean $\nabla_{(\Lambda)}$, the covariant derivative constructed from $g_{\mu\nu}(\Lambda)$ given by (\ref{ghydroscaleInf}). Furthermore, it is crucial that $T(\Lambda)$ and $c_s^2(\Lambda)$ have \textit{local thermodynamic interpretations}. This means that an equation-of-state exists should exist at each $\Lambda$, such that we can define $P(\Lambda, T^\infty)$ and $\epsilon(\Lambda, T^\infty)$,  so that 
\begin{equation}
t^\mu_{\phantom{\mu}\nu}(\Lambda) = \epsilon(\Lambda, T^\infty) \, u^\mu(\Lambda) u_\nu (\Lambda) + P(\Lambda, T^\infty)\, \Delta^\mu_{\phantom{\mu}\nu}(\Lambda) + \mathcal{O}(\epsilon^2), 
\end{equation}
with $u_\mu(\Lambda) = g_{\mu\nu}(\Lambda) u^\nu(\Lambda)$ and the effective Euler equations (\ref{Euler-cov}) should follow from $\nabla_\mu t^\mu_{\phantom{\mu}\nu}(\Lambda) = 0$. This is possible only if
$T(\Lambda) \equiv T(\Lambda, T^\infty)$ can be determined  from the thermodynamic identities:
\begin{eqnarray}\label{thermo-ids}
\epsilon(\Lambda, T^\infty) + P(\Lambda, T^\infty) = T(\Lambda, T^\infty) \, s(\Lambda, T^\infty), \quad \frac{\partial\epsilon(\Lambda, T^\infty) }{\partial T^\infty}= T(\Lambda, T^\infty)\frac{\partial s(\Lambda, T^\infty)}{\partial T^\infty},
\end{eqnarray}
at each $\Lambda$ (up to an overall numerical constant to be fixed later), and furthermore $c_s^2(\Lambda)\equiv c_s^2(\Lambda, T^\infty)$ is the scale-dependent speed of thermodynamic sound, i.e.
\begin{equation}\label{cs}
c_s^2 (\Lambda, T^\infty)= \frac{\frac{\partial P(\Lambda, T^\infty)}{\partial T^\infty}}{\frac{\partial \epsilon (\Lambda, T^\infty)}{\partial T^\infty}}
\end{equation}
at each $\Lambda$. The reader can note that using (\ref{thermo-ids}) and (\ref{cs}), one can also go in the reverse direction, i.e. determine $\epsilon(\Lambda)$ and $P(\Lambda)$ from $T(\Lambda)$ and $c_s^2(\Lambda)$.

Clearly, since $T(\Lambda) = T(\Lambda, T^\infty)$ can be found from the thermodynamic relations as a function of $T^\infty$, we can eliminate $T^\infty$ in favour of $T(\Lambda)$ at a fixed $\Lambda$. In this case $\epsilon$, $P$, $c_s^2$, and also $f$ and $g$ appearing in $g_{\mu\nu}(\Lambda)$ can be regarded as functions of $T(\Lambda)$ and $\Lambda$, instead of $T^\infty$ and $\Lambda$ -- the partial differentiation with respect to $T(\Lambda)$ is to be understood as being done with $\Lambda$ held fixed.

We can readily write
\begin{eqnarray}\label{scalar-conv}
\partial_\mu u^\mu (\Lambda) &=& \nabla_\alpha u^\mu(\Lambda) - \Gamma^\alpha_{\beta\alpha}[g] u^\beta(\Lambda)\nonumber\\
&=& \nabla_\alpha u^\alpha (\Lambda) - (u^\alpha(\Lambda)\partial_\alpha)\ln \sqrt{-{\rm det}(g)} +\mathcal{O}(\epsilon^2)\nonumber\\
&=&\nabla_\alpha u^\alpha (\Lambda) - \frac{1}{2} (u^\alpha(\Lambda)\nabla_\alpha)\left(\ln f + (d-1) \ln g\right)+\mathcal{O}(\epsilon^2)\nonumber\\
&=&\nabla_\alpha u^\alpha(\Lambda) - \frac{c_s^2(\Lambda)}{2}\left(\frac{\partial \ln f}{\partial \ln T(\Lambda)} + (d-1) \frac{\partial \ln g}{\partial \ln T(\Lambda)}\right)(u^\alpha(\Lambda)\nabla_\alpha)\ln T(\Lambda)+\mathcal{O}(\epsilon^2)\nonumber\\
&=& \Big[1+ \frac{c_s^2(\Lambda)}{2}\left(\frac{\partial \ln f}{\partial \ln T(\Lambda)} + (d-1) \frac{\partial \ln g}{\partial \ln T(\Lambda)}\right)\Big](\nabla_\alpha u^\alpha(\Lambda))+\mathcal{O}(\epsilon^2)\,.
\end{eqnarray}
In the final step above we have used the effective Euler equations (\ref{Euler-cov}). 

Similarly, after a lot of tedious steps we obtain using (\ref{Euler-cov}), (\ref{thermo-ids}) and (\ref{cs}) that
\begin{eqnarray}\label{vector-conv}
(u^\alpha(\Lambda)\partial_\alpha) u^\beta (\Lambda) 
&=& \frac{c_s^2(\Lambda)}{2} \frac{\partial\ln f}{\partial \ln T(\Lambda)}(\nabla_\alpha u^\alpha(\Lambda)) u^\mu (\Lambda)+ \nonumber\\&&+ \frac{g}{f}\Big[1 + \frac{1}{2}\frac{\partial\ln f}{\partial \ln T(\Lambda)}\Big](u^\alpha(\Lambda) \nabla_\alpha) u^\mu(\Lambda) + \mathcal{O}(\epsilon^2).
\end{eqnarray}

With (\ref{scalar-conv}) and (\ref{vector-conv}), we are now ready to rewrite the flow equations (\ref{uTevol}) for $u^\mu(\Lambda)$ and $T(\Lambda)$ obtained from field theory in the form of the flow equations (\ref{uTevolgrav}) which capture the classical gravity equations. The relations between these two equations are given by those between $\alpha^{(n,m)}_{\rm s,v}$ and $\beta^{(n,m)}_{\rm s,v}$ appearing in (\ref{uTevolgrav}), and $a^{(n,m)}_{\rm s,v}$ and $b^{(n,m)}_{\rm s,v}$ appearing in (\ref{uTevol}). These are as follows:
\begin{eqnarray}\label{alpha-beta-a-b}
\alpha^{(0)} &=& a^{(0)},\nonumber\\
\alpha_{\rm s}^{(1)} &=& \Big[1+ \frac{c_s^2(\Lambda)}{2}\left(\frac{\partial \ln f}{\partial \ln T(\Lambda)} + (d-1) \frac{\partial \ln g}{\partial \ln T(\Lambda)}\right)\Big]a_{\rm s}^{(1)}+\nonumber\\&&
+\frac{c_s^2(\Lambda)}{2}\frac{\partial\ln f}{\partial \ln T(\Lambda)}a_{\rm v}^{(1)}, \nonumber\\
\alpha_{\rm v}^{(1)} &=& \frac{g}{f}\Big[1 + \frac{1}{2}\frac{\partial\ln f}{\partial \ln T(\Lambda)}\Big]a_{\rm v}^{(1)}, \nonumber\\
\beta^{(0)} &=& b^{(0)}, \nonumber\\
\beta_{\rm s}^{(1)} &=& \Big[1+ \frac{c_s^2(\Lambda)}{2}\left(\frac{\partial \ln f}{\partial \ln T(\Lambda)} + (d-1) \frac{\partial \ln g}{\partial \ln T(\Lambda)}\right)\Big] b_{\rm s}^{(1)}.
\end{eqnarray}

Recalling our discussion from Section \ref{Kuperstein-Mukhopadhyay} we can see that in order to satisfy our second and third criteria, we require that 
\begin{itemize}
\item the flow coefficients $\alpha^{(n,m)}_{\rm s,v}$ and $\beta^{(n,m)}_{\rm s,v}$ in (\ref{uTevolgrav}) are determined in terms of $\epsilon(\Lambda)$, $P(\Lambda)$ and the scale-dependent transport coefficients via \textit{algebraic} identities at each $\Lambda$
, and  
\item $\epsilon(\Lambda)$, $P(\Lambda)$ and the scale-dependent transport coefficients should satisfy  precise \textit{first order} ordinary differential equations for scale evolution.
\end{itemize}
Indeed, the second and third criteria require that the scale evolution of $t^\mu_{\phantom{\mu}\nu}(\Lambda)$ in the fixed Minkowski space background should satisfy (\ref{t-rg-example}). Nevertheless, this equation is only expressed in the UV expansion in $T(\Lambda)/\Lambda$, so using the method  of \cite{Kuperstein:2013hqa}, we need to resum all orders in $T(\Lambda)/\Lambda$  at a fixed order in derivatives to obtain (\ref{EMTensorEoM-Full}), as outlined in Section \ref{Kuperstein-Mukhopadhyay}. The latter readily gives both , the algebraic relations that determine $\alpha^{(n,m)}_{\rm s,v}$ and $\beta^{(n,m)}_{\rm s,v}$ in terms of $\epsilon(\Lambda)$, $P(\Lambda)$ and the scale-dependent transport coefficients, and the first order ordinary differential equations for the scale evolution of the latter. Furthermore, to obtain these we do not need to know $g_{\mu\nu}(\Lambda)$ explicitly, which can nevertheless be determined next from appropriately resumming (\ref{g-example}) to all orders in $T(\Lambda)/\Lambda$ at a fixed order in derivatives as mentioned in Section \ref{Kuperstein-Mukhopadhyay}.

Up to first order, 
\begin{eqnarray}\label{thydroscalechap4}
t^\mu_{\phantom{\mu}\nu}(\Lambda) &=& \epsilon(\Lambda) \, u^\mu(\Lambda) u_\nu (\Lambda) + P(\Lambda)\, \Delta^\mu_{\phantom{\mu}\nu}(\Lambda) -\nonumber\\&&-\zeta(\Lambda) \, \left(\nabla\cdot u(\Lambda)\right)\, \Delta^\mu_{\phantom{\mu}\nu}(\Lambda)- 2\eta(\Lambda)\, \sigma^{\mu}_{\phantom{\mu}\nu}(\Lambda) + O(\epsilon^2),
\end{eqnarray}
with $u_\nu(\Lambda) =u^\mu(\Lambda) g_{\mu\nu}(\Lambda)$, and all hydrodynamic scalars and tensors constructed in the background $g_{\mu\nu}(\Lambda)$.

The algebraic relations that determine $\alpha^{(n,m)}_{\rm s, v}$ and $\beta^{(n,m)}_{\rm s}$ appearing in (\ref{uTevolgrav}) in terms of $\epsilon(\Lambda)$, $P(\Lambda)$ and the scale-dependent transport coefficients appearing in (\ref{thydroscalechap4}), at zeroth and first orders in derivative expansion are \cite{}:
\begin{eqnarray}\label{coeffs-algebraic}
\alpha^{(0)} &=& -\frac{1}{2\Lambda^{d+1}}\left(P(\Lambda) + \frac{d-2}{d-1}\epsilon(\Lambda)\right),\nonumber\\
\alpha_{\rm s}^{(1)} &=&\, \alpha_{\rm v}^{(1)} \, =\, \beta_{\rm s}^{(1)} \, =\, 0.
\end{eqnarray} 
Also, $T(\Lambda)$ and hence $\beta^{(0)}$ can be determined from $\epsilon(\Lambda)$ and $P(\Lambda)$, using (\ref{thermo-ids}) as discussed below. Indeed the vanishing of $\alpha_{\rm s}^{(1)}$, $\alpha_{\rm v}^{(1)}$ and $\beta_{\rm s}^{(1)}$ are remarkable consequences of diffeomorphism invariance, although we are mapping to gravity equations in Fefferman-Graham gauge. It is related to the fact that in these equations, which are explicitly (\ref{tensor-equation}), $\nabla$-derivatives can appear through ${\rm Ricci}[g]$ only and therefore can occur only even number of times. Secondly, in order that (\ref{thydroscalechap4}) satisfies the first order flow equations (\ref{t-rg-example}), we require that $\epsilon(\Lambda)$, $P(\Lambda)$ and the transport coefficients, namely $\zeta(\Lambda)$ and $\eta(\Lambda)$, satisfy the following first-order ordinary differential equations for scale-evolution \cite{Kuperstein:2013hqa}: 
\begin{eqnarray}\label{diff-eqns-transport}
\frac{\partial \epsilon(\Lambda)}{\partial \Lambda} &=& - \left(\frac{1}{2(d-1)\Lambda^{d+1}}\epsilon(\Lambda) - \frac{1}{\Lambda}\right) {\rm Tr}\, t(\Lambda), \nonumber\\
\frac{\partial {\rm Tr}\, t(\Lambda)}{\partial \Lambda} &=& - \left(\frac{1}{2(d-1)\Lambda^{d+1}}{\rm Tr}\, t(\Lambda) + \frac{d}{\Lambda}\right) {\rm Tr}\, t(\Lambda),\nonumber\\
\frac{\partial \zeta(\Lambda)}{\partial \Lambda} &=& -  \left(\frac{d-1}{\Lambda}\left(c_s^2(\Lambda)+1\right))+\frac{1}{\Lambda^{d+1}}\left(P(\Lambda) - \frac{c_s^2(\Lambda)}{2}\epsilon(\Lambda)\right)\right)\zeta(\Lambda),\nonumber\\
\frac{\partial \eta(\Lambda)}{\partial \Lambda} &=& \frac{1}{2\Lambda^{d+1}}\, \epsilon(\Lambda) \, \eta(\Lambda). \nonumber\
\end{eqnarray}
Above, for the sake of convenience, we have replaced the first order scale-evolution equation for $P(\Lambda)$ by that of ${\rm Tr}\, t(\Lambda) \equiv (d-1) P(\Lambda) - \epsilon(\Lambda)$, and $c_s^2(\Lambda)$ is determined from (\ref{cs}). As we will see below, $T^\infty$ appears as a constant of integration. Also in the above equations, $\epsilon$, $P$, $\zeta$ and $\eta$ have been regarded as functions of $\Lambda$ and $x$, and in the partial differentiation with respect to $\Lambda$, the spacetime position $x$ has been held fixed, therefore $T^\infty$ has been kept fixed as well.

The most general solutions for ${\rm Tr} \, t(\Lambda)$ and $\epsilon(\Lambda)$ are:
\begin{eqnarray}\label{solutions}
{\rm Tr} \, t(\Lambda) &=& 4d(d-1) \frac{\Lambda^{d}}{\left(\frac{\Lambda}{\Lambda_{\rm IR}}\right)^{2d} -1}, \nonumber\\
\epsilon(\Lambda) &=& 4(d-1) \frac{ \Lambda^d \left(\left(\frac{\Lambda}{\tilde\Lambda}\right)^d -1\right)}{\left(\frac{\Lambda}{\Lambda_{\rm IR}}\right)^{2d} -1}.
\end{eqnarray}
Above, $\Lambda_{\rm IR}$ and $\tilde\Lambda$ arise as the two constants of integration, which are $x-$dependent. According to our third requirement, namely the good endpoint behaviour condition, $P(\Lambda)$ can diverge at an infrared scale $\Lambda_{\rm IR}$ as $(\Lambda - \Lambda_{\rm IR})^{-1}$, thus $\Lambda_{\rm IR}$ is identified with the scale that marks the endpoint of the flow equations \cite{Kuperstein:2013hqa}. Remarkably, as we will see below, the location of the horizon in Fefferman-Graham coordinates is $r_{\rm H} = \Lambda_{\rm IR}^{-1}$. Thus the endpoint of the RG flow, emerging as an integration constant, gives the location of the horizon. Furthermore, $\epsilon(\Lambda)$ has to be finite at $\Lambda_{\rm IR}$ so that after rescaling the flow equations, the fluid at the endpoint follows the incompressible non-relativistic Navier-Stokes equations \cite{}.  The latter is possible, as clear from (\ref{solutions}), only if
\begin{equation}
\tilde{\Lambda} = \Lambda_{\rm IR}.
\end{equation}
Thus, we obtain
\begin{eqnarray}\label{epsilon-P}
P(\Lambda) &=& 4\, \Lambda^{d}\,\frac{\left(\frac{\Lambda}{\Lambda_{\rm IR}}\right)^{d}+ (d-1)}{\left(\frac{\Lambda}{\Lambda_{\rm IR}}\right)^{2d} -1}, \nonumber\\
\epsilon(\Lambda) &=& 4(d-1) \, \Lambda^d \, \frac{1}{\left(\frac{\Lambda}{\Lambda_{\rm IR}}\right)^{d} + 1}.
\end{eqnarray}
Note in the $UV$, i.e. at $\Lambda = \infty$, $\epsilon(\Lambda)$ and $P(\Lambda)$ both become finite and satisfy $\epsilon = (d-1) P$, as should be the case in a $d-$dimensional CFT, and furthermore these reproduce the familiar thermodynamics of the $(d+1)-AdS-$Schwarzschild black brane. This is just a special instance of the general fact that the good endpoint behaviour of the first order scale-evolution of $\epsilon$, $P$ and the transport coefficients  gives us the desired values of UV data, which not only satisfies the requirements demanded by the (anomalous) Weyl covariance of CFT, but also exactly have those numerical values that ensure that the dual solutions in gravity have regular future horizons. Thus we can use the criterion of good endpoint behaviour of the first-order RG flow to determine the CFT data in the UV, without explicitly knowing the bulk spacetime metric as in the standard holographic approach.

The thermodynamic identities (\ref{thermo-ids}), where we can legitimately replace $T^\infty$ by $\Lambda_{\rm IR}$ for the moment, imply:
\begin{eqnarray}\label{T-s}
T(\Lambda) &=& k \cdot4d \cdot\Lambda \cdot \frac{\Lambda}{\Lambda_{\rm IR}} \cdot\frac{\left(\left(\frac{\Lambda}{\Lambda_{\rm IR}}\right)^{d} + 1\right)^{\frac{d-2}{d}}}{\left(\frac{\Lambda}{\Lambda_{\rm IR}}\right)^{d} - 1},\nonumber\\
s(\Lambda) &=& \frac{1}{k}  \cdot\Lambda^{d-1} \cdot \left(\frac{\Lambda}{\Lambda_{\rm IR}}\right)^{d-1} \cdot\frac{1}{\left(\left(\frac{\Lambda}{\Lambda_{\rm IR}}\right)^{d} +1\right)^{2\frac{d-1}{d}}},
\end{eqnarray}
with $k$ being a numerical constant that is independent of $x$. This numerical constant $k$ can be fixed by relating $\Lambda_{\rm IR}$ with $T(\Lambda = \infty) \equiv T^\infty$. This can be fixed, a posteriori, by reconstructing the bulk space-time metric using the method discussed before, and demanding the Euclidean contribution at the leading order in derivatives, has no conical singularity. This gives us:
\begin{equation}\label{k-Tinf}
k = \frac{2^{\frac{2}{d}-4}}{\pi} , \quad {\rm i.e.} \quad T^\infty = \frac{2^{\frac{2}{d}-2}d}{\pi } \Lambda_{\rm IR}.
\end{equation}
Of course, using the above we can now determine $\epsilon(\Lambda, T^\infty)$, $P(\Lambda, T^\infty)$, $T(\Lambda, T^\infty)$, etc. which we required before. However, it is convenient to actually keep using $\Lambda_{\rm IR}$ instead of $T^\infty$. Finally, we obtain from (\ref{cs}) that
\begin{equation}\label{csLambda}
c_s^2(\Lambda) = \frac{\left(\frac{\Lambda}{\Lambda_{\rm IR}}\right)^{2d}+2(d-1)\left(\frac{\Lambda}{\Lambda_{\rm IR}}\right)^{d}+1}{(d-1)\left(\left(\frac{\Lambda}{\Lambda_{\rm IR}}\right)^{2d}- 1\right)}\, .
\end{equation}
Clearly, at $\Lambda=\infty$, we recover $c_s^2 = 1/(d-1)$, the standard result for a CFT.

Furthermore, the requirement of the incompressible non-relativistic Navier-Stokes endpoint, imposes that $\zeta$ and $\eta$ should be finite at $\Lambda= \Lambda_{\rm IR}$ \cite{Kuperstein:2013hqa}. The only solution for $\zeta$ in (\ref{diff-eqns-transport}) that is finite at $\Lambda = \Lambda_{\rm IR}$, is 
\begin{equation}
\zeta(\Lambda) = 0,
\end{equation}
thus it vanishes at all scales. On the other hand, all solutions for $\eta$ are finite at $\Lambda = \Lambda_{\rm IR}$, because (\ref{diff-eqns-transport}) implies:
\begin{equation}
\frac{\partial}{\partial \Lambda}\left(\frac{\eta(\Lambda)}{s(\Lambda)}\right) = 0,
\end{equation}
and $s$ is finite at $\Lambda = \Lambda_{\rm IR}$. Nevertheless, $\eta$ appears in the \textit{source-terms} in the inhomogeneous first-order scale-evolution equations of the second-order tensor transport coefficients, and it has been shown that in particular the infrared criterion for the tensor transport coefficient associated with $(\nabla\cdot u) \sigma^\mu_{\phantom{\mu}\nu}$ is satisfied only if the integration constant in the scale evolution of $\eta(\Lambda)$ is fixed to a specific value (recall the detailed discussion in Section \ref{Kuperstein-Mukhopadhyay}). Thus, we obtain \cite{Kuperstein:2013hqa}:
\begin{equation}
\eta(\Lambda) = \frac{1}{4\pi}\cdot s(\Lambda)\, .
\end{equation}
In $d=4$,
\begin{equation}
\eta(\Lambda) =   2\sqrt{2}  \cdot\Lambda^{3} \cdot \frac{\left(\frac{\Lambda}{\Lambda_{\rm IR}}\right)^{3}}{\left(\left(\frac{\Lambda}{\Lambda_{\rm IR}}\right)^{4} +1\right)^{\frac{3}{2}}}\, .
\end{equation}
Once again, the good endpoint behaviour reproduces the numerical UV data which results in regular future horizons, without requiring us to determine the bulk spacetime metric explicitly. In particular, this not only determines the UV value of $\eta$ but also other second order transport coefficients \cite{Kuperstein:2013hqa}. 

At each order in derivative expansion, once the transport coefficients are determined uniquely by the criterion of good endpoint behavior, the $\alpha^{(n,m)}_{\rm s, v}$ and the  $\beta^{(n,m)}_{\rm s}$ in the first order flow equations (\ref{uTevolgrav}) for $u^\mu(\Lambda)$ and $T(\Lambda)$, which are algebraically related to these transport coefficients, are also determined. Up to second order in derivatives, we can obtain $\alpha^{(0)}$, $\alpha^{(1)}_{\rm s,v}$ and $\beta_{\rm s}^{(1)}$ from (\ref{coeffs-algebraic}), with $\epsilon(\Lambda)$ and $P(\Lambda)$ given by (\ref{epsilon-P}). Specifically in $d=4$,
\begin{equation}\label{alpha0}
\alpha^{(0)} = - \frac{2}{\Lambda}\frac{\Lambda_{\rm IR}^4 (3\Lambda^4 + \Lambda_{\rm IR}^4)}{\Lambda^8 - \Lambda_{\rm IR}^8}\,.
\end{equation}
Similarly, $\beta^{(0)}$ can be readily determined from (\ref{T-s}):
\begin{equation}\label{beta0}
\beta^{(0)} = -\frac{2\sqrt{2}}{\pi}\cdot\frac{1}{\sqrt{1 + \left(\frac{\Lambda}{\Lambda_{\rm IR}}\right)^4}} \cdot\frac{\Lambda \Lambda_{\rm IR}^3(3\Lambda^4 + \Lambda_{\rm IR}^4)}{(\Lambda^4 - \Lambda_{\rm IR}^4)^2}\, .
\end{equation}

Finally, $g_{\mu\nu}(\Lambda)$ is given by (\ref{g-example}) and can be readily determined from the known values of transport coefficients. Alternatively, it can determined from the first-order equation (\ref{t-z4d})  following \cite{Kuperstein:2013hqa}. This first-order equation has unique solution for the boundary condition $g_{\mu\nu}(\Lambda = \infty) = \eta_{\mu\nu}$, which means we can determine all the functions in (\ref{ghydroscaleInf}). Without going into further details, using manipulations described in Section \ref{Kuperstein-Mukhopadhyay}, we obtain in $d=4$
\begin{eqnarray}\label{f-g}
f &=& - \frac{\left(\left(\frac{\Lambda}{\Lambda_{\rm IR}}\right)^4- 1\right)^2}{ \left(\frac{\Lambda}{\Lambda_{\rm IR}}\right)^4 \left(\left(\frac{\Lambda}{\Lambda_{\rm IR}}\right)^4 +1\right)}, \quad g \, = 1 + \left(\frac{\Lambda}{\Lambda_{\rm IR}}\right)^4,\nonumber\\
h_{\rm s(1)} &=& h_{\rm s(2)} \, = h_{\rm v} \, =0, \nonumber\\
h_{\rm t} &=& \frac{1}{\pi}\left(1 + \left(\frac{\Lambda}{\Lambda_{\rm IR}}\right)^4\right) \ln \left(\frac{1 - \left(\frac{\Lambda}{\Lambda_{\rm IR}}\right)^4}{1 + \left(\frac{\Lambda}{\Lambda_{\rm IR}}\right)^4}\right)\,.
\end{eqnarray}
Once again, we have used $\Lambda_{\rm IR}$ in lieu of $T^\infty$. This reproduces the explicit bulk metric \cite{Gupta:2008th} of fluid/gravity correspondence in the Fefferman-Graham gauge up to second order in derivatives.

Having determined the functions $f$ and $g$ in the metric $g_{\mu\nu}(\Lambda)$ (as in (\ref{f-g})) along with $\alpha^{(0)}$, $\beta^{(0)}$, $\alpha_{\rm s,v}^{(1)}$ and $\beta_{\rm s}^{(1)}$ (as in (\ref{coeffs-algebraic}, \ref{alpha0}, \ref{beta0})), and with $\epsilon(\Lambda)$, $P(\Lambda)$, $T(\Lambda)$ and $c_s^2(\Lambda)$ also determined (as in (\ref{epsilon-P}, \ref{csLambda})), we can readily obtain $a^{(0)}$, $b^{(0)}$, $a_{\rm s,v}^{(1)}$ and $b_{\rm s}^{(1)}$ from (\ref{alpha-beta-a-b}). Thus we reach our final goal of obtaining first order scale-evolution equations for $v^{(0)}$, $w^{(0)}$, $v_{\rm s,v}^{(1)}$ and $w_{\rm s}^{(1)}$, by substituting $a^{(0)}$, $b^{(0)}$, $a_{\rm s,v}^{(1)}$ and $b_{\rm s}^{(1)}$ as has been determined above, in (\ref{uTevol-coeffs}). These equations have unique solutions given that at $\Lambda = \Lambda^\infty$,
\begin{equation}
v^{(0)} = w^{(0)} = 1, \quad {\rm and}\quad v_{\rm s,v}^{(1)}= w_{\rm s}^{(1)}=0\,,
\end{equation}
 so that $u^\mu(\Lambda)$ and $T(\Lambda)$ coincide with ${u^\mu}^\infty$ and $T^\infty$. The required solutions in $d=4$ are
 \begin{eqnarray}\label{final-solution-cgf}
 v^{(0)} &=& \frac{\sqrt{1 + \left(\frac{\Lambda_{\rm IR}}{\Lambda}\right)^4}}{1 - \left(\frac{\Lambda_{\rm IR}}{\Lambda}\right)^4}, \nonumber\\
 v_{\rm v}^{(1)} &=&  v_{\rm s}^{(1)} \, = 0, \nonumber\\
 w^{(0)} &=&  \frac{\sqrt{1 + \left(\frac{\Lambda_{\rm IR}}{\Lambda}\right)^4}}{1 - \left(\frac{\Lambda_{\rm IR}}{\Lambda}\right)^4},\nonumber\\
 w_{\rm s}^{(1)} &=& 0\,.
 \end{eqnarray}
 Remarkably $v^{(0)} = w^{(0)}$. Above, $\Lambda_{\rm IR}$ above is related to $T^\infty$ via (\ref{k-Tinf}). Finally we obtain the unique solutions for the coarse-graining functions in (\ref{cguT}), up to second order in derivatives, which satisfy all the three criteria of the highly efficient RG flow, and lead to a unique $g_{\mu\nu}(\Lambda)$, as given by (\ref{ghydroscaleInf}) and (\ref{f-g}) that satisfy specific $(d+1)-$classical gravity equations, namely those of Einstein's gravity. Clearly, our method extends to higher order in derivatives.
 
 It is to be mentioned here that our derivation rests on two main caveats:
 \begin{enumerate}
 \item We do not have an explicit proof that the criterion of good endpoint behaviour reproduces exactly those values for the transport coefficients, which are required for the regularity of the future horizon, although we have strong support in its favour. Indeed, it has been shown in \cite{Kuperstein:2013hqa} that we recover the desired values of the first and second order transport coefficients using this criterion.
 \item We also do not have explicit proof that the criterion of good endpoint behaviour fixes the structural ambiguities in the RG flow, that are related to the choice of gravitational counter-terms, in a manner that we have exactly as many potentially dangerous terms in the scale-evolution equations that behave badly in the IR, as there are integration constants to fix. Once again the available evidence in \cite{Kuperstein:2013hqa} is only at the leading and sub-leading orders in derivatives.
\end{enumerate}
 In the future, we would like to at least check the validity of these statements at higher orders in derivatives, if not prove them. To this end, the analysis of special sectors in the field theory dual to algebraically special solutions of gravity \cite{Mukhopadhyay:2013gja,deFreitas:2014lia,Bakas:2014kfa,Ortaggio:2014gpa,Gath:2015nxa,Petropoulos:2015fba} could be immensely valuable.

We conclude this subsection with brief comments on why our RG flow construction \textit{is constructive field theory}, beyond the already established fact that it can determine all the microscopic transport coefficients (in the UV). Firstly, as we will discuss in the next section, our RG flow construction extends beyond the  hydrodynamic limit, so we can reconstruct all the UV data that which completely characterises $\langle{t^\mu_{\phantom{\mu}\nu}}^\infty\rangle$ beyond the transport coefficients, particularly using the third criterion. Secondly, we can repeat our RG flow construction in \textit{any} fixed background metric instead of flat Minkowski space. This leads us to obtain $\langle{t^\mu_{\phantom{\mu}\nu}}^\infty\rangle$ in background metrics infinitesimally away from flat Minkowski space rather easily, and this also directly leads us to obtain the correlation functions of ${t^\mu_{\phantom{\mu}\nu}}^\infty$. Thus, we can demonstrate that the microscopic theory gets defined by the RG flow construction. 

\subsection{Other gauges and the lifted Weyl symmetry}\label{diffeo}
The derivation of the classical gravity equations from the RG flow in the Fefferman-Graham gauge has the advantage that the scale-evolution equations for $t^\mu_{\phantom{\mu}\nu}(\Lambda)$, namely (\ref{schematic1}), and the scale-dependent background metric $g_{\mu\nu}(\Lambda)$, namely (\ref{schematic2}), are state-independent. In other gauges, both (\ref{schematic1}) and (\ref{schematic2}) will contain two auxiliary variables, $\alpha$ and $\beta^\mu$ which can be identified with the pseudo-lapse function and pseudo-shift vector on the gravity side. In practice, both $\alpha$ and $\beta^\mu$  can be state-dependent also. As for example, in the fluid-gravity context, the choices corresponding to the Eddington-Finkelstein gauge are $\alpha(\Lambda, x) = 0$, and $\beta^\mu(\Lambda, x) = {u^\mu}^\infty(x)$. Thus the corresponding equations (\ref{schematic1}) and (\ref{schematic2}) become state-dependent, although the state-dependence is via the auxiliary variables $\alpha$ and $\beta^\mu$ which have no intrinsic dynamics.

Naturally two important questions arise:
\begin{enumerate}
\item Can we generalise the second criterion, namely state-independence, which defines the highly-efficient RG flow in order to obtain the classical gravity equations in other gauges?
\item If this is the case, how do we directly understand from the RG flow equations to which gauge-fixing it corresponds to on the gravity side, or equivalently, which time-like hypersurface foliations we have to chose in the emergent $(d+1)-$geometry to represent the RG flow?
\end{enumerate}

The answers to both the above questions have essentially been provided in the first part of our work \cite{Behr:2015yna}. Here we briefly revisit them in the context of the specific coarse-graining procedure in the hydrodynamic limit.

The answer to the first question is that the generalised second criterion is simply that it should always be possible to find four-parameter transformations of the form:
\begin{eqnarray}\label{diffeos}
\tilde{\Lambda} = \Lambda + \rho(\Lambda, x), \quad \tilde{x}^\mu = x^\mu + \chi^\mu(\Lambda, x),
\end{eqnarray}
by performing which the scale-evolution equations for $t^\mu_{\phantom{\mu}\nu}(\Lambda)$, namely (\ref{schematic1}), and the scale-dependent background metric $g_{\mu\nu}(\Lambda)$, namely (\ref{schematic2}), can be made state-independent. This simply corresponds to the change from the arbitrary coordinate system to the Fefferman-Graham gauge for appropriate $\rho(\Lambda, x)$ and $\chi^\mu(\Lambda, x)$, which as discussed in \cite{Behr:2015yna}, are determined by the corresponding $\alpha$ and $\beta^\mu$ in the gauge-fixing corresponding to the RG flow. Furthermore, $\rho$ and $\chi^\mu$ should vanish at $\Lambda = \infty$ so that the UV data remain unchanged.
 
From the point of the RG flow, we are supposed to find the transformation of $t^\mu_{\phantom{\mu}\nu}(\Lambda)$ under (\ref{diffeos}) first and then find the appropriate transformation for $g_{\mu\nu}(\Lambda)$, so that  $\nabla_{(\Lambda)\mu}t^\mu_{\phantom{\mu}\nu}(\Lambda) = 0$ is preserved at each $\Lambda$. These can be done as shown in \cite{Behr:2015yna}. The \textit{infinitesimal} transformation for $g_{\mu\nu}(\Lambda)$ is given by 
\begin{equation}\label{gdiffeogen}
\tilde{g}_{\mu\nu} = g_{\mu\nu} + \rho \, \frac{\partial g_{\mu\nu}}{\partial \Lambda} + 2\,\frac{\rho}{ \Lambda} \, g_{\mu\nu}-\mathcal{L}_\chi g_{\mu\nu},
\end{equation}
and is universal, meaning independent of the UV data of the RG flow and the corresponding dual gravity theory (with $\mathcal{L}_\chi$ denotes the Lie-derivative along $\chi^\mu$). This is easily understood as diffeomorphism from a gauge \textit{infinitesimally} away from Fefferman-Graham gauge to the Fefferman-Graham gauge. Also the induced metric $\gamma_{\mu\nu}$ in the gauge corresponding to the RG flow equations prior to the transformations, at $r = \Lambda^{-1} = \text{constant}$ hypersurfaces is given by $\gamma_{\mu\nu} = l^2 \Lambda^2 g_{\mu\nu}$ following (\ref{gphys}) (note in \cite{Behr:2015yna} the discussion has been in the opposite direction, i.e. considering a transformation from Fefferman-Graham gauge to one infinitesimally far away). The \textit{infinitesimal} transformation for $t^\mu_{\phantom{\mu}\nu}(\Lambda)$ under (\ref{diffeos}) is
\begin{eqnarray}\label{tdiffeogen}
\tilde{t}^\mu_{\phantom{\mu}\nu} &=& t^\mu_{\phantom{\mu}\nu} + \rho \,\Lambda\, \frac{\partial t^\mu_{\phantom{\mu}\nu}}{\partial \Lambda} + d\,\frac{\rho}{\Lambda} \, t^\mu_{\phantom{\mu}\nu}- \mathcal{L}_\chi t^\mu_{\phantom{\mu}\nu} \nonumber\\&&+ \text{non-universal corrections with at least two derivatives,}
\end{eqnarray}
again for a change from a gauge infinitesimally far away from Fefferman-Graham gauge to the Fefferman-Graham gauge (see \cite{Behr:2015yna} for the details about the non-universal terms containing two-derivatives). After doing the transformations (\ref{gdiffeogen}) and (\ref{tdiffeogen}) one can obtain RG flow equations which satisfies all the three desired criteria including the \textit{strict state-independence} as discussed in the previous subsection, and the corresponding gravity equations are then in the Fefferman-Graham gauge. 

At this point one can ask how the coarse-graining functions (\ref{cguT}) are related in the RG flows corresponding to the two gauges. This can be readily obtained as follows. Note, the definition of $u^\mu(\Lambda)$ and $T(\Lambda)$ are given by $t^\mu_{\phantom{\mu}\nu}(\Lambda)u^\nu(\Lambda) = -\epsilon(T(\Lambda),\Lambda) \, u^\mu (\Lambda)$, so that $u^\mu(\Lambda)$ is the time-like eigenvector and $T(\Lambda)$ is related to the eigenvalue of $t^\mu_{\phantom{\mu}\nu}(\Lambda)$, and the normalisation condition is $u^\mu(\Lambda)g_{\mu\nu}(\Lambda)u^\nu(\Lambda) = -1$ in any RG flow. These relations and the transformations of $g_{\mu\nu}(\Lambda)$ and $t^\mu_{\phantom{\mu}\nu}(\Lambda)$ as given by (\ref{gdiffeogen}) and (\ref{tdiffeogen}) respectively, imply the following transformations:
\begin{eqnarray}\label{uTdiffeogen}
\tilde{u}^\mu &=&  + \rho \,\Lambda\, \frac{u^\mu}{\partial \Lambda} + \frac{\rho}{\Lambda} \, u^\mu- \mathcal{L}_\chi u^\mu \nonumber\\&&+ \text{non-universal corrections with at least two derivatives,}, \nonumber\\
\tilde{T}&=&T + \rho \,\Lambda\, \frac{\partial T}{\partial \Lambda} + \,\frac{\rho}{\Lambda} \,T- \mathcal{L}_\chi T \nonumber\\&&+ \text{non-universal corrections with at least two derivatives,}
\end{eqnarray}
where all the non-universal terms can be readily found. Furthermore, the coarse-graining functions in (\ref{cguT}) can be obtained from the scale evolution equations of $u^\mu(\Lambda)$ and $T(\Lambda)$, namely (\ref{uTevol}) as discussed before, so from these one can readily find the relations between the corresponding coarse-graining functions in the two RG flows corresponding to the two different gauges. This completes the answer to the first question, as indeed there is a unique RG flow corresponding to a given gauge-fixing of the $(d+1)-$diffeomorphisms of the corresponding gravity equations, and the new RG flows can be found using the modified second criterion and the transformations, as mentioned above.

To answer the second question, as discussed in the first part of our work \cite{}, we first need to construct the RG flow in a fixed conformally flat metric $\eta_{\mu\nu}e^{2\sigma(x)}$, and then find the one-parameter symmetry under $\sigma(x) \rightarrow\sigma(x) +\delta\sigma(x)$, under which the coarse-graining functions in (\ref{cguT}) remain invariant, and so do the scale-evolution equations (\ref{schematic1}) for $t^\mu_{\phantom{\mu}\nu}(\Lambda)$, and the expression (\ref{schematic2}) for $g_{\mu\nu}(\Lambda)$ as a function of $t^\mu_{\phantom{\mu}\nu}(\Lambda)$ and $\Lambda$. This automorphism of the RG flow reduces to Weyl transformation in the UV, but at an arbitrary scale $\Lambda$ is complicated, and has been called by us the \textit{lifted Weyl symmetry} in \cite{Behr:2015yna}. The scale-evolution equations (\ref{schematic1}) for $t^\mu_{\phantom{\mu}\nu}(\Lambda)$, which generalises (\ref{t-rg-example}) to any conformally flat background, have been constructed in \cite{Behr:2015yna} for the Fefferman-Graham gauge, while the relation (\ref{schematic2}) between $g_{\mu\nu}(\Lambda)$ and $t^\mu_{\phantom{\mu}\nu}(\Lambda)$ in Fefferman-gauge is as in (\ref{t-z}). This symmetry corresponds to the \textit{residual gauge symmetries} of the Fefferman-Graham gauge \cite{Penrose,Brown:1986nw}, which reduces to Weyl transformations in the UV and are given by:
\begin{eqnarray}\label{PBH4}
\Lambda = \tilde{\Lambda} + \rho(\tilde{\Lambda},  \tilde{x}), \quad  x^\mu = \tilde{x}^\mu + \chi^\mu(\tilde{\Lambda},  \tilde{x}), \quad \text{with}\nonumber\\
\rho = - \tilde{\Lambda}\,\delta\sigma(\tilde{x}), \quad \chi^\mu = -\int_{\tilde{\Lambda}}^\infty d\hat{\Lambda} \, \frac{1}{\hat{\Lambda}^3}g^{\mu\nu}(\hat{\Lambda}, \tilde{x}) \frac{\partial \delta\sigma(\tilde{x})}{\partial \tilde{x}^\nu},
\end{eqnarray}
at arbitrary $\Lambda$. The automorphism of the RG flow in the Fefferman-Graham gauge are given by the transformations (\ref{gdiffeogen}), (\ref{tdiffeogen}) and (\ref{uTdiffeogen}), with $\rho$ and $\chi^\mu$ determined by $\delta\sigma(x)$ as in (\ref{PBH4}), although now we are also changing the UV data by a Weyl transformation.

In another gauge, the automorphism group is as follows. Let $\mathcal{G}$ be the \textit{unique} transformation which takes the RG flow variables corresponding to the new gauge to that corresponding to the Fefferman-Graham gauge \textit{without} modifying UV data. This corresponds to a diffeomorphism on the gravity side as discussed above. Then the automorphism group of the RG flow in the new gauge is $\mathcal{G}^{-1}\mathcal{P} \mathcal{G}$, with $\mathcal{P}$ being the automorphisms (\ref{PBH4}) of the RG flow corresponding to the Fefferman-Graham gauge. Conversely knowing the automorphisms, namely $\mathcal{G}^{-1}\mathcal{P} \mathcal{G}$, corresponding to the residual gauge symmetries that map to Weyl transformations in the UV in the new gauge, we can also find $\mathcal{G}$ as $\mathcal{P}$ is explicitly known. Finally, the knowledge of $\mathcal{G}$ directly allows us to find the gauge-fixing of the diffeomorphisms in the corresponding gravity equations, and also the corresponding choice of foliations in the bulk spacetime that represents the RG flow. This completes the answer to the second question via the one-to-one correspondence between the special automorphisms of the RG flow, also called here the lifted Weyl symmetry, and the corresponding gauge fixings of the diffeomorphisms in the dual classical gravity equations.

\section{The beta function for the highly efficient RG flow}\label{beta}

In our discussion so far, we have focused on states in the CFT in which the energy-momentum tensor is the only non-trivial operator with an expectation value. We take a first look here into a highly efficient RG flow reconstruction of the vacuum state in a strongly coupled large $N$ CFT deformed by a relevant coupling in the UV. We need to therefore understand the scale-evolution of the operators $\langle t^\mu_{\phantom{\mu}\nu}\rangle = (1/d)\,\langle{\rm Tr}\,t\rangle\delta^\mu_{\phantom{\mu}\nu}$ and $\langle O\rangle$ (to which the relevant coupling couples) -- these will be the only ones to have non-trivial vacuum expectation values in the simplest examples. We would like to understand the beta function of the running coupling and how does it compare with the case of a Wilsonian RG flow. 

We begin with bringing out conceptual differences between Wilsonian RG flow and the highly efficient RG flow. In case of the Wilsonian RG flow, we look for the elementary fields which constitute the composite operators ${\rm Tr}\,t$ and $O$, and coarse-grain by integrating out the higher momentum modes of the elementary fields. In our highly efficient RG flow construction, we use collective variables to label expectation values of operators, and then coarse-grain these collective variables to define the RG flow. This implies that we cannot just look at the vacuum state in order to define the collective variable(s). If $g$ is the relevant coupling which deforms the CFT in the UV, we need to actually consider the case where this coupling is a slowly varying function $g(x)$ as in local Wilsonian RG flow. The derivative is so slowly varying that we can ignore all derivatives of  $g(x)$ -- obviously we can systematically correct this via a derivative expansion as in the case of the hydrodynamic limit. We can then think about $g(\Lambda, x)$ as the coarse-grained collective variable defining ${\rm Tr}\,t[\Lambda, g(\Lambda)]$ and $O[\Lambda, g(\Lambda)]$ -- this $g(\Lambda)$ is the analogue of scale-dependent transport coefficients in the hydrodynamic limit considered before, and its first order evolution gives the \textit{beta function} of the highly efficient RG flow. We will find that although the beta function in highly efficient RG flow has a different conceptual meaning, structurally it is very similar to the case of local Wilsonian RG flow at the leading order in derivatives. We will leave a more detailed investigation to higher order in derivatives for the future.

To be concrete, let us take a simple example where the $4-$dimensional large $N$ strongly coupled CFT is deformed by a coupling $g$ of mass dimension $1$, and the corresponding operator $O$ to which it couples to has therefore mass dimension $3$. Assuming $g(x)$ to be a slowly varying function, the goal is to construct the highly efficient RG flow and recover dual classical gravity equations. We require to define a coarse-graining where at each scale we will satisfy the Ward identity,
\begin{equation}\label{WIO}
\nabla_{(\Lambda)\mu}\langle t^\mu_{\phantom{\mu}\nu}(\Lambda)\rangle = \langle O(\Lambda)\rangle \nabla_{(\Lambda)\nu}g(\Lambda).
\end{equation}
We can further assume that,
\begin{equation}\label{sf}
g_{\mu\nu}(\Lambda,x) = e^{2\sigma(\Lambda, x)} \eta_{\mu\nu}.
\end{equation}
Since we ignore the derivatives of $g(\Lambda)$, $O(\Lambda)$, $t^\mu_{\phantom{\mu}\nu}(\Lambda)$ and $\sigma(\Lambda)$ (except when imposing the Ward identity at the leading order), the above simplifications must work. 

As we have seen before, when we ignore derivatives of the collective variables, we merely need to consider field-redefinitions. In the case of the hydrodynamic variables, we may recall that the role of $v^{(0)}$ and $w^{(0)}$ in Eq. (\ref{cguT}) were mere field-redefinitions at leading order in derivatives. However, it is an extremely non-trivial task to find the right field redefinitions such that we can satisfy the Ward identity (\ref{WIO}) at each scale -- we have already seen in the hydrodynamic limit that finding the right scale-dependent equation of state $\epsilon(P(\Lambda),\Lambda)$ such that the Ward identity can be satisfied has been very non-trivial. We therefore take advantage of our theorem discussed in Section \ref{gravity-flow-UV-expansion} -- that the Ward identity (\ref{WIO}) can only be solved by mapping $O(\Lambda)$ and $t^\mu_{\phantom{\mu}\nu}(\Lambda)$ to variables of a diffeomorphism invariant classical gravity theory in $(d+1)-$dimensions. So we just need to find out a state-independent map between the variables of gravity and the scale-dependent operators.

The simplest possible gravity theory (in 5-dimensions) is Einstein's gravity minimally coupled to a scalar given by the action:
\begin{equation}
S = \frac{1}{\kappa_5^2}\int_{\mathcal{M}} {\rm d}^5X\,\sqrt{-G}\left(\frac{1}{2}R-\frac{1}{2}G^{MN}\partial_M \chi \partial_N \chi - V(\chi)\right) + \frac{1}{\kappa_5^2}\int_{\partial\mathcal{M}}{\rm d}^4x\,\sqrt{-\gamma} K,
\end{equation}
where $\kappa_5 = 8\pi G_N$. Einstein's equations of motion read:
\begin{equation}\label{ES}
R_{MN}- \frac{1}{2}R G_{MN} = \partial_M \chi \,\partial_N \chi - \frac{1}{2}G_{MN}\,G^{PQ}\,\partial_P \chi \,\partial_Q \chi - G_{MN}V(\chi).
\end{equation}
The Bianchi identity implies the scalar equation of motion which we do not write separately. The vacuum solution we are looking for is of the type:
\begin{equation}
{\rm d}s^2 = \frac{1}{r^2}\left({\rm d}r^2 + e^{2\rho(r)} \eta_{\mu\nu}{\rm d}x^\mu {\rm d}x^\nu\right), \quad \chi = \chi(r)
\end{equation}
in the Fefferman-Graham coordinates. We have set $l_{\rm AdS} = 1$ at the asymptotic boundary (UV). For sake of convenience we define:
\begin{equation}
A(r) = \rho(r) - {\rm ln}r.
\end{equation}
It is well known that solving (\ref{ES}) and the scalar field equations are equivalent to solving the following first order system \cite{Freedman:2003ax}:
\begin{eqnarray}\label{foeom}
\dot{A} =- \frac{1}{3}W(\chi), \quad \dot{\chi} =  W'(\chi),
\end{eqnarray}
where $\dot{\phantom{A}}$ denotes derivative with respect to $\ln r$ and ${\phantom{A}}\prime$ denotes derivative with respect to $\chi$, and the \textit{fake superpotential} $W(\chi)$ is defined via:
\begin{equation}\label{V}
V = -\frac{2}{3}W^2 +\frac{1}{2}W^{'2}.
\end{equation} 
In order to reproduce $l_{\rm AdS} = 1$ and the scaling dimension of the dual operator $\Delta = 3$ we can choose,
\begin{equation}\label{W}
W = 3 + \frac{1}{2}\chi^2 + C \chi^4,
\end{equation}
with $C$ being an arbitrary dimensionful parameter. Note, we do not need to assume the AdS/CFT dictionary -- for the moment we can just proceed by thinking the above form of $W$ as an assumption. The asymptotic expansions of $\chi$ and $A$ are:
\begin{eqnarray}\label{asymp}
\chi  &=& g\, r + 2C g^3 \, r^3 + 6 C^2 g^5 \, r^5+ \mathcal{O}(r^7), \nonumber\\
A &=&- {\rm log}\, r - \frac{1}{12}g^2 r^2-\frac{1}{4}Cg^4 r^4 -\frac{8}{9}C^2 g^6 r^6 +\mathcal{O}(r^8).
\end{eqnarray}
It is to be noted that if $g$ is promoted to a space-time dependent function $g(x)$, all the above statements regarding the vacuum solution remain true as long as we ignore spacetime derivatives of $g(x)$.

The Ward identity (\ref{WIO}) should follow from the constraints of Einstein's equations. Following the logic of Section \ref{gravity-flow-UV-expansion}, we argue that $\chi$ and $\rho$ should be related to the scale-dependent coupling $g(\Lambda)$ and the scale factor $\sigma(\Lambda)$ in (\ref{sf}) only via functions of $r$. Furthermore, we do not want $l_{\rm AdS}$ to appear in the scale evolution equations of motion for $g(\Lambda)$ and  $\sigma(\Lambda)$. Therefore, we obtain that at all scales:
\begin{equation}\label{sourceg}
r = \Lambda^{-1}, \quad g(\Lambda) = \frac{\chi}{r}, \quad g_{\mu\nu}(\Lambda)  = r^2\gamma_{\mu\nu}(r),
\end{equation}
with $\gamma_{\mu\nu}$ being the induced metric on $r =\Lambda^{-1}$. In the UV limit, we recover the traditional AdS/CFT dictionary. We require $\sigma(\Lambda=\infty) = 1$, so that the field theory lives in Minkowski space. Note the above identifications are valid in any coordinate system, and not only in the Fefferman-Graham coordinates. In the latter case, we obtain
\begin{equation}
\rho(r = \Lambda^{-1}) = \sigma(r).
\end{equation}

The map between $t^\mu_{\phantom{\mu}\nu}(\Lambda)$ and the variables of gravity (which should also be coordinate-independent) should be given by Eqs. (\ref{tgravity1}) and (\ref{BY}), but in our case:
\begin{equation}
{T^\mu_{\phantom{\mu}\nu}}^{\rm ct} = \mathcal{P}(\chi)\delta^\mu_{\phantom{\mu}\nu}= -(d-1) \delta^\mu_{\phantom{\mu}\nu}+ \tilde{\mathcal{P}}(\chi)\delta^\mu_{\phantom{\mu}\nu},
\end{equation}
if we ignore derivatives of $g(x)$. The term $-(d-1)\delta^\mu_{\phantom{\mu}\nu}$ above is required to cancel the leading volume divergence and should be there even if the potential $V(\chi)$ is a constant -- this is something we have already derived using the third criterion of good IR behaviour of the RG flow. The above is the most general form which preserves diffeomorphism invariance (and therefore valid in any coordinate system). In the Fefferman-Graham coordinates (restoring the normalisation factor $1/(2\kappa_5^2)$ and setting $d= 4$),
\begin{eqnarray}\label{tsLambda}
t^\mu_{\phantom{\mu}\nu}(\Lambda = r^{-1}) &=&\frac{1}{2\kappa_5^2}\left( \frac{1}{r^3}\left(z^\mu_{\phantom{\mu}\nu} - {\rm Tr}\, z \delta^\mu_{\phantom{\mu}\nu}\right)+ \frac{2}{r^4}\tilde{\mathcal{P}}(\chi)\delta^\mu_{\phantom{\mu}\nu}\right),\nonumber\\
&=& \frac{1}{\kappa_5^2}\left( -\frac{3}{r^4}\left(1 - \frac{W}{3}\right)+ \frac{1}{r^4}\tilde{\mathcal{P}}(\chi)\right)\delta^\mu_{\phantom{\mu}\nu} = \frac{1}{\kappa_5^2}\Lambda^4 \left(W(\chi(\Lambda)) - \mathcal{P}(\chi(\Lambda))\right)\delta^\mu_{\phantom{\mu}\nu}.
\end{eqnarray}
Above, we have used the equations of motion (\ref{foeom}). It is easy to see from Eqs. (\ref{W}) and (\ref{asymp}) that in order to remove a $1/r^2$ (i.e. $\Lambda^2$) divergence, we should have
\begin{equation}
\tilde{\mathcal{P}}(\chi) = -\frac{1}{2} \chi^2 - K \chi^4 +\mathcal{O}(\chi^6),
\end{equation}
with $K$ left undetermined. The counterterms of $\mathcal{O}(\chi^6)$ do not affect UV quantities as we will see soon -- this is quite reminiscent of the case of pure gravity. In principle, we should derive all the above counterterms from the good IR behaviour of the RG flow and even the leading term need not be fixed by cancellation of the UV divergence -- we leave this investigation for future work. From the asymptotic expansion (\ref{asymp}), we obtain:
\begin{equation}
\langle t^\mu_{\phantom{\mu}\nu} (\Lambda = \infty)\rangle = \frac{1}{\kappa_5^2}(C-K){g^{\infty}}^4\delta^\mu_{\phantom{\mu}\nu},
\end{equation}
with ${g^{\infty}} = g(\Lambda = \infty) = g$. We are now left with the task of defining $O(\Lambda)$. To find this, we promote $g$ to be a function of $x$ and substitute (\ref{tsLambda}) in the Ward identity (\ref{WIO}). Using the identfication of the coupling $g(\Lambda)$ as given by (\ref{sourceg}), we obtain:
\begin{equation}\label{OLambda}
\langle O(\Lambda = r^{-1}) \rangle = \frac{1}{\kappa_5^2}\frac{1}{r^3}\frac{{\rm d}}{{\rm d}\chi}\left(W(\chi) - \mathcal{P}(\chi)\right) = \frac{1}{\kappa_5^2}\Lambda^4\frac{{\rm d}}{{\rm d}g(\Lambda)}\left(W(g(\Lambda)/\Lambda) - \mathcal{P}(g(\Lambda)/ \Lambda)\right).
\end{equation}
We obtain from the asymptotic expansion (\ref{asymp}) that:
\begin{equation}
\langle O(\Lambda= \infty) \rangle = 4\frac{1}{\kappa_5^2}(C-A){g^{\infty}}^3.
\end{equation}

Finally, we can put all the results compactly by defining a \textit{partition function} for the highly efficient RG flow:
\begin{equation}
\ln Z(\Lambda) = \frac{1}{\kappa_5^2}\int {\rm d}^4 x \,\, \Lambda^4(\left(W(g( \Lambda)/\Lambda) - \mathcal{P}(g( \Lambda)/\Lambda)\right) =\frac{1}{\kappa_5^2} \int {\rm d}^4 x \,\, \frac{1}{r^4}(\left(W(\chi( r)) - \mathcal{P}(\chi(r))\right).
\end{equation}
We can readily promote the above \textit{effective low energy action} to a fixed scale-independent background metric $h_{\mu\nu}$ (once again replacing $\eta_{\mu\nu}$ by a weakly curved $h_{\mu\nu}$ has no effect on the solution at leading order in derivatives) for the sake of defining functional derivatives:
\begin{equation}\label{Zh}
\ln Z(\Lambda) = \frac{1}{\kappa_5^2}\int {\rm d}^4 x \,\sqrt{-h}\, \Lambda^4(\left(W(g( \Lambda)/\Lambda) - \mathcal{P}(g( \Lambda)/\Lambda)\right) =\frac{1}{\kappa_5^2} \int {\rm d}^4 x \,\sqrt{-h}\, \frac{1}{r^4}(\left(W(\chi( r)) - \mathcal{P}(\chi(r))\right).
\end{equation}
In this case, we can define the beta function by requiring that we satisfy the familiar identity
\begin{equation}\label{Zid1}
\frac{{\rm d}}{{\rm d}\, \ln \, \Lambda}\ln\, Z(\Lambda) = \langle{\rm Tr}\, t(\Lambda)\rangle + \beta(\Lambda) \langle O(\Lambda)\rangle,
\end{equation}
with
\begin{equation}\label{Zid2}
\langle O(\Lambda)\rangle =\frac{1}{\sqrt{-h}} \frac{\delta\ln Z(\Lambda)}{\delta g(\Lambda)}, \quad \langle t_{\mu\nu}(\Lambda)\rangle = -\frac{2}{\sqrt{-h}} \frac{\delta\ln Z(\Lambda)}{\delta h^{\mu\nu}}.
\end{equation}
From Eqs (\ref{tsLambda}) and (\ref{OLambda}), it follows that
\begin{equation}\label{Zid3}
\beta(\Lambda) = \frac{{\rm d}g(\Lambda)}{{\rm d}\, \ln \Lambda},
\end{equation}
satisfying the usual definition of the field theory. Unlike Wilsonian RG flow however, the left hand side of (\ref{Zid1}) does not vanish in highly efficient RG flow. Firstly we are not integrating out but rather \textit{projecting out} high momentum modes of \textit{directly measurable collective variables} and this projection is also field-dependent, therefore we do not expect a scale-independent partition function. The map of the highly efficient RG flow to gravity then preserves the Ward identitiy (\ref{WIO}) at each scale provided:
\begin{equation}
\beta(\Lambda) = \frac{{\rm d}g(\Lambda)}{{\rm d}\, \ln \Lambda} =\Lambda  \frac{{\rm d}\chi}{{\rm d}\, \ln \Lambda} + \Lambda \chi = g(\Lambda) - \frac{{\rm d}}{{\rm d} g(\Lambda)}W(g(\Lambda)/\Lambda),
\end{equation}
as directly follows from the equations of motion (\ref{foeom}) if we use the Fefferman-Graham frame. The first term originates from the mass dimension $1$ of $g$ (being a purely classical effect) and the second term (which is of quantum origin in the field theory) implies gradient flow of $g(\Lambda)$. Clearly, the beta function of the highly efficient RG flow shares common properties with that of the local Wilsonian RG flow. Note $\ln Z$ can be defined in any coordinate system, in which the identities Eq. (\ref{Zid1}), (\ref{Zid2}) and (\ref{Zid3}) are also satisfied.

We leave a more detailed exploration of the beta function of the highly efficient RG flow involving operators of arbitrary scaling dimensions and its consistency conditions to future work. Our derivation is somewhat similar to that of \cite{Erdmenger:2001ja} (see also \cite{Kiritsis:2014kua}), however in this work the scale-dependent Ward identity is not the main guiding principle. Furthermore, in the latter work, the derivation is not coordinate invariant, in particular the scale is not identified in a coordinate invariant way.

It is to be noted that $\ln Z(\Lambda)$ coincides with the on-shell renormalised gravitational action at $\Lambda = \infty$, however at a finite $\Lambda$ this is \textit{not} the same as the on-shell renormalised gravitational action with a cut-off at $r = \Lambda^{-1}$, nor is it the same on-shell renormalised gravitational action without a cut-off evaluated with different variables. The cut-off dependent on-shell gravitational action in Fefferman-Graham gauge is:
\begin{equation}
S^{\rm grav}(r_{\rm c}= \Lambda^{-1}) = =\frac{1}{\kappa_5^2} \int {\rm d}^4 x \,\sqrt{-h}\, \frac{e^{4\rho(r_{\rm c})}}{r_{\rm c}^4}(\left(W(\chi( r_{\rm c})) - \mathcal{P}(\chi(r_{\rm c}))\right),
\end{equation}
which is different from (\ref{Zh}) with the latter being more manifestly coordinate invariant. Note, as noted above in highly efficient RG flow the partition function need to be scale invariant if it exists. It is nevertheless interesting that we can define a partition function for the highly efficient RG flow at least in this simple toy example.

\section{Outlook}\label{Outlook}
We mention here some of the developments which can be readily pursued.
\subsection{Completing the reconstruction of the pure gravity sector and beyond}\label{Outlook1}
Although our discussion here has been limited to the hydrodynamic limit, it is in principle possible to go beyond to reconstruct the operator ${t^\mu_{\phantom{\mu}\nu}}^\infty$ and its correlation functions in the most general case, thus completing the reconstruction of the entire dual pure gravity sector. The first observation is that recently it has been established that quasi-normal mode dynamics can be obtained from a \textit{resurgent trans-series} \cite{Dunne:2015eaa} that gives non-perturbative completion to the perturbative microscopic asymptotic hydrodynamic expansion \cite{Heller:2013fn,Basar:2015ava}. Indeed the hydrodynamic expansion is perturbative in $\epsilon =1/(T^\infty t^{\rm var})$ where $t^{\rm var}$ is the typical time scale of variation of the perturbations about the thermal state. The quasi-normal modes indeed behave as $\exp(-a/\epsilon)$ at late time, giving a natural instanton series in $\epsilon$, with $a$ being constants, which can be determined from the requirement that the instanton series cancels the Borel-pole singularity of the perturbative hydrodynamic expansion, etc.  Secondly, the full dynamics can be captured by introducing new collective variables ${\pi^{\mu}_{\phantom{\mu}\nu}}^{\rm(nh)\infty}$, so that
\begin{equation}\label{tfull}
{t^\mu_{\phantom{\mu}\nu}}^\infty = {t^\mu_{\phantom{\mu}\nu}}^{\rm(hydro)\infty}+{\pi^{\mu}_{\phantom{\mu}\nu}}^{\rm(nh)\infty},
\end{equation}
with ${\pi^{\mu}_{\phantom{\mu}\nu}}^{\rm(nh)\infty}$ being not perturbative in $\epsilon$, and thus capturing dynamics beyond the hydrodynamic limit. Furthermore, we can use generalised Israel-Stewart theory \cite{Iyer:2009in,Iyer:2011qc} to write dynamical equations for ${\pi^{\mu}_{\phantom{\mu}\nu}}^{\rm(nh)\infty}$, in a manner which can provide the necessary resurgent trans-series that defines (\ref{tfull}) completely and non-perturbatively.

Finally, we can use the same resurgent trans-series construction at any value of $\Lambda$ by coarse-graining ${\pi^{\mu}_{\phantom{\mu}\nu}}^{\rm(nh)\infty}$ via generalisation of (\ref{cguT}), while still requiring that $\nabla_{(\Lambda)\mu}t^\mu_{\phantom{\mu}\nu} (\Lambda)= 0$ is preserved at each $\Lambda$, the second criterion of our RG flow construction, namely upliftability to operator dynamics is also satisfied and that the endpoint of the RG flow can still be described by incompressible non-relativistic Navier-Stokes equations after the rescaling (\ref{rescale}). Indeed our RG flow construction then indicates that not only the transport coefficients but the complete UV data that characterises ${t^{\mu}_{\phantom{\mu}\nu}}^{\infty}$ in (\ref{tfull}), can be determined from the existence of the right endpoint behaviour of the RG flow, by requiring that the latter can be described by the incompressible non-relativistic Navier-Stokes equations, which has only one parameter, namely the shear-viscosity of the horizon in the emergent geometry. 

The same route can be followed to go beyond the pure gravity sector, by including other single-trace operators as observed in Section \ref{beta} (see \cite{Kiritsis:2014kua} also).  The difficulty here is mainly that we do not know how to characterise the endpoint of the highly efficient RG flow and state the third criterion in full generality. On the gravity side, clearly we can have solutions like  domain walls \cite{Kraus:1999it} or stars \cite{Buchel:2013uba}, which do not even have horizons. Therefore, it is possible that the endpoint is another CFT, or characterised by the generic behaviour of  the core of the stars. Indeed a complete characterisation of endpoint behaviour is a difficult task, but we may be able to learn enough from the sector of states at high temperature that do thermalise and is described by black hole formation in the dual geometry. In this case, we expect that the end-point can be described by appropriate forced incompressible non-relativistic Navier-Stokes equations \cite{Donos:2015gia}.

\subsection{Towards a new general framework for constructive field theory at large $N$}\label{semihol}

The crucial advantage of our approach of reconstructing holography as RG flow is that we have a precise definition of the effective operator $t^\mu_{\phantom{\mu}\nu}(\Lambda)$ at \textit{any} scale $\Lambda$, and our construction can work, in principle, even in situations where the dual geometry has no well-defined UV -- all physical data at any given scale can be determined from our third criterion of good infrared behaviour. Specifically, an ambiguity-free definition of $t^\mu_{\phantom{\mu}\nu}(\Lambda)$ at a finite scale is absent in the traditional  way of stating the holographic correspondence. This opens the door to constructing new frameworks for non-conformal theories, even in cases where they are asymptotically free in the UV, by matching traditional Wilsonian RG flow with our highly efficient RG flow construction at an appropriate scale. It is necessary that we do this in the large $N$ limit where our RG flow construction works.

In more details, we may proceed as follows in case of an asymptotically free theory like QCD. We can follow Wilsonian RG flow from the UV up to a scale $Q$, where the effective coupling $\alpha_s(Q)$ becomes large. More precisely, at this scale we can calculate the scaling dimensions of various single-trace operators, and find that only a few have small scaling dimensions -- the gap in the spectrum of scaling dimensions can be thought of as a measure of how strongly interacting the system is at the scale $Q$. Crudely assuming this gap to be infinite (this assumption is to be refined below), we can simply match the Wilsonian effective single-trace operators, like ${t^\mu_{\phantom{\mu}\nu}}^{\rm W}(Q)$ with the ${t^\mu_{\phantom{\mu}\nu}}^{\rm Hol}(Q)$ of a highly efficient RG flow. The matching implies that at the scale Q:
\begin{enumerate}
\item we switch from scale-dependent elementary fields of the Wilsonian RG flow to collective variables of the highly efficient RG flow, and
\item we match the parameters like transport coefficients, etc. in the two RG flows at the scale Q,  with these determined by Kubo-like formulae from the UV side and by the criterion of good endpoint behaviour from the IR side.
\end{enumerate}
Thus we can say that the highly efficient RG flow completes the non-perturbative definition of the theory below a certain scale, where perturbative Wilsonian RG flow becomes unusable due to strong coupling.

However, the above expectation is naive, because the third criterion of good endpoint behaviour in the highly efficient RG flow is difficult to satisfy and the required data perhaps cannot be interpolated to an asymptotically free theory. In this case, we may need to amend the perturbative Wilsonian action at the scale $Q$ by a non-perturbative \textit{semi-holographic} completion, such that the full effective action at scale $Q$ takes the form:
\begin{equation}\label{semiholaction}
S^{\rm tot} = S^{\rm W}[A_\mu(Q), \alpha_s(Q)] + S^{\rm Hol}\left[g_{\mu\nu}^{\rm (b)} = \eta_{\mu\nu} + H_{\mu\nu}[A_\mu(Q), \alpha_s(Q)], \text{etc.}\right]\,.
\end{equation}
The crucial point is that the holographic action is supposed to give the right non-perturbative completion of the Wilsonian action (in the form of a resurgent trans-series \cite{Dunne:2015eaa}) at the scale $Q$, and yet it should \textit{live} in a different \textit{fictitious} background metric $g^{\rm (b)}_{\mu\nu}$ determined by the perturbative degrees of freedom. In principle, the holographic action is determined by the matching with highly efficient RG flow that reconstructs the gravity theory, as discussed below. Nevertheless, string-theoretic top-down models like Sakai-Sugimoto \cite{Sakai:2004cn,Sakai:2005yt,Rebhan:2014rxa} for QCD, can offer good hints about this holographic theory (more comments below). One can readily show that \cite{Mukhopadhyay:2015smb} the full action (\ref{semiholaction}) has a conserved ${t^\mu_{\phantom{\mu}\nu}}^{\rm full}$ that satisfies:
\begin{equation}\label{tfullcons}
\partial_{\mu}{t^\mu_{\phantom{\mu}\nu}}^{\rm full} = 0
\end{equation}
in flat Minkowski space, when the perturbative fields $A_\mu[Q]$ satisfy the equations of motion obtained from the same full action (\ref{semiholaction}). Furthermore, the holographic action has appropriate initial conditions which makes it a functional of the perturbative fields $A_\mu[Q]$ only, so the full non-perturbative operators like ${t^\mu_{\phantom{\mu}\nu}}^{\rm full}[Q]$ are also functionals of perturbative fields $A_\mu[Q]$.

What remains now is to determine are the functionals $H_{\mu\nu}[A_\mu(Q), \alpha_s(Q)]$ which tell us how the perturbative fields modify the background metric and sources for the holographic theory. This can be simply obtained from the RG flow matching condition
\begin{equation}\label{matching}
{t^\mu_{\phantom{\mu}\nu}}^{\rm full} = {t^\mu_{\phantom{\mu}\nu}}^{\rm Hol}[Q],
\end{equation}
which implies that the dynamics below the scale $Q$ is given by the emergent holographic geometry, that is determined by the highly efficient RG flow that reconstructs the holographic theory. This matching also allows us to determine the functionals $H_{\mu\nu}[A_\mu(Q), \alpha_s(Q)]$ such that the requirement of good endpoint behaviour is satisfied in the infrared. For each single trace operator, we will have a function like $H_{\mu\nu}[A_\mu(Q), \alpha_s(Q)]$ which determines the background source for that operator in the CFT, as in (\ref{semiholaction}) for the case of the energy-momentum tensor, and a matching condition like (\ref{matching}) to determine this function.

Indeed for the matching (\ref{matching}) to work, one needs to have a precise definition of ${t^\mu_{\phantom{\mu}\nu}}^{\rm Hol}[Q]$, which is given by the highly efficient RG flow reconstruction, provided we also specify the choice of the gauge fixing of the $(d+1)-$diffeomorphism symmetry,  that gives us a precise bulk hypersurface and specific coordinates on this hypersurface also. This is automatically determined by the matching condition. Indeed ${t^\mu_{\phantom{\mu}\nu}}^{\rm Hol}[Q]$ is subject to the constraints of the holographic theory which follow from the constraints of $(d+1)-$dimensional diffeomorhism invariant classical gravity theory. Firstly, it follows a conservation equation in a different background metric, while ${t^\mu_{\phantom{\mu}\nu}}^{\rm full}$ is conserved in flat Minkowki space (cf. (\ref{tfullcons})). Secondly, its trace is also determined by the Hamiltonian constraint, therefore it implies the same for ${t^\mu_{\phantom{\mu}\nu}}^{\rm full}$. This simply means that, we should choose the bulk hypersurface $\hat{r}(r, x) = Q^{-1}$ and also the coordinates $\hat{x} = \hat{x}(x)$ on this hypersurface  for evaluating ${t^\mu_{\phantom{\mu}\nu}}^{\rm Hol}[Q]$ carefully, so that the matching (\ref{matching}) can be done successfully while satisfying the $(d+1)-$constraints. Indeed, as discussed in Section \ref{diffeo}, we can implement our RG flow construction on any hypersurface related to an arbitrary radial foliation of spacetime that is determined by the choice of gauge-fixing, or equivalently that of the bulk coordinate system.

If $Q$ is sufficiently larger than $\Lambda_{\rm IR}$, the endpoint of the RG flow which is related to the confinement scale or the temperature scale, then we can expect that the functionals $H_{\mu\nu}[A_\mu(Q), \alpha_s(Q)]$ can have an expansion of the form
\begin{equation}
H_{\mu\nu}[A_\mu(Q), \alpha_s(Q)] = \gamma\left(\Lambda_{\rm IR}/Q\right)\cdot \frac{1}{Q^4} \,t_{\mu\nu}^{\rm W}[A_\mu(Q), \alpha_s(Q)] + \mathcal{O}(Q^{-6}),
\end{equation}
where $\gamma$ can be determined from the matching condition (\ref{matching}). In this case, one can expect our construction to also be an effective theory for both perturbative and non-perturbative dynamics for those states where $\langle t_{\mu\nu}^{\rm W}[Q] \rangle < Q^4$. It then reduces to the phenomenological semi-holographic models of the form proposed in \cite{Iancu:2014ava} with the added advantage that the hard-soft couplings like $\gamma$ can be determined now from first principles (see also \cite{Faulkner:2010tq,Mukhopadhyay:2013dqa} in the context of non-Fermi liquids). 

The interesting point is that the entire non-perturbative definition of the theory, as evident from (\ref{semiholaction}) becomes \textit{manifest} at a scale $Q$, which is a semi-hard scale where non-perturbative effects start getting significant. We can go beyond the case of matching only few single-trace operators as in (\ref{matching}) systematically, by studying $\alpha'(\alpha_s(Q))$ corrections where $\alpha'$ is square of the string length in the holographic theory. Indeed stringy degrees of freedom are necessary to obtain the dynamics of all other single trace operators in the traditional holographic correspondence as well \cite{Aharony:1999ti}.
Once again, although the matching with highly efficient RG flow can determine the holographic theory that describes the infrared, we may be also tempted to appeal to top-down string theoretic models to guess the relevant holographic theory as (consistent truncations of) a holographic closed string theory that reproduces the infrared behaviour of QCD like the Sakai-Sugimoto model \cite{Sakai:2004cn,Sakai:2005yt,Rebhan:2014rxa} with an appropriate UV cut-off. 

It may be interesting to pursue examples like higher spin theories which are holographically dual to Vasiliev like gravity theories \cite{Gaberdiel:2012uj} in this context also, mainly because in this case the field theory is solvable. Thus we may get special insights by learning how the Wilsonian RG and the highly efficient RG flow can be mutually translated in an efficient manner (in this context please see \cite{Douglas:2010rc,Sachs:2013pca,Leigh:2014qca,Mintun:2014gua,Jin:2015aba}). This should be useful in the more generic contexts where we need to match an entire Regge trajectory of operators while passing from the weakly coupled to the strongly coupled side. Perhaps, existence of effective Ward identities that are related to higher spin generalisations of diffeomorphism invariance will turn out to be important.

\subsection{The unknown territory}\label{final}
The major question, to which we do not have any possible answer to at this moment, is how to extend our construction  to the case when $N$ is finite, meaning when the quantum effects should become visible on the gravity side. In fact, this should be a major challenge as this directly leads us to reconstruct quantum gravity with appropriate boundary conditions as a RG flow in a field theory with finite $N$, and perhaps even to resolve some outstanding puzzles \cite{Hawking:1976ra,Mathur:2009hf,Braunstein:2009my,Almheiri:2012rt}. Nevertheless, we expect that the three criteria proposed here should still hold with appropriate modifications.

It is likely that the generalisation of our construction to finite $N$ will use ideas from quantum information theory, specially those related to how we can efficiently coarse-grain quantum information while retaining the knowledge about long range entanglement properties of the system as maximally as possible. Indeed it has been pointed out in recent literature that the features related to quantum entanglement are key to the reconstruction of the dual geometry \cite{Swingle:2012wq,Lashkari:2013koa,MIyaji:2015mia}. We can start from understanding how our three criteria for constructing highly efficient RG flow serve to coarse-grain quantum information efficiently in the large $N$ regime. 

Generalisation of our constructive approach to finite $N$ will also lead us towards developing concrete and novel non-perturbative frameworks for at least a large class of quantum field theories.

\begin{acknowledgments}
We thank B. P. Dolan for collaboration during the initial stages of this project. We thank H. Liu, Y. Nakayama, F. Nitti, F. Preis and A. Rebhan for useful discussions. We also thank Y. Nakayama, G. Policastro, A. Rebhan and O. Rosten for comments on the manuscript, and S. Banerjee and N. Gaddam for collaboration on Section \ref{beta}. N.B. and A.M. acknowledge the ``Research in Groups'' grant sponsored by ESPRC of UK and managed by ICMS, Edinburgh, which has kickstarted the present collaboration. A.M. thanks ICMS and Heriot-Watt University for hospitality at Edinburgh, where a large portion of the work in this project has been done. A.M. also acknowledges support from a Lise-Meitner fellowship
of the Austrian Science Fund (FWF), project no. M 1893-N27. We are dedicating this work as a tribute to the occasion of the 100th anniversary of General Relativity.
\end{acknowledgments}



\bibliography{HolographyAsRGFLOW2-refs}
\end{document}